\documentclass[dissertation, copyright]{uathesis}

\usepackage{amsmath}
\usepackage{amssymb}
\usepackage{braket}
\usepackage{float}
\usepackage{graphicx,color}
\usepackage{epstopdf}
\usepackage{textcomp, gensymb}
\usepackage[utf8]{inputenc}
\usepackage{mathrsfs}
\usepackage{mathtools}
\usepackage{mwe}
\usepackage[numbers,sort&compress]{natbib}
\usepackage[percent]{overpic}	
\usepackage{rotating}
\usepackage[section]{placeins} 	
\usepackage{pdfpages}
\usepackage{physics}
\usepackage[section]{placeins}
\usepackage{subfigure}
\usepackage{tabularx}
\usepackage{tensor}
\usepackage[numbered,autolinebreaks,useliterate]{mcode}	
\usepackage{listings}	
 
\usepackage[pdftex,bookmarks,colorlinks]{hyperref}
\usepackage{cleveref}
\usepackage[numbered]{bookmark}

\usepackage{notoccite}

\usepackage{calligra}
\DeclareMathAlphabet{\mathcalligra}{T1}{calligra}{m}{n}
\DeclareFontShape{T1}{calligra}{m}{n}{<->s*[2.2]callig15}{}

\newcommand{\p}[1]{\, #1 \,} 
\renewcommand{\=}{\, = \,} 
\newcommand{\bk}[1]{\langle #1 \rangle}

\completetitle{Physical Emulation of Nonlinear Spin System Hamiltonians via Closed Loop Feedforward Control of a Collective Atomic Spin}
\fullname{Ian M. Pannemarsh}        
\degreename{Doctor of Philosophy}   

\begin{document}

\maketitlepage
{Department of Physics}   
{2~0~2~5}                 


\includepdf[pages=-]{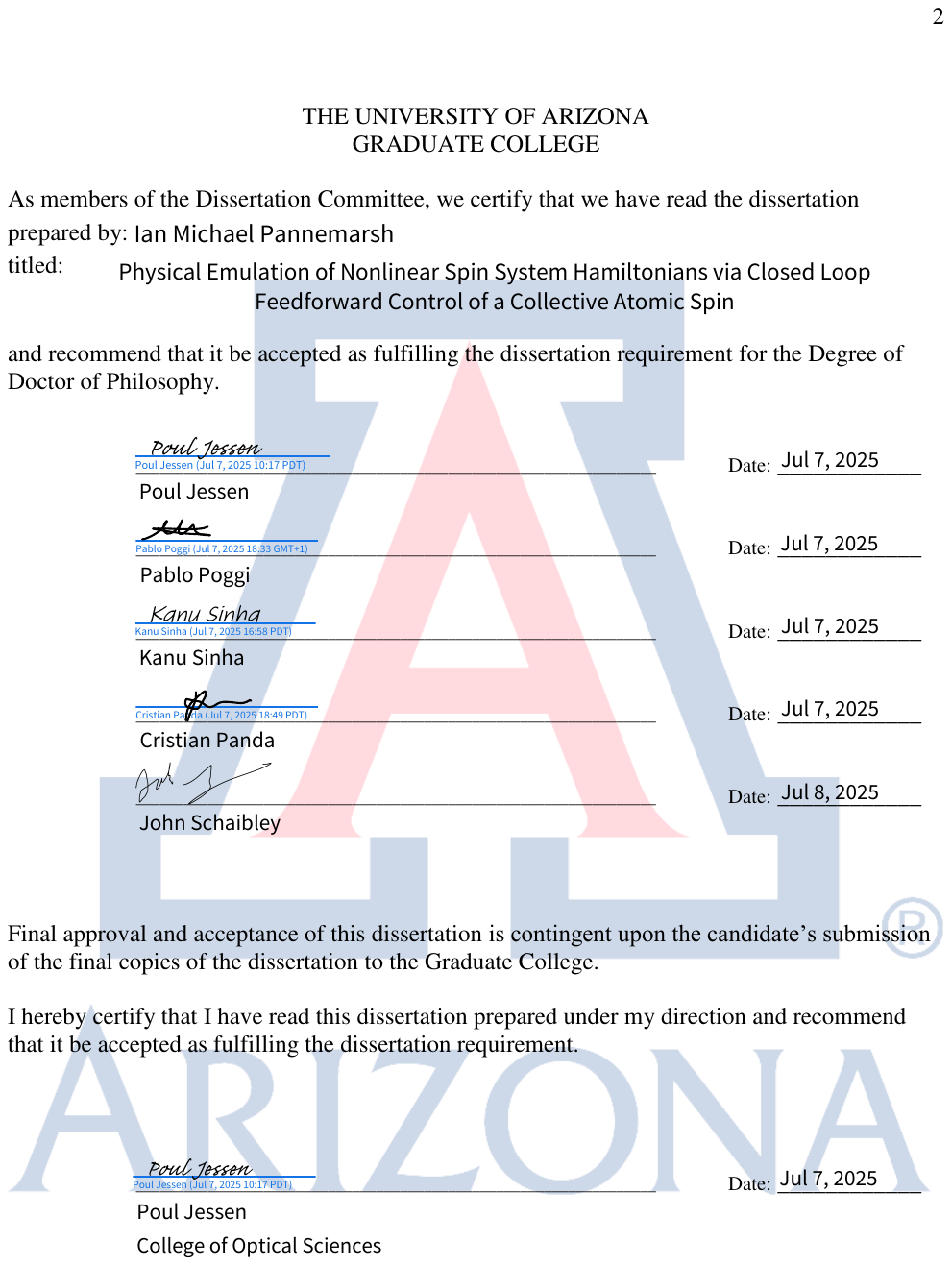}
%


\incacknowledgements{subtex/0b-acknowledgement}

\incdedication{subtex/0c-dedication}

{
\hypersetup{linkcolor=black}	
\tableofcontents

\listoffigures
}


\incabstract{subtex/0d-abstract}

\chapter{Introduction}
\label{chapter:ch1}
\paragraph{}

From its inception, the guiding light of scientific endeavors has always been to understand the physical nature of reality. The history of science is one of technological innovation aimed at achieving this goal. New tools allow greater understanding of our world, affording greater control and the ability to make more accurate predictions. The invention of the computer began an altogether new revolution in scientific study, ushering in what we refer to today as the information age. It enabled the rapid development of numerical simulations of physical systems, impacting nearly all fields of study, from high energy physics, to chemistry, to cosmology. All the while, the processing power of computers has grown exponentially at a steady rate, a trend commonly known as Moore's Law \cite{Moore1965}. This growth, in turn, provides the capabilities for better, faster, and more detailed simulations, driving the wheel of scientific progress. 

Despite this, modern computers still struggle with simulations of large quantum systems. The primary reason for this is the exponential size of the state space of quantum systems: A general quantum system consisting of $n$ 2-state particles has $2^{n}$ complex probability amplitudes that must be stored and manipulated to carry out a fully quantum $n$-body simulation. To make matters worse, modern computer processors are fast approaching hard physical limits on the size and density of their building blocks, transistors \cite{Waldrop2016, Kumar2015FundamentalLT}. 

So, if classical computers cannot cope with the resources required to simulate large quantum systems, why not develop a quantum counterpart? This idea, made famous by Nobel laureate Richard Feynman in 1982 \cite{Feynman1982}, spurred on a burgeoning research effort that was theoretically validated by Seth Lloyd in 1996 \cite{Lloyd1996}. A universal quantum computer, where classical bits are replaced by quantum 2-state systems called \textit{qubits}, can indeed simulate an arbitrary quantum system without worrying about exponential resource overhead, at least in principle. In such a paradigm, the state of the simulated system is encoded into the state of the qubits of the quantum computer platform of choice, and a dynamical simulation is achieved by repeated application of a finite set of universal quantum gates that manipulate the qubits appropriately.

All of that sounds very attractive, and, in addition to quantum many-body simulations, many other potential applications for a universal quantum computer have been explored. Despite this, a fully-digital, fault-tolerant, and scalable quantum computer has yet to be realized. Current quantum processors house on the order of 100 qubits, and while some demonstrations claiming to achieve a "quantum advantage" in some computational task have been reported, it has been argued that we have yet to see any notable applications of quantum computers towards practical problems \cite{BrooksNature2023}. 

So what can we do to answer questions about many-body quantum systems while a universal quantum computer is still out of reach? One option is to construct a physical emulation of that system which can be controlled in the lab. The key difference between computational simulation of a system and physical emulation lies in the former's numerical abstraction of reality, of working with a numerical rather than a physical representation of a system.

Using analogous physical systems to study the dynamical properties of another system is not a new idea. For instance, undergraduates are often taught that one can draw analogy between electrical and mechanical systems because the mathematical description of the components of one system can be mapped to an equivalent component in the other: inductors and inertia, capacitors and springs, resistors and friction, etc. This idea, first proposed by James Clerk Maxwell in the mid-19th century, enables one to use analysis techniques common to one domain to study the other, such as the use of filter analysis in the design of a speaker's vibrating membrane. However, the type of emulation most promising to us is a direct emulation, where we take the system of interest itself, rather than an analogous system, and engineer the influences on it to generate the desired dynamics directly. Moreover, in order to be competitive in the face of early quantum processors, we would like our emulator to be flexible enough that we can easily change the engineered controls, allowing us to study different models by simply ``plugging them in", so to speak. 

The main idea of this work is to experimentally demonstrate that a given nonlinear Hamiltonian system satisfying certain conditions can be emulated by performing a mean field approximation to the nonlinear terms, recasting them as linear interactions that are conditioned on a measurement outcome. This process is referred to as \textit{quantum measurement and feedback} (QMF), and was first explored in \cite{Lloyd2000} for general quantum systems. 

In our case, we are specifically interested in the evolution of a collective spin state of arbitrary size. The application of the QMF protocol to systems described by collective spin variables was first discussed in \cite{Munoz2020_QMFsim} with the goal of applying it to study the emergence of dynamical chaos in a class of models referred to as kicked $p$-spin models, a generalization of the well known Quantum Kicked Top (QKT) model \cite{Haake1987}. In that paper, a numerical model was constructed to study the protocol, which found good agreement between the simulated quantum trajectories and regions of the classical phase space characterized by positive Lyapunov exponents.

The QKT is not a bad place to start if our aim is to provide a demonstration of nonlinear spin system emulation via the QMF protocol. However, the fact that the main point of interest for this model is as a case study for quantum chaos makes it a little less appealing. It seems prudent that the first model we attempt to emulate should be more well behaved, so as to more readily admit comparison between theory and experiment. 

Fortunately, we do not have to look far. The QKT model, which is a stroboscopic, or discrete time system, has a continuous time cousin, the Lipkin-Meshkov-Glick (LMG) Hamiltonian \cite{Lipkin1965}. The interaction terms for the two models are the same, a linear and a quadratic rotation, but in the QKT the nonlinear rotation is a periodic impulse, rather than a continuous torque as it is in the LMG. Those in the quantum computation community see the QKT as a \textit{Trotterized} version of the LMG \cite{Sieberer2019}, wherein the two interactions are handled separately in a quantum circuit, rather than as a single unitary. This is done out of necessity, as the nature of a universal quantum computer makes it so that arbitrary unitary evolution must be constructed by a sequence of a finite set of fundamental logic gates. This perspective has offered insight into the observation of chaos in quantum simulations that use the Trotterization protocol. 

The LMG is a well-known model in its own right, and it has the advantage that the dynamics it produces is fully regular, in spite of its nonlinearity. Moreover, its main characteristic is a symmetry breaking phase transition, wherein one of the fixed points in phase space bifurcates into a pair. Emulating this model would then provide a means to make an experimental observation of spontaneous symmetry breaking in a quantum system.

While the main goal of this work is to demonstrate the QMF technique in a collective spin system, there are other broad questions we are interested in exploring. Our collective spin is realized as a dilute gas of cold neutral atoms, and, as such, we can easily control the number of atoms that make up the collective spin. In a measurement of the spin projection of $n$ atoms, the signal scales linearly with $n$ while the quantum variance scales as $\sqrt{n}$. Therefore, the signal-to-noise ratio also scales as $\sqrt{n}$. At smaller atom numbers, quantum projection noise is more relevant, and so it is a natural thing to ask how the dynamical evolution induced by our emulator is affected when quantum projection noise is more relevant. Having a way to slide between a quantum and a classical system is an attractive thing. For example, the emergence of classical dynamical chaos from quantum systems is still an active field of research, especially regarding finite dimensional many-body systems such as ours \cite{Pavel2018}. For this reason, we attempt to emulate both the LMG and the QKT models on our platform.

The structure of this document is as follows. We will start in Chapter \ref{chapter:ch2} by laying the theoretical foundation for this work, describing the nature of our collective spin and how we can coherently manipulate it. We will lay out the QMF protocol, which naturally leads us to a discussion of the spin projection measurement. We then close out Chapter \ref{chapter:ch2} with a treatment of some problematic side effects of our measurement method which we will have to work to avoid. In Chapter \ref{chapter:ch3}, we will provide comprehensive detail for the major systems that make up our experimental apparatus, as well as our techniques for addressing the aforementioned undesirable side effects of our spin measurement.

In Chapter \ref{chapter:ch4} we lay out, step by step, the way in which we form and purify the initial collective spin coherent state. Chapter \ref{chapter:ch4} also contains a discussion of control errors and our methods for detecting and diagnosing them. The final, and perhaps the most crucial piece of this experiment is discussed in Chapter \ref{chapter:ch5}: the feedback controller and our implementation of the control laws that generate the desired dynamics.

All this will culminate in Chapter \ref{chapter:ch6} with a presentation of the results of our experiments, alongside comparisons to known theory. For the LMG, we will make measurements of some order parameters associated with the dynamical phase transition, and we will set up the conditions in order to make an observation of spontaneous symmetry breaking. For the QKT, we will explore its chaotic aspects by attempting to estimate the maximal Lyapunov exponent, which measures the degree of chaoticity in classical systems. We will also explore a configuration of the model that produces a rather interesting and relatively novel phase of matter: a time crystal \cite{Munoz2022_TC}.

\chapter{Theoretical Background}
\label{chapter:ch2}
\paragraph{}
In this chapter we discuss the theoretical background for the work presented in this thesis. We will begin with a description of the collective spin, including some details concerning the particular species of atom we use, $^{133}$Cs. This leads us to an introduction to the dynamics of collective spin states, with a focus on the nonlinear Hamiltonian models whose dynamics we are interested in emulating. We then discuss the theoretical ideas that allow us to realize those dynamics by combining a continuous weak measurement with carefully designed closed loop feedback conditioned on the measurement outcome. It seems natural then to discuss the details of the measurement itself, which encodes the spin orientation onto the polarization state of probing light, as well as some potential negative side effects of that method that we must work to mitigate.

\section{Physical System}
\label{chapter:ch2p1}
\begin{figure}[ht]
\renewcommand{\baselinestretch}{1}
\centering
\includegraphics[width=0.9\columnwidth]{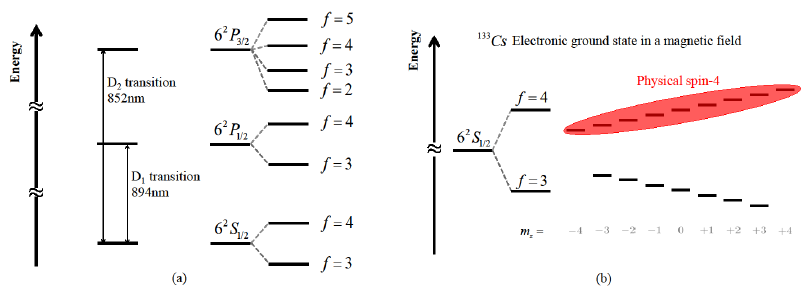}
\caption[Hyperfine structure of $^{133}$Cs]{(a) The hyperfine structure of the ground and 1st two excited states of $^{133}$Cs are shown. (b) A magnetic field splits the magnetic sublevels in the ground state of $^{133}$Cs. The physical spin-4 system comprised of the f=4 hyperfine manifold is highlighted.}
\label{fig:fig2p1}
\end{figure}

\paragraph{}
For this experiment the particular species of atom we use is $^{133}$Cs. As an alkali atom, cesium has only one valence electron and is therefore relatively easy to characterize and work with. Detailed information about its properties can be found in \cite{Steck2010}. We work with the atoms prepared in their electronic ground state, $6^{2}S_{1/2}$, using transitions to the second excited state, $6^{2}P_{3/2}$ to cool and spin polarize the atoms via optical pumping. This transition is commonly referred to as the D2 line. The methods for trapping, cooling, and pumping the atoms are discussed in chapter \ref{chapter:ch3}. When probing the internal spin state of the atoms we use both the first  and second excited state transitions, the former of which $\qty(6^{2}S_{1/2} \, \rightarrow \, 6^{2}P_{1/2})$ is referred to as the D1 line. The relevant energy levels and transitions are shown in Fig. \ref{fig:fig2p1}(a).
	
 The ground state of cesium has an orbital angular momentum of $l = 0$. Note that here we are using lower case letters to refer to the eigenvalues of the associated spin operator, for which we use capital letters. The nuclear spin is $i = 7/2$, and the spin of the valence electron is $s = 1/2$. The total angular momentum in the ground state can therefore take on the values $f = \{3, 4\}$, resulting in two hyperfine manifolds separated by 9.192631770 GHz. This frequency is in fact exact, as it has been used as the definition of the second since 1967, with the most recent revision in 2018 \cite{BIPM2018}. Each hyperfine manifold has $2f+1$ degenerate magnetic sublevels with quantum numbers $m_f = \{-f, -f+1, ..., +f\}$. Throughout this thesis we exclusively work with atoms in the $f=4$ ground state, and so $m_f$ can range between $\pm 4$. 

 In order to spin polarize the atoms, we must lift the degeneracy of these sublevels, which is achieved by applying a constant magnetic field. The field axis, which we take to be the $\hat{z}$ direction, establishes the quantization axis, and so hereafter we refer to this magnetic field as the Bias field. 
	
 For a single atom in the ground state, the internal atomic spin can be described by
	\begin{equation}
		\label{eq:eq2p1_spinstate}
		\ket{\psi} = \sum_{f,m} A_{f,m} \ket{f,m},
	\end{equation}
where $A_{f,m}$ is the probability amplitude for the basis state $\ket{f,m}$ with total angular momentum $f$ and z-component $m$. The commutation relations for the internal spin operators $\hat{\mathbf{f}}$ are given by
	\begin{equation}
		\label{eq:eq2p1_comm}
		\comm{\hat{f_i}}{\hat{f_j}} = i\epsilon_{ijk}\hat{f_k},
	\end{equation}
where $i,j,k$ denote the three orthogonal coordinate axes. Recall that the variance in measurements of an observable $\hat{O}$ is $\Delta\hat{O}^{2} \, = \, \expval{\hat{O}^{2}}{\psi} - \bk{\hat{O}}^{2}$. The Heisenberg uncertainty relation for measurements of the spin operators is then
	\begin{equation}
		\label{eq:eq2p1_heis}
		\Delta\hat{f_i}\Delta\hat{f_j} \geq \frac{1}{2}\abs{\langle\hat{f_k}\rangle},
	\end{equation}
for $i \neq j \neq k$. The equality is symmetrically satisfied for a class of states known as spin coherent states (SCS), which have maximum projection along a particular direction in space. One can think of a SCS as an angular momentum vector pointing in some direction, with symmetric uncertainty in the orthogonal projections. 

In our experiments we typically work with a large collection of individual atomic spins ($\sim 10^6$), which we can describe using the collective operators
	\begin{equation}
	\label{eq:eq2p1_cSpin}
		\hat{F}_{z} \p{=} \sum_{n}^{N} \hat{f}_{z}^{(n)} , 
	\end{equation}
where $\hat{f}_{z}^{(n)}$ is the projection of the $n^{th}$ atom's spin onto the z-axis. 

In addition to demonstrating the feasibility of the feedforward emulation method, one of the things we are interested in making statements about is the effect of quantum noise on our experiments. As such, we must find the variance in measurements of our collective spin operators. For a state symmetric under particle exchange and where all of the atoms have been prepared identically, the variance in measurements of the z-projection of the collective spin operator, which we refer to as the quantum projection noise (QPN), is given by
\begin{equation}
	\label{eq:eq2p1_Fvar}
		\Delta \hat{F}_z^2 = N\Delta \hat{f}_z^2 + N(N-1) \left\langle \Delta \hat{f}_z^{(n)} \Delta \hat{f}_z^{(q)} \right\rangle_{n \neq q}.
\end{equation}
If there are no quantum mechanical correlations between atoms, the second term in \ref{eq:eq2p1_Fvar} vanishes, but it can also become negative for highly entangled ``squeezed" states, leading to reduced QPN.

\section{Magnetic Control of Collective Spin States}
\label{chapter:ch2p2}
\paragraph{}
The collective spin states that we have described in the previous section can be manipulated by the careful application of magnetic fields. As our experiment hinges on our ability to exert accurate control over the atomic spins, here we will briefly review both the classical and quantum pictures of the interaction of atomic spins with magnetic fields.

\subsection{Classical Interaction Picture}
\label{chapter:ch2p2p1}
\paragraph{}
For an atom with magnetic moment $\boldsymbol{\mu}$ in the presence of an external magnetic field $\mathbf{B}$, the magnetic potential is
\begin{equation}
	\label{eq:eq2p2p1_cBpotential}
	U \= -\boldsymbol{\mu}\cdot\mathbf{B}.
\end{equation}
$\boldsymbol{\mu}$ is proportional to the atoms angular momentum, $\mathbf{F}$, given by 
\begin{equation}
	\boldsymbol{\mu} \= \gamma\mathbf{F} = -g_{F}\frac{\mu_{B}}{\hbar}\mathbf{F}
\end{equation}
where $\gamma$ is the gyromagnetic ratio, $g_{F}$ is the Land\'{e} g-factor for a given hyperfine state of the atom, and $\mu_{B} = \frac{ e \hbar}{2 m}$ is the Bohr magneton.

The nature of the dynamics resulting from this interaction depends on the nature of the magnetic field. For the purposes of this work we will consider only fields which are spatially uniform over the atomic ensemble. Although inhomogeneous fields are used in the initial trapping step, for the purposes of control, the fields must be uniform to maintain coherence. In this case, the potential in Eq. \ref{eq:eq2p2p1_cBpotential} depends only on the relative orientation of the atomic spin and the field, resulting in a torque
\begin{equation}
	\label{eq:eq2p2p1_cTorque}
	\mathbf{\tau} \= \boldsymbol{\mu}\times\mathbf{B}
\end{equation}
According to Newton's laws, the torque is equal to the rate of change of $\mathbf{F}$, leading to the equation of motion
\begin{equation}
	\label{eq:eq2p2p1_cEOM_const}
	\frac{d\mathbf{F}}{dt} \= \gamma\mathbf{F}\times\mathbf{B}
\end{equation}
In the simplest case, where $\mathbf{B}$ is constant in time, the atoms will tend to precess around the magnetic field direction at a constant rate $\omega_{L} = \gamma\qty|B|$, called the Larmor precession frequency. On the other hand, for magnetic fields which vary in time Eq. \ref{eq:eq2p2p1_cEOM_const} must in general be solved numerically. 


\subsection{Quantum Interaction Picture}
\label{chapter:ch2p2p2}
\paragraph{}
In the quantum picture, the angular momentum of the atom is replaced by its corresponding operator with discrete values as discussed in Section \ref{chapter:ch2p1}. Mirroring the previous section, we can describe the interaction with magnetic fields by the Hamiltonian
\begin{equation}
	\hat{H} \= g_{F}\frac{\mu_{B}}{\hbar}\hat{\mathbf{F}}\cdot\mathbf{B}.
	\label{eq:eq2p2p2_qHam_spinB}
\end{equation} 
For a constant magnetic field along the $\hat{n}$ direction, this reduces to 
\begin{equation}
	\hat{H} \= g_{F}\frac{\mu_{B}\qty|B|}{\hbar}\hat{\mathbf{F}}\cdot\hat{e} \= \omega_{L}\hat{\mathbf{F}}\cdot\hat{n}.
\end{equation}
According to the Schrodinger equation, the quantum state $\ket{\psi}$ will then evolve under the action of this Hamiltonian as $\ket{\psi(t)} \= \hat{U}(t)\ket{\psi_0}$, where the unitary operator
\begin{equation}
\hat{U}(t) \= e^{-i\hat{H}t/\hbar} \= e^{-i\omega_{L}t\hat{\mathbf{F}}\cdot\hat{n}/\hbar} \= \hat{R}\qty(\omega_{L}t, \hat{n})
\label{eq:eq2p2p2_qRot}
\end{equation}
is the rotation operator acting on the atomic angular momentum state. This mirrors the classical picture, except now it is a quantum state that is being rotated instead of a classical vector. For the class of spin states we are primarily concerned with in this thesis, namely spin coherent states, the correspondence is more direct. A SCS rotated according to Equation \ref{eq:eq2p2p2_qRot} will remain a SCS, and has its mean rotated exactly like its classical counterpart.

\section{Nonlinear Spin System Dynamics}
\label{chapter:ch2p3}
\paragraph{}
In Section 2.2, the interaction between a spin angular momentum $\mathbf{F}$ and a magnetic field in Equation \ref{eq:eq2p2p1_cBpotential} and its quantum counterpart, \ref{eq:eq2p2p2_qHam_spinB}, were both linear in $\mathbf{F}$. Nature is rarely ever so kind as to admit a truly linear system. In fact, systems whose Hamiltonians contain terms that are nonlinear in the spin have proven to be a rich source of interesting physical phenomenon, including phase transitions \cite{Sachdev2011}, spontaneous symmetry breaking \cite{Malomed2016}, out-of-equilibrium phases of matter \cite{Else2016}, and dynamical chaos \cite{Chaudhury2009, Gubin2012}. From a quantum perspective, a term of order $n$ in the spin can be interpreted as representing $n$-body interactions.

There are many models one can explore to study these phenomenon, but a prime example is the Lipkin-Meshkov-Glick (LMG) model and its closely related cousin the Kicked Top (KT). The LMG has a well defined phase transition in which one of the two stable fixed points bifurcates into a symmetric pair of fixed points, with an unstable point taking its place \cite{Lipkin1965}. The KT model, on the other hand, exhibits dynamical chaos in certain regions of its parameter space \cite{Haake1987}. Both models are well-studied and serve as good test-beds for the feedforward emulation method we will demonstrate in this work.

\subsection{The Lipkin-Meshkov-Glick (LMG) Model}
\label{chapter:ch2p3p1}
\paragraph{}
The Hamiltonian for the LMG model is 
\begin{equation}
	\label{eq:eq2p3p1_LMGham}
	\frac{H_{LMG}}{\Lambda} \= -\qty(1-s)J_{x} -\frac{s}{2J}J_{z}^{2},
\end{equation}
where $\Lambda$ is a parameter with units of $[s]^{-1}$ which determines the evolution time scale. Physically, this model is describing a spin system undergoing two rotations. The first is about the $\hat{x}$ axis at a constant rate. The second is a rotation about the $\hat{z}$ axis where the rotation rate is proportional to $J_{z}$. On its own, this  results in a sort of twisting of the phase space about the $\hat{z}$-axis. The dynamical structure of this model is set by $s$ parameter, which is bounded in the interval $\qty[0,1]$. 

To understand the behavior of the model, a good starting point is to determine whether it has any conserved quantities. Since both terms are only rotations of the spin, the total angular momentum $J^{2}$ must be a conserved. As such, the spin will always lie on a sphere of radius $J$,  meaning we only need two parameters to describe the state. An obvious choice are the polar and azimuthal angles, $(\theta, \phi)$, with $J_{X} \= J\cos(\phi)\sin(\theta)$ and $J_{z} \= J\cos(\theta)$. We can calculate a state's time evolution under Equation \ref{eq:eq2p3p1_LMGham} via Hamilton's equations:

\begin{equation} \label{eq:eq2p3p1_LMGeom}
\begin{split}
	\dot{\theta} \= -\frac{\partial H}{\partial \phi} &\= -\qty(1-s)\Lambda J \sin(\theta)\sin(\phi) \\
	\dot{\phi} \= \phantom{-} \frac{\partial H}{\partial \theta} &\= -\qty(1-s)\Lambda J \cos(\theta)\cos(\phi) + s\Lambda J \cos(\theta) \sin(\theta).
\end{split}
\end{equation} 

This constitutes a set of nonlinear coupled differential equations. Analytical solutions to such a system are hard to come by, and so numerical methods are often used. That said, because of the nonlinearity numerical solutions may be subject to chaotic evolution, where numerical error would grow until the solutions are unreliable. We are, in fact, interested in studying the nature of chaotic systems in this context, but it seems prudent that the first model we choose should be well behaved over its parameter space for the sake of benchmarking performance. Fortunately, because the LMG carries no explicit time dependence, there is one more conserved quantity: the total energy. As such, the system is integrable, and the Poincaré–Bendixson theorem guarantees that continuous dynamical systems of dimension 2 or less cannot be chaotic. As such, we can rely on numerical methods. 

\begin{figure}[ht]
\renewcommand{\baselinestretch}{1}
\centering
\includegraphics[width=0.95\columnwidth]{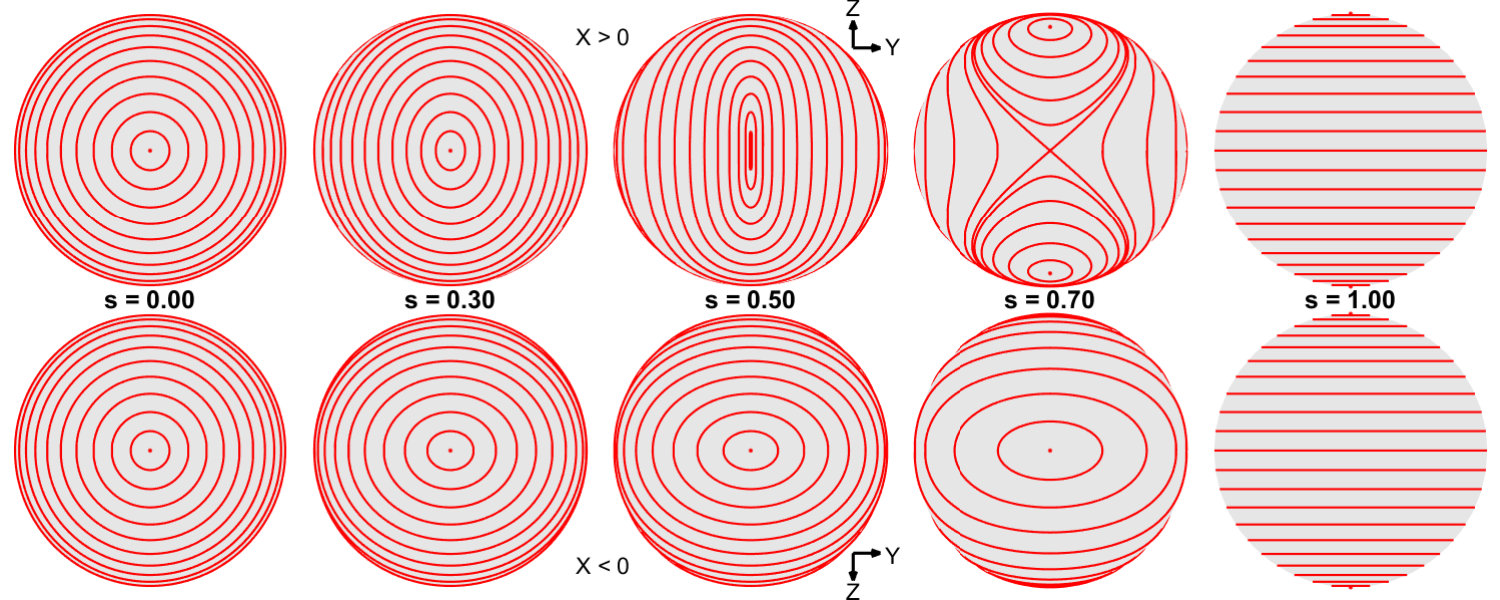}
\caption[Phase space examples for LMG model]{Phase space portraits of the LMG model for several values of $s$. Top row shows the $x \, >\, 0$ hemisphere and bottom shows $x \, <\, 0$. At $s \, =\, 0$ we have pure rotation about $\hat{x}$, and at $s \, =\, 1$ we have pure rotation about $\hat{z}$. In between, there is a dynamical phase transition that changes the distribution of fixed points.}  \label{fig:fig2p3p1_LMG_PhasePlots}
\end{figure}

For a given value of $s$, we can visualize the solutions to Equations \ref{eq:eq2p3p1_LMGeom} by plotting the orbits $(\theta(t), \phi(t))$ for a number of different initial conditions. Some examples of these phase space diagrams for different values of $s$ are given in figure \ref{fig:fig2p3p1_LMG_PhasePlots}. Some notable features of the phase space are immediately apparent. We can see that for small $s$, the constant rotation about $\hat{x}$ dominates, yielding two stable fixed points at $\pm J\hat{x}$. Such points constitute states which do not change over time. A fixed point is stable when nearby points in the phase space remain in a localized neighborhood, orbiting periodically. 

\begin{figure}[htp]
\renewcommand{\baselinestretch}{1}
\centering
\includegraphics[width=0.8\columnwidth]{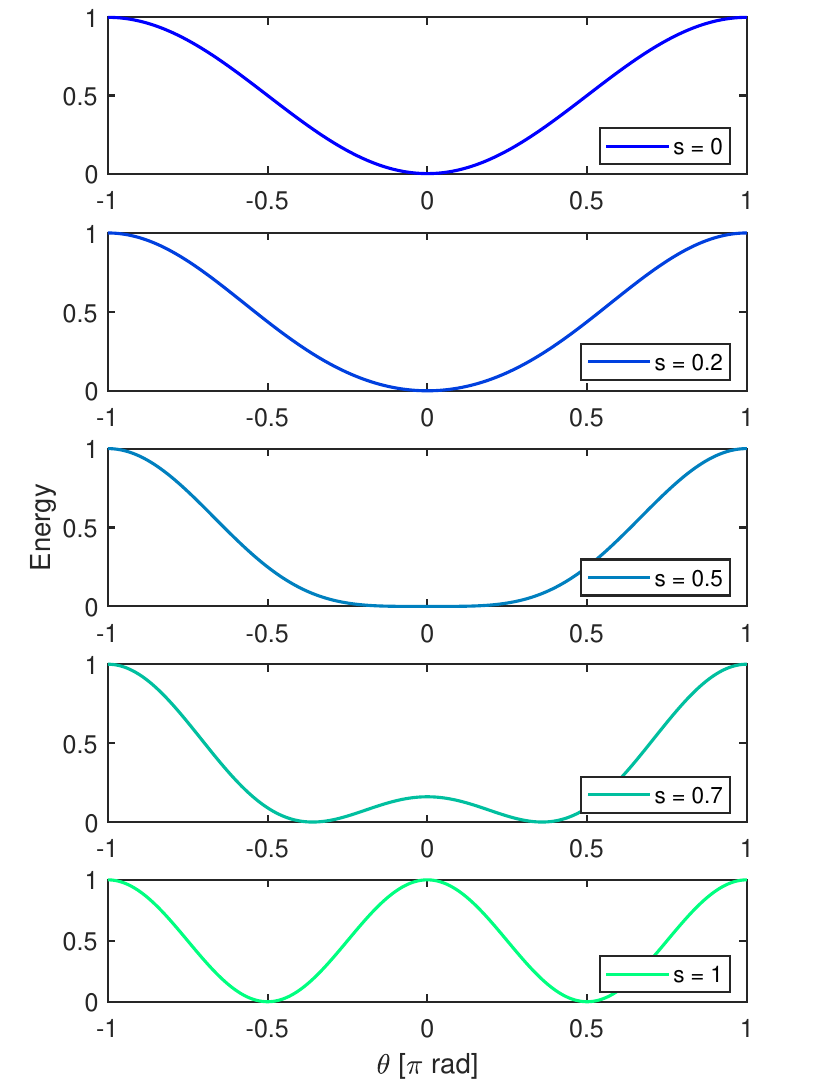}
\caption[Classical energy for LMG Hamiltonian]{Classical energy for the  LMG model along ($\phi$ = 0, $\theta$) for several values of $s$. Each plot has been shifted and scaled to lie in the range [0, 1]. Note that for $s > 0.5$ the global minimum splits into a pair of local minima. A state initially prepared at ($\phi$ = 0, $\theta$ = 0) would undergo spontaneous symmetry breaking in the presence of noise.} \label{fig:fig2p3p1_LMGenergy}
\end{figure}

On the other hand, when $s > 0.5$ the stable fixed point at $+ J\hat{x}$ bifurcates into a pair of stable points located symmetrically about the XY plane, with an unstable fixed point in its place. The precise locations of the new fixed points can we determined by considering the total energy along the great circle with $\phi \= 0$,
\begin{equation} \label{eq:eq2p3p1_LMGenergy}
	\frac{E(\theta; s)}{J\Lambda}  \= -\qty(1-s)\sin(\theta) -\frac{s}{2}\cos^{2}(\theta).
\end{equation}
\indent The energy is plotted in Figure \ref{fig:fig2p3p1_LMGenergy} for several different values of $s$. Taking the derivative of Equation \ref{eq:eq2p3p1_LMGenergy} with respect to $\theta$ and setting it equal to 0, we find that our fixed points satisfy
\begin{equation} \label{eq:eq2p3p1_LMGfixPts}
	\sin(\theta) \= \frac{\qty(1-s)}{s},
\end{equation}
which has real solutions only for $s > 0.5$, as expected.

\FloatBarrier
\subsection{The Kicked Top (KT) Model}
\label{chapter:ch2p3p2}
\paragraph{}
The Kicked Top Hamiltonian has much the same form as the LMG, with one notable difference: the continuous nonlinear rotation is broken up into a periodic train of instantaneous impulses. It is given by \cite{Haake1987}
\begin{equation}
	\label{eq:eq2p3p2_KTham}
	H_{KT} \= -\frac{\alpha}{\tau}J_{x} -\frac{k}{2J}J_{z}^{2}\sum_{n}^{\infty} \delta \qty( t - n\tau ),
\end{equation}
where $\alpha$ and $k$ are angles in radians, and $\tau$ is the kick periodicity. It is also possible to think of this model as applying an alternating train of x- and z-rotations; in a time $\tau$, the spin is rotated by $\alpha$ about $\hat{x}$, and then after the impulse is has been rotated by $\dfrac{kJ_{z}}{2J}$ about $\hat{z}$. We can also see this by looking at the corresponding quantum evolution operator
\begin{equation}
	\label{eq:eq2p3p2_Uqkt}
	\hat{U}_{QKT} = \exp[i\frac{k}{2J}\hat{J}_{z}^{2}] \, \exp[i\alpha\hat{J}_{x}].
\end{equation}

\begin{figure}[h]
\renewcommand{\baselinestretch}{1}
\centering
\includegraphics[width=1.0\columnwidth]{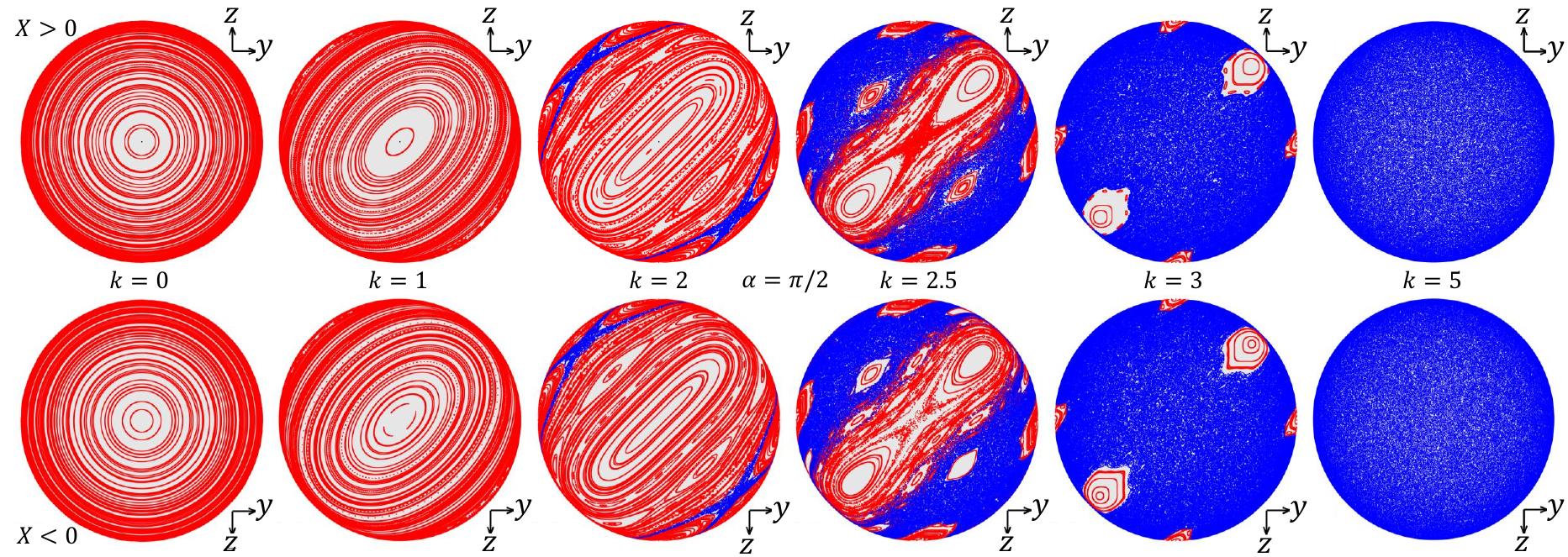}
\caption[Phase space examples for KT model]{Phase space portraits of the KT model with $\alpha = \pi/2$ for several values of $k$. Top row shows the $x \, >\, 0$ hemisphere and bottom shows $x \, <\, 0$. Regular orbits are colored red and chaotic orbits are blue.}  \label{fig:fig2p3p2_KTphaseEx}
\end{figure}

It therefore makes sense to look at the dynamics of this model stroboscopically, rather than continuously. As such, the time evolution is best described by a discrete map $P\qty[\mathbf{X}_{i}]$, where $\mathbf{X} = \mathbf{J}/J$ the normalized spin vector. This map, known as a Poincar\'{e} map because of its area-preserving properties, is found by simply applying the two corresponding rotation matrices successively. It is given by \cite{Munoz2021_pSpin}
\begin{align}
	\label{eq:eq2p3p2_KTeom}
	X_{m+1} &\=           - \sin(k W_{m})\qty[\cos(\alpha)Y_{m} + \sin(\alpha)Z_{m}] + \cos(k W_{m})X_{m}, \\
	Y_{m+1} &\=  \phantom{-}\cos(k W_{m})\qty[\cos(\alpha)Y_{m} + \sin(\alpha)Z_{m}] + \sin(k W_{m})X_{m}, \\
	Z_{m+1} &\= -\sin(\alpha)Y_{m} + \cos(\alpha)Z_{m},
\end{align}
where $W_{m} = \qty[\cos(\alpha)Z_{m} - \sin(\alpha)Y_{m}]$. We show in Figure \ref{fig:fig2p3p2_KTphaseEx} some example phase space portraits of the KT with $\alpha \, =\, \pi/2$. Trajectories corresponding to regular, periodic motion are colored red, while chaotic trajectories are colored blue. For these plots, the distinction is made by calculating the spectral entropy for the z-component of each trajectory, which is a measure of the broadness of the frequency content of a signal \cite{Ray2016, bajkova2025, kathpalia2023}. It is given by 
\begin{equation}
	S \= \frac{1}{\ln(n_{\nu})} \sum_{\nu} \mathcal{F}\qty[Z_{m}] \ln(\mathcal{F}\qty[Z_{m}]),
	\label{eq:eq2p3p2_specEntropy}
\end{equation}
where $\mathcal{F}\qty[Z_{m}]$ is the discrete Fourier transform, and the sum is taken over the frequency content of the transform spectrum. Periodic motion will have finite support on only a few frequencies, which results in a low spectral entropy. Chaotic motion, on the other hand, tends to have broad frequency content due to its irregular character, and therefore has a high spectral entropy. 

Note that due to the stroboscopic nature of the evolution, a single trajectory consists of a set of disconnected points. However, a set of these points may, over many iterations, fill in a line (or a manifold consisting of several disconnected regions). Points on a manifold are mapped to points on the same manifold, but nearby points are not necessarily consecutive in time. Because of this fact it is possible for nearby manifolds to intersect. The intersection points $\mathbf{x}_{int}$ belong to both manifolds and must therefore map either to themselves or to another point of intersection between the two manifolds. As the map is area-preserving, the regions between intersecting manifolds  must also be mapped to each other. If the intersection point happens to be an unstable fixed point, points on the manifold begin to bunch up as they approach it, and the boundary of a region of phase space that lies partially on the manifold will begin to spread out, creating longer and thinner loops due to area preservation. These loops eventually become distorted, spreading the points inside its boundary across phase space in a way that appears stochastic, leading to chaos \cite{Tabor1989}. 
	
As we increase $k$, new fixed points appear and bifurcate throughout phase space, leading to more and more intersections. Eventually, the chaotic region spreads throughout the entire phase space. On small scales, we can still find individual fixed points in chaotic regions. However, in the presence of noise such points are irrelevant, as we must consider the evolution of regions of phase space.

\FloatBarrier
\section{Emulation of Nonlinear Dynamics via Measurement and Feedback}
\label{chapter:ch2p4}
\paragraph{}
In this work, we wish to explore a method of directly emulating the dynamical evolution of a given Hamiltonian model by using the outcome of a weak measurement of the system state to condition the subsequent unitary evolution of that system, a process referred to as quantum measurement and feedback (QMF). This technique was first explored in \cite{Lloyd2000} for general quantum systems. 

We are, of course, interested in the evolution of a collective spin state of arbitrary size. The specific application of the QMF technique to systems described by collective spin variables was first discussed in \cite{Munoz2020_QMFsim} with the goal of applying it to study the emergence of dynamical chaos in the QKT model. In that paper, a numerical model was constructed to study the protocol, which found good agreement between the simulated quantum trajectories and regions of the classical phase space characterized by positive Lyapunov exponents. 

Here we will give an overview of the QMF protocol in the context of our collective spin system. Consider an ensemble of N noninteracting systems described by the collective spin operators as defined in Eq. \ref{eq:eq2p1_cSpin}. The state is initially prepared in the SCS $\ket{J_{\hat{e}_{\mathbf{n}}}}$, i.e. spin polarized along some direction given by $\hat{e}_{\mathbf{n}} \leftrightarrow (\theta, \phi)$. The protocol is performed through repeated application of a two-step process:
\begin{enumerate}
	\item[(i)] A nonprojective measurement of the $\hat{J}_{z}$ component of the collective spin with measurement outcome $m$, followed by
	\item[(ii)] the application of a unitary map $\hat{U}\qty[f(m)]$ which is conditioned by the measurement outcome. 
\end{enumerate}

This technique should work for any Hamiltonian as long as we limit ourselves to operators associated to the physical influences we can exert over the system, and the nonlinear terms are associated with measurable quantities. The measurement type considered in \cite{Munoz2020_QMFsim} is a nonprojective measurement with Gaussian noise, which can be described by the Kraus operators
\begin{equation}
	\label{eq:eq2p3_kraus}
	\hat{K}_{m} \= \frac{1}{\qty(2\pi\sigma^{2})^{1/4}} e^{-\frac{\qty(\hat{J}_{z}-m)^{2}}{4\sigma^{2}}},
\end{equation}
where $\sigma$ is the measurement resolution. A given outcome $m$ is sampled with probability $P_{m} = \tensor[_i]{\ev{\hat{K}_{m}^{\dagger}\hat{K}_{m}}{\Psi}}{_i}$, where $\ket{\Psi}_{i}$ is the state of the system at the $i$-th iteration of the process.

In the second step, the state is updated according to the quantum Bayes rule \cite{Jacobs2010},
\begin{equation}
	\label{eq:eq2p3_qfmUpdate}
	\ket{\Psi}_{i+1} \= \frac{1}{\sqrt{P_{m}}} \, \hat{U}\qty[f(m)] \, \hat{K}_{m}\ket{\Psi}_{i}.
\end{equation}
The function $f(m)$, called the feedback policy, can in fact be chosen freely, and this choice determines the dynamical evolution. The art of this technique is then to find an $f(m)$ that will reproduce some target Hamiltonian.

In the case of the LMG and QKT models, a simple way to achieve this is to linearize the Hamiltonian through a mean-field approximation to the nonlinear terms. We replace the nonlinear terms with their expected values so that $\hat{J}_{z}^{2} \rightarrow \bk{\hat{J}_{z}}\hat{J}_{z}$. In the case of the LMG, for example, the mean-field Hamiltonian is
\begin{equation}
	\Hat{H}_{LMG}^{MF} \= -\gamma\qty(1-s)\hat{J}_{x} - \frac{\gamma\, s}{2}\frac{\bk{\hat{J}_{z}}}{\bk{\hat{J}}} \hat{J}_{z},
\end{equation}
and then the corresponding unitary would be 
\begin{equation}
	\label{eq:eq2p3_qfm_ULMG}
	\Hat{U}\qty[f_{LMG}(m)] \= \exp[i\qty(\gamma\qty(1-s)\hat{J}_{x} + \frac{\gamma\, s}{2}\frac{m}{\bk{\hat{J}}} \hat{J}_{z})\frac{t}{\hbar}],
\end{equation}
where we have replaced $\bk{\hat{J}_{z}}$ with the measurement outcome $m$. This approximation is valid because the large system size ($\sim 10^{6}$) means we are working near the classical limit. More generally, this technique will work with any Hamiltonian whose nonlinear terms are polynomial in the chosen measurement operator, which in our case is $\hat{J}_{z}$.

We can see that Eq. \ref{eq:eq2p3_qfm_ULMG} now appears to consist of two simultaneous linear rotations, one about $\hat{x}$ and one about $\hat{z}$. The first is constant, as before, but the strength of the second rotation is modulated by the measurement outcome. A natural way to realize this is with a feedback controller that modulates the strength of a magnetic field along the $\hat{z}$ direction according to some real-time measurement of $\hat{J}_{z}$.

We will describe in detail the properties of the feedback controller we use in this experiment in Chapter \ref{chapter:ch5}, along with details concerning the implementation of the LMG and QKT models. The only other piece of this puzzle is the nature of the measurement itself, which we will discuss now.

\section{QND Spin Measurements via the Faraday Interaction}
\label{chapter:ch2p5}
\paragraph{}
As we saw in the previous section, in order to perform our emulation experiments we must be able to measure some quantity of our system without damaging the coherence of its initial state. In our case, we make use of the Faraday interaction between the collective spin and the polarization state of a probe laser. With this, we are able to make measurements of the projection of the collective spin onto the propagation direction of the probe, which we set to be the $z$-axis. Moreover, we operate in the limit of low photon scattering rate, where the probe intensity is much lower than the saturation intensity ($I/I_{sat} \ll 1$), and the probe detuning is much greater than than the natural linewidth ($\Gamma/\Delta \ll 1$), which qualifies this as a quantum non-demolition (QND) measurement. 

To get a sense for the nature of the measurement, we will first consider a 1D homogeneous model for the atom-light interaction \cite{Vasilyev2012, Kupriyanov2005, Smith2003}, in which a linearly-polarized plane wave propagating along the $z$-direction passes through an ensemble of atoms with homogeneous density $\rho$. Depending on the atomic polarizability and the projection of the collective spin onto the $z$-direction, the ensemble exhibits a degree of birefringence between left- and right-handed circular polarization. The difference in the refractive indices is given by the diagonal terms of the atomic polarizability tensor in spherical coordinates,
	\begin{equation}
		n_{+} - n_{-} \= \rho \frac{\alpha(\Delta)}{6\epsilon_{0}} \frac{\bk{\hat{F}_{z}}}{F} ,
	\end{equation}
where $\alpha(\Delta) \= -3\epsilon_{0}\lambda^{3}/8\pi^{2}\Delta$ is the scalar polarizability for a two-level atom. The interaction imposes a differential phase shift between the circular components of the probes polarization, given by
	\begin{equation}
		\phi \= \frac{C^{(1)}}{2} \frac{\sigma_{0}}{A} \frac{\Gamma}{\Delta} \bk{\hat{F}_{z}} ,
	\end{equation}
where $\sigma_{0} = 3\lambda^{2}/2\pi$ is the on-resonant interaction cross-section and A is the cross sectional area of the probe. The full tensor coefficients $C^{(k)}$ are derived from the Wigner-Eckart theorem \cite{Deutsch2010}, but in the limit of large detunings, for a probe tuned near the D2 transition, $C^{(1)} \approx -1/(3f)$. This phase shift, of course, results in a net rotation of the probe polarization. 

We can think of this process as entangling the atomic spin state with the probe polarization. In fact, if we represent the probe polarization by its Stokes vector $\hat{\mathbf{S}}$ we can see this explicitly in the unitary that describes the Faraday effect \cite{Deutsch2010, Hammerer2010},
	\begin{equation}
		\label{eq:eq2p4_Ufarad}
		\hat{U}_{Faraday} \= \exp(-i \chi \hat{F}_{z} \hat{S}_{3}),
	\end{equation}
which tells us that the Stokes vector will rotate about the 3-axis by an angle $\chi \bk{\hat{F}_{z}}$, which we interpret as linearly polarized light being rotating through that same angle. 

\begin{figure}[H]
\centering
\includegraphics[width=0.9\columnwidth]{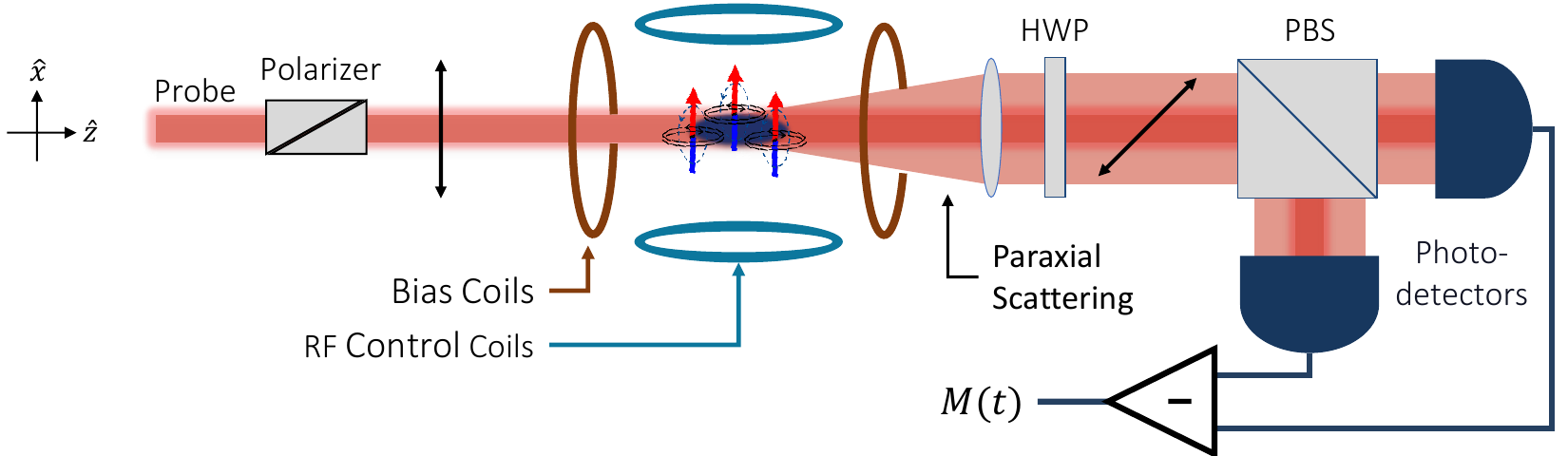}
\caption[Diagram of polarimeter for Faraday measurement ]{Schematic of polarimeter used to measure the $z$-component of the collective spin. A linearly polarized probe laser is passed through the atomic ensemble parallel to the bias magnetic field. After passing through the beam is collimated and its polarization rotated so that we are measuring in a 45 degree basis when there are no atoms present. The orthogonal polarization components are split using a polarizing beamsplitter, and the optical power is measured for each beam. The probe polarization rotates by an angle proportional to $J_{z}$, and the difference in power gives a measurement of the net rotation.}
\label{fig:fig2p4_polarimeter}
\end{figure}

After the interaction, the probe is split into two orthogonal polarization components by passing through a polarizing beamsplitter, which are each then sent into a pair of balanced photodetectors to measure the difference in optical power between them. The beamsplitter is oriented in a $45^{\circ}$ basis relative to the probes initial polarization so that there would be equal power in each beam if one of three conditions are met: there are no atoms present, they are uniformly unpolarized (known as a maximally mixed state, see Section \ref{chapter:ch4p5}), or the mean polarization is orthogonal to the probe. A diagram of this arrangement is shown in figure \ref{fig:fig2p4_polarimeter}.

For small rotations of the probe polarization, the measurement outcome of the detector is given by 
	\begin{equation}
		\label{eq:eq2p4_meas1d}
		\mathcal{M}_{F} \p{\approx} P\phi \= P \chi \bk{\hat{F}_{z}} 
	\end{equation}
where $P$ is the total probe power and $\qty|\chi| = \frac{\sigma_{0}}{6 f A} \frac{\Gamma}{|\Delta|}$ is the measurement strength. We can see that $\mathcal{M}_{F}$ is proportional to the mean value of the z-projection of the collective spin, and therefore serves as a measurement of the same. 

This simplified model, while instructive, does not accurately describe the experiment qualitatively. To do so, we must extend to a 3D model of paraxial scattering which accounts for the necessarily inhomogeneous atom distribution and probe intensity profile \cite{Enrique2015, Baragiola2014}. The atom cloud occupies an ellipsoidal region of space with its long axis along the z direction, whose spatial density profile is well approximated as Gaussian. The probe is also Gaussian, and the contribution of each atom to the collective spin measurement depends on the local probe intensity. The geometry of the cloud is chosen such that the spatial mode of the collective atomic radiation matches closely with the probe field mode, which promotes better atom-light coupling.

The probe is focused so that its waist lies at the center of the cloud, and is initially polarized in the x-direction. We can write this field as
	\begin{equation}
		\vec{E}_{probe}(\vec{r}) \= \vec{\epsilon}_{x} \, E_{0} \, \mathcal{U}_{00}(\vec{r}) \, e^{ikz}
	\end{equation}
where $\mathcal{U}_{00}(\vec{r})$ is the TEM$_{00}$ mode of the probe. Through the Faraday effect, some probe photons are scattered into the orthogonal polarization in the y-direction, with each atoms contribution dependent on both its spin projection along $z$ and the local probe intensity. Moreover, because the probe is far off resonance and the density of the trapped atoms is low, the overall change to the number of photons in the original polarization mode is small. Therefore we can approximate it as being unchanged by the interaction, and so we can write the total probe field as
	\begin{equation}
		\vec{E}_{total}(\vec{r}\,) \p{\approx} \vec{\varepsilon}_{x}E_{probe}(\vec{r}\,)+\vec{\varepsilon}_{y} E_{scatter}^{(y)}(\vec{r}\,) 
	\end{equation}
	
with 
	\begin{equation}
		\label{eq:eq2p4_scat}
		E_{scatter}^{(y)}(\vec{r}\,)\propto \left\langle \sum_n E_{probe}(\vec{r}_{n}\,) \hat{f}_{z}^{(n)} \right\rangle.
	\end{equation}
	
As before, the probe field intensity is measured differentially in the $45^{\circ}$ basis by the polarimeter, which is equivalent to a homodyne detection of the scattered field $E_{scatter}^{(y)}$ with the probe field as the local oscillator. This measurement is given by	
	\begin{equation}
		\label{eq:eq2p4_meas3d}
		\mathcal{M}_{F} = 2 Re \qty{ \int d^2 \vec{r}_{\perp} \; E_{probe}^*(\vec{r}\,) E_{scatter}^{(y)}(\vec{r}\,) } ,
	\end{equation}
integrated over the area of the detector.
	
	We can write the measurement in terms of individual spins by combining Equations \ref{eq:eq2p4_scat} and \ref{eq:eq2p4_meas3d}, then performing the integration, as
	\begin{equation}
	\label{eq:eq2p4_meas3dbeta}
		\mathcal{M}_{F} \= P \chi \sum_n \beta(\vec{r}_n) \bk{ \hat{f}_z^{(n)} } ,
	\end{equation}
	
where $\beta(\vec{r}_{n}) = \frac{I(\vec{r}_{(n)}}{I_{max}}$ is a weighting factor that depends on the relative local probe intensity. Comparing Equations \ref{eq:eq2p4_meas3dbeta} and \ref{eq:eq2p4_meas1d}, we can can see that we have defined the collective spin operator as a weighted sum over the individual atoms,

 	\begin{equation}
 		\label{eq:eq2p4_F3d}
		\hat{F}_z \= \sum_n \beta(\vec{r}_n) \hat{f}_z^{(n)}.
	\end{equation}

If we have prepared all individual spins identically, we have $\hat{f}_z^{(n)} = \hat{f}_z$, and so we can define an effective number of atoms:
	\begin{equation}
	\label{eq:eq2p4_N1eff}
		N_{eff}^{(1)} \= \sum_n \beta(\vec{r}_n)
	\end{equation}
	
Our polarimetry measurement scales with $N_{eff}^{(1)}$, and when we talk about the number of atoms in our collective spin we are referring to this weighted sum. 

 We can rewrite Equation \ref{eq:eq2p1_Fvar} in terms of the weights to find the variance in measurements of $\hat{F}_z$:
	\begin{equation}
		\Delta F_z^2 \= \sum_n \beta^2(\vec{r}_n) \bk{ \qty(\Delta \hat{f}_z^{(n)})^2 } + \sum_{n\neq m}\beta(\vec{r}_n)\beta(\vec{r}_m) \bk{ \Delta \hat{f}_z^{(n)} \Delta \hat{f}_z^{(m)} }. 
	\end{equation}
	
As in Equation \ref{eq:eq2p4_N1eff}, if all atoms are identical and there are initially no quantum correlations between atoms, we can reduce this to 
  	\begin{equation}
  		\label{eq:eq2p4_F3dVarN2eff}
		\Delta F_{z}^{2} \= \qty(\sum_n \beta^2(\vec{r}_n)) \Delta f_{z}^{2} \= N_{eff}^{(2)} \Delta f_{z}^{2},
	\end{equation}

	The quantum projection noise is then determined by this new effective atom number, $N_{eff}^{(2)}$. Note that $N_{eff}^{(2)} \leq N_{eff}^{(1)} \leq N$, with both equalities being satisfied only if the probe intensity is uniform over the atoms. Also, for a fixed probe-cloud geometry, the ratio $N_{eff}^{(2)} \,/\, N_{eff}^{(1)}$ is independent of the overall probe intensity, allowing us to express both the measurement and its variance in terms of $N_{eff}^{(1)}$. 
	
Each measurement M is an average of the Faraday rotation signal over some time T. We can describe a single measurement outcome as the sum of three terms:
	\begin{equation}
	\label{eq:eq2p4_measout}
		\mathcal{M} \= \mathcal{M}_{F} \p{+} \mathcal{M}_{QPN} \p{+} \mathcal{M}_{SN}
	\end{equation}

The first is the measurement of the collective spin as in Eq. \ref{eq:eq2p4_meas3dbeta}. The second and third terms are normally distributed stochastic quantities with zero mean and variances given by 
	\begin{equation}
		\label{eq:eq2p4_qpn}
		\Delta\mathcal{M}_{QPN}^{2} \= P^{2} \chi^{2} \Delta F_z^2
	\end{equation}
and
	\begin{equation}
		\label{eq:eq2p4_sn}
		\Delta\mathcal{M}_{SN}^{2} \= \frac{\hbar \omega}{T} P
	\end{equation}
This latter term represents the contribution of intensity noise from the probe light, also known as shot noise (SN). In our experiments we perform continuous weak measurements throughout the emulation at a specified rate $1/T$. This process generates some backaction onto the state, generating atom-atom entanglement which localizes the state in z up to the shot noise uncertainty while also squeezing out the uncertainty in x and y. One consequence is that while the QPN only plays a role in the initial measurements, the SN contributes equally to each point in the measurement record.

\section{Light Shift Hamiltonian}
\label{chapter:ch2p6}
\paragraph{}
	The optical fields used to trap and probe the atoms in this experiment can have damaging effects on the prepared collective states if care is not taken to mitigate them. Full details of these so-called light shifts can be found in \cite{Deutsch2010}, but due to the specifics of our experiment, we can express the light shift Hamiltonian as a combination of three distinct terms,
	\begin{equation}
	\label{eq:eq2p35}
	\begin{split}
			\hat{H}_{\text{LS}} &= \sum_{j'f'} V \left[ \Bigg \{ C_{j'f'f}^{(0)} + C_{j'f'f}^{(2)}\frac{\hat{f}^2}{6} \Bigg \} - \Bigg \{ C_{j'f'f}^{(1)} \frac{|\epsilon_l|^2-|\epsilon_r|^2}{|\epsilon_l|^2+|\epsilon_r|^2} \hat{f}_z \Bigg \} - \Bigg \{ C_{j'f'f}^{(2)}\frac{\hat{f}_z^2}{2}  \Bigg \}  \right] \\
			&= \hat{H}_{\text{SLS}} + \hat{H}_{\text{VLS}} + \hat{H}_{\text{TLS}}, 
	\end{split}
	\end{equation} 
which we call the scalar, vector, and tensor light shifts, respectively. Here, $f$ and $f'$ refer to the ground ($f = {3, 4}$) and excited  ($f' = {2, 3, 4, 5}$) state manifolds, respectively, while $j'$ refers to either the D1 ($j' = 1/2$) or the D2 ($j' = 3/2$) lines. $V$ is the ac-Stark shift associated with an optical field of intensity $I$ acting on a transition with unit oscillator strength and saturation intensity $I_{sat}=2 \pi^2 \hbar c \Gamma / 3 \lambda^3$, given by
	\begin{equation}
			V= \left( \frac{\hbar \Gamma}{8} \frac{I}{I_{sat}} \right) \frac{\Gamma}{\Delta_{j'f'f}}.
	\end{equation}	
As discussed previously, we have taken the probe propagation direction to be the $\hat{z}$-axis, and we have assumed that, due to the presence of the large bias magnetic field parallel to the probe, the $\hat{f}_x$ and $\hat{f}_y$ components average to zero over the measurement duration. 

In the limit of large detunings as compared to the hyperfine splitting, the tensor coefficients $ C_{j'f'f}^{(n)} $ for probes tuned near the D1 and D2 transitions can be approximated by
	\begin{align}
	\label{eq:eq2p5_LS_coeffs}
	\begin{split}
		D_1: & \quad C_{D1}^{(0)}=\frac{1}{3}, \quad C_{D1}^{(1)} = +\frac{1}{3f}, \quad C_{D1}^{(2)} =\frac{\beta_{D1} \Gamma_{D1}}{\Delta_{D1}}, \quad \beta_{D1}>0 ,	 \\
		D_2: & \quad C_{D2}^{(0)}=\frac{2}{3}, \quad C_{D2}^{(1)} =- \frac{1}{3f}, \quad C_{D2}^{(2)} =\frac{\beta_{D2} \Gamma_{D2}}{\Delta_{D2}}, \quad \beta_{D2}<0 .
	\end{split}	
	\end{align}

The three terms in \ref{eq:eq2p35} all have distinct influences on the magnetic sublevels which need to be carefully understood and managed. In section \ref{chapter:ch3p5} we will go into more detail regarding the specific strategies we have developed to eliminate them when possible, but for now we will give a brief conceptual description of their effects, starting with the scalar light shift. 

The strength of the SLS depends on the optical intensity and detuning, and the quantum number $f$. It therefore acts on all magnetic sublevels equally, serving as a trapping potential. In the case of the dipole trap (Section \ref{chapter:ch3p3}) this is exactly what we are taking advantage of to hold the atoms. The probe, however, also induces a small scalar light shift, exerting a force on the atoms which can lead to center of mass motion of the atom cloud. This is problematic for two reasons. First, if any of the magnetic fields used for control are inhomogeneous, as the atom cloud moves about it will experience different splittings at different times, leading to dephasing and control errors. Second, as we have seen the local intensity of the probe is very much inhomogeneous across the trapping region. If the atoms are sloshing about we may see it as a small oscillation in the measurement, giving us an inaccurate estimate of the real-time length of the collective spin. 

The form of the vector light shift (VLS) is analogous to an interaction between the spin and a magnetic field along z. It is proportional to the quantity $(|\epsilon_l| - |\epsilon_r|)$, which is the difference in amplitude between the left- and right-handed circular components to the probe polarization, characterizing its degree of ellipticity. If it is non-zero, then in the rotating frame, the VLS will manifest as a ficticious magnetic field, detuning the level splitting. Again, due to the inhomogeneous laser intensity profile, atoms in different regions will precess at different rates, leading to inhomogeneous rotations of the prepared spin coherent state which we would see as increased classical noise, as well as a faster rate of decay.

The tensor light shift (TLS) is actually similar in structure to the nonlinear term in the LMG and KT models. In fact, in the past it has been used to induce KT dynamics in a fully quantum regime at the level of the $F=3$ hyperfine ground state of cesium \cite{Chaudhury2009}. However, for our purposes its presence is undesirable because it leads to a much faster rate of decay. In essence, since the unitary associated with the tensor component induces a rotation of the SCS which is proportional to $\bk{J_{z}}$, under its action the prepared spin coherent state will shear about the equatorial plane of the Bloch sphere. As the uncertainty patch spreads out, the mean length of the spin decreases. Moreover, like the VLS it depends on the local intensity of the probe and so is inhomogeneous across the cloud, amplifying the effect.

\chapter{Experimental Apparatus and Techniques}
\label{chapter:ch3}
\paragraph{}

In this chapter we will discuss the fundamental platform on which this experiment is based. Previously we discussed the details of the particular species of atom, $^{133}$Cs, that we use as the core of what we refer to as the physics package, as well as the basic theory underpinning the atom-light interaction. Here we will describe the methods used to trap and cool the atoms, as well as techniques used to benchmark and diagnose problems with the trap. We will then provide details regarding the trapped atom cloud itself. Next we discuss the polarimeter, which we use to measure the collective spin polarization of the ensemble of atoms. Finally, we will discuss our methods for eliminating harmful aspects of the atom-light interaction, known as light shifts. 

\section{Laser Trapping and Cooling}
\label{chapter:ch3p1}
\paragraph{}
The first step in the experiment is, of course, to trap the atoms. There is a rich literature related to techniques for trapping and cooling of atoms  \cite{Raab1987,Wineland1979,Migdall1985,Lett1989,Kuppens2000,OHara2001}, and more details for the approaches used by our group can be found in \cite{Enrique2015,SosaThesis2017,LeeThesis2012,AndersonThesis2013,SmithThesis2012, SoumaThesis2008,OThesis2008}. As such, we will not go into extreme detail regarding this part of the experimental apparatus. At the heart of the experiment package is a glass vacuum cell containing a dilute gas of Cs. The atoms are initially room-temperature and held at a pressure of $~8\times10^{-8}$ Torr. We use a standard magneto-optical trap (MOT) architecture of three crossed, counter-propagating, circularly polarized lasers, overlapped in the center of a magnetic quadrupole field created by a pair of circular coils in an anti-Helmholtz configuration. The MOT beams consist of two light sources: the primary laser, red-detuned from the $f=4 \rightarrow f'=5$ transition by just over one linewidth (6 MHz), and a second, much weaker laser tuned to $f=3 \rightarrow f'=4$, which we call the repumper. Typically the MOT beams would be mutually orthogonal, but out of necessity from space constraints the two lateral beams are at a shallower angle to each other relative to the measurement z-axis set by the bias. This results in a somewhat flattened MOT, but this is not a problem because the MOT is only used for initial trapping and cooling of the atoms.

\begin{figure}[H]
\centering
\includegraphics[width=0.9\columnwidth]{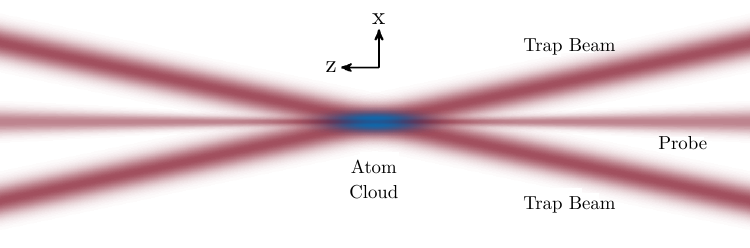}
\caption[Diagram of the far off-resonance crossed dipole trap and probe arrangement]{Relative sizes of the dipole trap, probe, and trapped atom cloud (blue). The waists of the two arms of the dipole trap cross at a shallow angle near $6^{\circ}$, and the probe passes directly through the middle, coplanar with the two beams. Note that for visibility, the x and z axes are not to scale.} \label{fig:fig3p1}
\end{figure}

\indent Instead, to hold the atoms during our experiments we use a dipole force trap consisting of two high power, far off-resonance (1064 nm) laser beams \cite{Miller1993,Takekoshi1996}, which are focused and aligned to have their waists overlap at the center of the MOT at a shallow relative angle of $6^{\circ}$. Based on previous work by our group \cite{Enrique2015}, the geometry of the dipole trap is designed to maximize the optical depth of the atomic cloud when probed by a laser coplanar to the crossed trapping beams, passing midway between them. A diagram of this arrangement is shown in figure \ref{fig:fig3p1}. The power in each beam is 15 W, and the $1/e^2$ waist diameter is $140 \mu m$. This results in a trapping potential of $200 \mu K$.

\indent Between experimental cycles, the MOT is enabled for 1 second, during which time atoms are being loaded into the trap. When a cycle begins, the dipole force trap is turned on, and atoms are loaded for an additional 145 ms. The second step is to put the atoms through optical molasses to cool them down. The quadrupole field turns off, and, using acousto-optical modulators, the primary MOT beam is increasingly red-detuned in a sequence of steps lasting 10 ms. We have found empirically that an additional sequence of pulses of MOT light, lasting 25 ms total with a period of 3.5 ms and a duty cycle of 0.1, considerably improves loading and cooling of atoms into the dipole force trap. At the end of this sequence we are left with a trapped cloud consisting of $2\times 10^7$ atoms at a temperature on the order of $15 \mu K$.

\section{Time of Flight and Stern-Gerlach Measurements}
\label{chapter:ch3p2}
\paragraph{}
In order to determine the temperature of the atoms, we use time-of-flight (TOF) measurements \cite{Salomon1990} in which the atoms are dropped from the trap and eventually fall through a flat laser beam resonant with the $4 \rightarrow 5'$ D2 transition, producing fluorescence which is then picked up by a photodetector. The photodetector signal will be a convolution of the cross-sectional profile of the flat beam and the instantaneous shape of the atom cloud as it passes through, which is expanding due to the finite temperature of the atoms. In principle, the integrated photodetector signal is a measurement of the total number of trapped atoms in the $f = 4$ manifold, and the width of the signal about the peak is a measurement of their temperature. In order to calibrate these measurements, we must determine the initial size of the cloud before it is dropped, which can be done by direct imaging on a camera which is sensitive to 852 nm light, as well as the thickness of the TOF beam sheet, which is wide enough to be laterally homogeneous. 

A full analysis of the expected form for the TOF signal can be found in \cite{Yavin2002ACO} and \cite{Hagman2009}, but the essential arguments for determining the temperature goes as follows. Assume the atom cloud distribution is a Gaussian ellipsoid with $1/e^2$ radii $\sigma_{i}$, where $i \in {x,y,z}$, and let the T be the temperature of the atoms. The most probable velocity of the trapped atoms is 
\begin{equation} \label{eq:eq3p2p1}
v_{mp} \p{=} \sqrt{\frac{2k_{B}T}{m}},
\end{equation}
where $m$ is the atomic mass and $k_{B}$ is Boltzmann's constant. Once the atoms are dropped, the cloud will begin to expand as it falls. The average size of the cloud grows as $\sigma_{i}(t) = \sigma_{i} + v_{mp} t$. The cloud's center of mass moves under free fall, and if it is initially at a distance $D$ above the center of the TOF beam, its position is given as $y(t) = D - 0.5gt^2$. We can then describe the atomic density with the expression
\begin{equation} \label{eq:eq3p2p2}
\rho(x',y',z'; t) \p{=} N \exp(-\frac{x'^2}{2\sigma_{x}^2(t)})\exp(-\frac{(y'-y(t))^2}{2\sigma_{y}^2(t)})\exp(-\frac{z'^2}{2\sigma_{z}^2(t)}),
\end{equation}
The intensity of the TOF beam is given by 
\begin{equation} \label{eq:eq3p2p3}
I(x',y',z'; t) \p{=} I_{0} \exp(-2\frac{2y'^2}{\omega^2}),
\end{equation}
where $\omega$ is the $1/e^2$ diameter of the beam in the vertical direction. Note that we have assumed that compared to the cloud the TOF beam is infinitely wide and long. This allows us to integrate out the x and z parts of the distribution. The total fluorescence will be a convolution of equations \ref{eq:eq3p2p1} and \ref{eq:eq3p2p2} with a combined temporal width of 
\begin{equation}
\label{eq:eq3p2p4}
\sigma_{TOF} \p{=} \frac{\omega^2}{4} + \sigma_{y}^2(t),
\end{equation}
\indent Of course, since the width of the cloud is growing as it passes through the TOF beam, this is not a true Gaussian in time, but if the temperature is small enough relative to the speed of the center of mass when it passes through the TOF beam and the initial size of the cloud is small compared to its initial height, we can approximate it as a Gaussian with width given by substituting in  $t_{arrival} = \sqrt{2D/g}$. A measurement of the TOF signal width can then be converted back into an estimate for the temperature. 

If we wish, we can also measure the population in the $f=3$ manifold by first blowing away $f=4$ atoms through radiation pressure from the same MOT light as before, but with no detuning from resonance. Once the blow pulse is finished, we turn the repump light on, pumping the $f=3$ atoms back up to 4. Typically, we use this technique to diagnose the efficiency of the optical pumping step of our state preparation procedure, discussed in chapter \ref{chapter:ch4}

\begin{figure}[H]
\centering
\includegraphics[width=0.7\columnwidth]{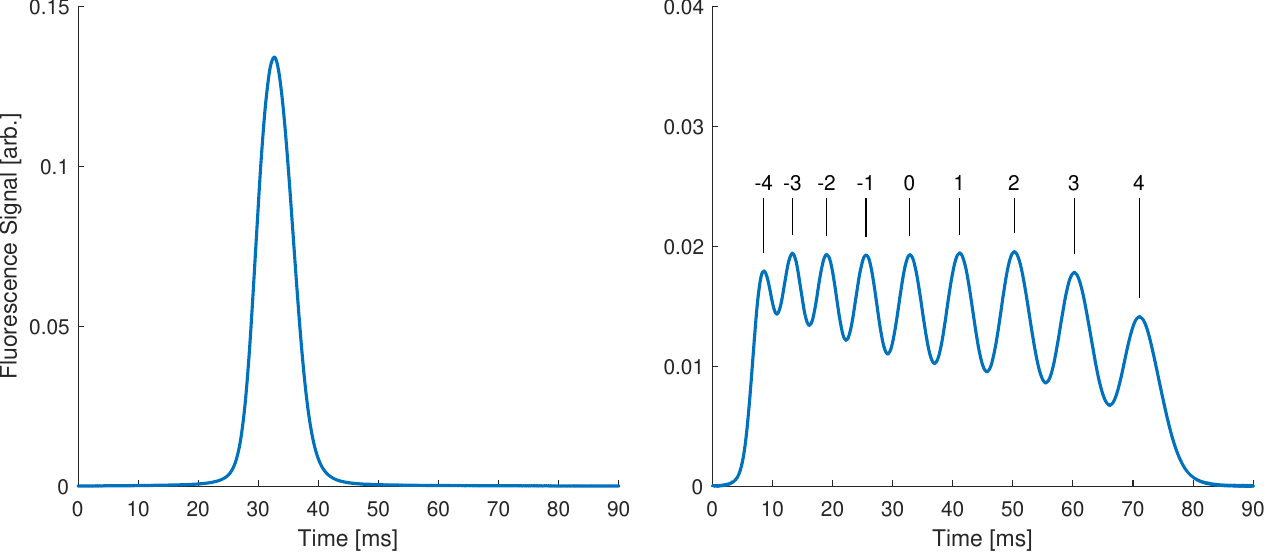}
\caption[Examples of Time-of-Flight measurements of trapped atom cloud]{Examples of TOF measurements of atom cloud after being dropped from the trap, normalized to the total area under each signal after subtracting the background offset. a) The width of the curve is related to the temperature of the atom cloud, and the total area is proportional to the number of atoms. b) TOF signal after separating the atomic populations with a Stern-Gerlach magnetic pulse. Each peak is labeled by its magnetic quantum number.} \label{fig:fig3p2}
\end{figure}

\indent We can also measure the relative populations in the magnetic sublevels in the ground state manifolds by performing a Stern-Gerlach measurement \cite{Marechal1998,Chormaic1994}. After the atoms have been dropped and any state preparation steps have completed, we drive a strong current pulse through the MOT coils, setting up a magnetic field gradient that the atoms fall through. The atoms each feel a force which is dependent on the magnetic sublevel $m$ it occupies as well as the local magnetic field gradient. The net result is that the different populations of atoms pass through the TOF beam at different times.

\section{Dipole Trap \& Atom Cloud Characterization}
\label{chapter:ch3p3}
\paragraph{}
One of the important features of this experiment is the ability to precisely control the magnitude of the collective spin, which is proportional to the number of atoms that participate in the measurement. In order to do this, we need precise knowledge of the geometry not only of the dipole trap and the probe lasers, but of the distribution of the trapped atoms as well. As mentioned in section \ref{chapter:ch3p1}, previous work by our group went into developing a model of the interaction of the probe and trapped atoms \cite{Enrique2015}, with the goal of maximizing the optical depth (OD) of the trap. 

\begin{figure}[H]
\centering
\includegraphics[width=0.7\columnwidth]{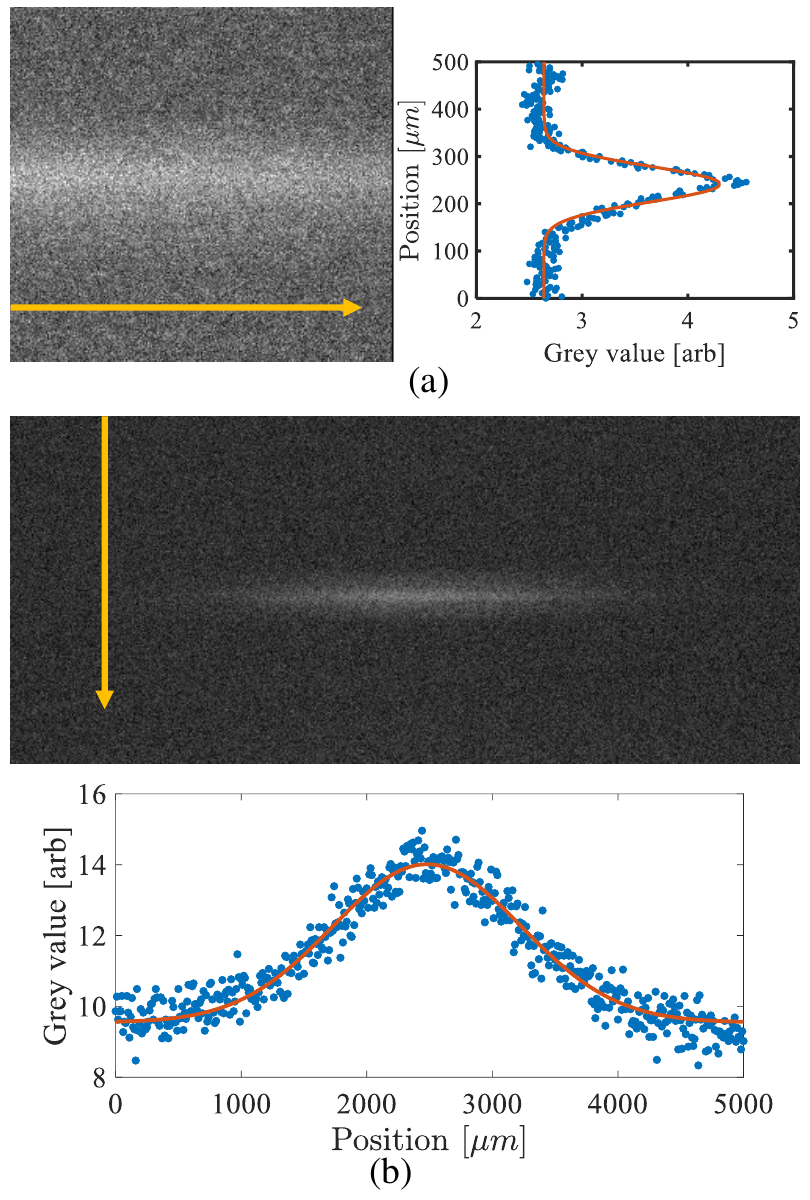}
\caption[Measurement of the size of the trapped atom cloud]{Average of 20 images of the atom cloud held by the dipole trap. The grey value of each image is averaged along the direction indicated by the arrow, and the resulting data is fit to a Gaussian profile. (a) Transverse profile, taken near the center of the lower image. (b) Longitudinal profile. Figure obtained from \cite{HemmerThesis2020}.} \label{fig:fig3p3a}
\end{figure}

\indent The cloud is expected to be a Gaussian ellipsoid, with the long axis oriented along the $\hat{z}$-direction. By calibrated imaging of the atom cloud using the scattered fluorescence (see Figure \ref{fig:fig3p3a}), we have measured the $1/e^{2}$ dimensions of the cloud to be 1500 $\mu$m along the long axis with a 75 $\mu$m radius in the transverse directions. Since the trap and probe beams are fixed in space, this geometry stays consistent from run to run.

\indent Another important aspect of the dipole trap is the lifetime of the atoms held in it, since it affects the maximum duration of the experiments we can perform. There are two main factors that affect this: collisions with background gas not held in the trap, and heating from trap light scattering off the atoms. We can measure the lifetime by dropping the atoms out from the trap at different times and performing a TOF measurement to determine the relative atom number. The number of atoms in the trap decays exponentially, with a $1/e$ time constant on the order of 200 ms, which is at least 2 orders of magnitude greater than the typical timescale of a single experiment.

\pagebreak
\indent The final aspect of the trap that can affect the quality of our experiments is motion of the atoms in the trap. The atoms are typically cooled to a temperature which is an order of magnitude smaller than the maximum depth of the trap. They are tightly confined in the radial direction, and in this regime the trap potential is approximately that of a harmonic oscillator. If the intensity of the trapping light is modulated at a frequency double that of the characteristic trap frequency $\omega_{0}$, we can excite parametric resonances of the motion, heating up the atoms and therefore causing them to escape the trap. Then via spectroscopy, the trap frequency has been measured to be $\omega_{0} = 2\pi \times 177$ Hz.

\begin{figure}[ht]
\centering
\includegraphics[width=1.0\columnwidth]{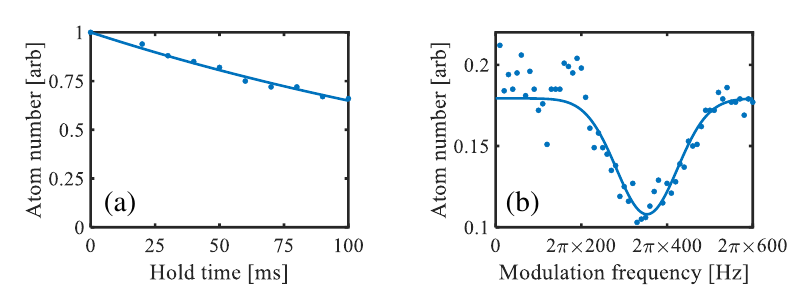}
\caption[Measurements of the dipole trap lifetime and characteristic frequency]{(a) Measurement of the lifetime of atoms held in the dipole trap. By varying the length of time atoms are held for and then measuring the magnitude of the resulting TOF signal, the lifetime is estimated to be approximately $230$ ms. (b) When the trap intensity is modulated at frequencies near double the characteristic trap frequency, the atoms heat up and leave the trap more readily. Figure obtained from \cite{HemmerThesis2020}.} \label{fig:fig3p3b}
\end{figure}

\indent Knowing this frequency allows us to place further limits on the timescales on which the atoms are usable for experiments. The maximum velocity of an atom oscillating in the trap will be on the order of $v_{max} = \omega_{0}r_{\perp}$, where $r_{\perp}$ is the $1/e^2$ transverse radius of the cloud. The time taken to traverse the probe region by an atom moving with this velocity is approximately $d/v_{max}$, where $d$ is the $1/e^2$ diameter of the probe. With a probe waist of 25 $\mu$m, the timescale for atomic motion in the trap is roughly $600 \mu$s. A typical feedforward experiment lasts for around 1-2 ms, so this estimate might be concerning, but we can account for this effect to some degree since the net effect is a slow and gentle ripple in the measurement signal that is consistent from run to run. As we shall see in chapter \ref{chapter:ch5}, the control signal is modulated in such a way that it accounts for the instantaneous magnitude of the collective spin. In principle, we can simply use a record of the average polarimetry signal when the spin is oriented up along z, but for the results in this thesis we have found it to be sufficient to account for the exponential decay alone.

\section{Polarimetry}
\label{chapter:ch3p4}
\paragraph{}
The primary measurement type used in this experiment is what allows us to estimate the projection of the collective spin onto the z-axis. We direct linearly polarized light from the probe through the center of the trapped atom cloud along the z-axis, and the Faraday-effect interaction with the atoms causes the probe polarization to rotate by an angle proportional to the z-component of the SCS. After exiting the cloud and the vacuum cell, the probe light passes through through an optical filter which removes any stray 1064 nm light from the dipole trap. A pair of lenses focuses the probe beam appropriately, and then it is passed through a polarizing beam splitter. A half-wave plate is placed before the PBS and is oriented such that the input polarization of the probe is at an angle of 45 degrees when no atoms are present in the trap. The two beams, consisting of the horizontal and vertical polarization components of the original probe, are sent into the two input photodetectors of a Thorlabs PDB450A balanced differential amplifier. 

The detector output is proportional to the difference in power between the two polarization components, which in turn is a measurement of the net rotation of the probe polarization. The gain of the detector is set to its highest setting, a factor of $10^{7}$. At this gain setting, the bandwidth of the detector is roughly 100 kHz. The detector is shot noise limited up to at least 1 mW of incident power, and in fact its electronic noise floor is nearly 30 dB lower than the probe shot noise at the typical probe powers used ($\sim$ 20-30 $\mu$W). 

\begin{figure}[ht]
\centering
\includegraphics[width=0.85\columnwidth]{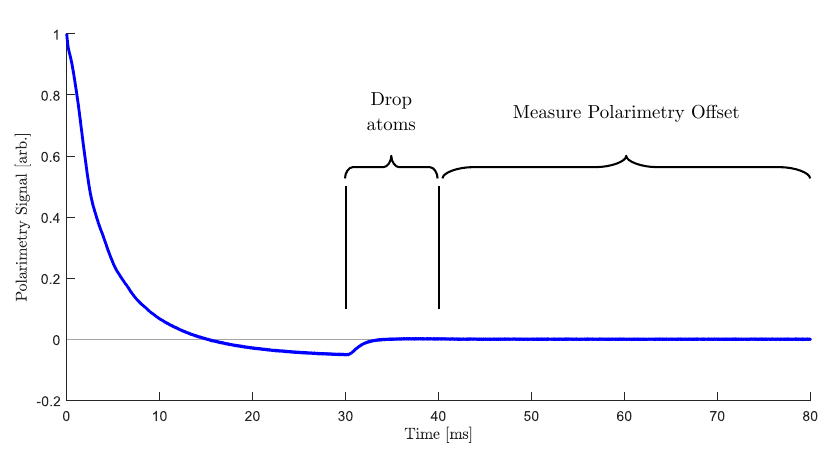}
\caption[Example of polarimetry measurement with spin up along z]{Example of a polarimetry signal when the collective spin is up along z. The atoms are present during the initial 30 ms, and are slowly decaying to the f=3 ground state. Experiments take place during the initial 1.5 ms. At 30 ms the atoms are dropped, and at 40 ms the dipole trap light is turned back on and the polarimetry signal is averaged to find the signal offset.} \label{fig:fig3p4}
\end{figure}

The polarimetry signal is routed to two different places. For closed loop experiments, the raw signal is sent directly to the FPGA controller, which will be discussed in detail in chapter \ref{chapter:ch5}. For all other purposes it is first sent through a Krohn-Hite 3362 4-pole Butterworth filter set to 200 kHz, and then is recorded by our National Instruments based data acquisition system. An example of a typical polarimetry signal can be seen in figure \ref{fig:fig3p4}.

\section{Light Shift Elimination Strategies}
\label{chapter:ch3p5}
\paragraph{}
In section \ref{chapter:ch2p5}, we discussed the theory behind problematic aspects of the interaction between the atoms and the light used for trapping and probing. As we saw, the light shift Hamiltonian can be grouped into three separate terms, each representing a different effect on the atomic energy levels. These three terms are identified by the tensor rank of the corresponding operator: scalar (rank 0), vector (rank 1), and tensor (rank 2). In this section, we will discuss the effects of each term on the collective spin state and the strategies we have developed to mitigate or altogether eliminate them.

\subsection{Scalar Light Shift}
\label{chapter:ch3p5p1}
\paragraph{}
The problematic center of mass motion of the atoms induced by the SLS is essentially an impulse of radiation pressure imparted to the cloud when we turn on the probe light. To avoid it, all we need to do is to turn the probe on adiabatically so that the atoms do not oscillate. This does slightly reduce the maximum usable atom number, since the probe light does cause the spin coherent state to decay while it is ramping, but only by a very small amount ($< 1\%$). We have found that ramping on the probe over a period of 1.5 ms is sufficient to mitigate any center of mass oscillation.

\subsection{Vector Light Shift}
\label{chapter:ch3p5p2}
\paragraph{}
The vector light shift (VLS) is perhaps the most important of the three as both the dipole trap and probe beams can contribute to the total shift, but fortunately it is also the most straightforward to eliminate. Recall that, as discussed in Chapter \ref{chapter:ch2p2}, the VLS is proportional to the degree of ellipticity in the optical field under consideration. If there is any ellipticity in the light polarization, then, in the co-rotating frame, the VLS will manifest as a ficticious magnetic field which becomes a detuning contribution to the level splitting. 

\indent The solution is then to make the probe and trap beam polarizations as linear as possible at the atoms. For the dipole trap, we use flat dielectic plates which we can tilt to adjust the beam polarization, and for the probe, we use a Glan-Thompson polarizing cube from ThorLABS. However, we cannot use these elements on their own, because the glass vacuum cell has some residual stress built up in the walls, resulting in birefringence. We therefore need to use atoms themselves to indicate when the VLS is completely nulled. 

\indent To eliminate the VLS from the dipole trap, we perform a series of microwave spectroscopy measurements both in and out of the trap field. Comparing these measurements allows us to estimate not only the VLS, but the SLS as well. Let $\omega_{clk}$ be the clock frequency for the ground states of cesium, $\ket{f=3, m_{f} = 0} \leftrightarrow \ket{f=4, m_{f} = 0}$. In the presence of an optical field, the clock frequency is changed by the scalar light shifts to both ground state manifolds to 
\begin{equation} \label{eq:eq3p5p1p1}
\omega_{clk}' \,=\, \omega_{clk} + \delta_{3}^{(0)} + \delta_{4}^{(0)} \,=\, \omega_{clk} + \Delta^{(0)},
\end{equation}
where $\Delta^{(0)}$ is the net SLS. If there is a constant magnetic field of amplitude $B$, the level splitting for ground state manifold $f$ is given by
\begin{equation} \label{eq:eq3p5p1p2}
\Delta_{f} \,=\, \frac{g_{f}\mu_{B}}{\hbar}B + \delta_{f}^{(1)}
\end{equation}
where $\delta_{f}^{(1)}$ is the VLS averaged over the atom cloud. For maximum sensitivity, we will perform spectroscopic measurements on the two extremal microwave transitions: the 'stretched' states, ( $\ket{4, +4} \leftrightarrow \ket{3,+3}$ ), and its counterpart which we refer to as the 'squished' states, ( $\ket{4, -4} \leftrightarrow \ket{3,-3}$ ). These transitions are so named because the g-factor for the two ground state manifolds are equal in magnitude but opposite in sign, so in the bias the stretched states spread apart while the squished states are brought closer to each other in energy. 

\indent Setting aside nonlinear effects for the moment, the shifted frequency for the stretched state is given by
\begin{align} \label{eq:eq3p5p1p3}
\begin{split}
\omega_{stretch}' \,&=\, \omega_{clk}' + 4\Delta_{4} - 3\Delta_{3}  \\
	&=\, \omega_{clk} + \Delta^{(0)} + 4(\frac{\mu_{B}}{4\hbar}B + \delta_{4}^{(1)}) - 3(\frac{-\mu_{B}}{4\hbar}B + \delta_{3}^{(1)}) \\
	&=\, \omega_{clk} + \Delta^{(0)} + 7(\frac{\mu_{B}}{4\hbar}B) + (4\delta_{4}^{(1)} - 3\delta_{3}^{(1)}) \\
	&=\, \omega_{clk} + \Delta^{(0)} + 7\Delta_{B} + \Delta^{(1)},\\
\end{split}
\end{align}
where $\Delta_{B}$ is the level splitting from the bias and $\Delta^{(1)}$ is the total shift from the VLS. Likewise, for the squished state we can show that
\begin{equation} \label{eq:eq3p5p1p4}
\omega_{squish}' \,=\, \omega_{clk} + \Delta^{(0)} - 7\Delta_{B} - \Delta^{(1)}. 
\end{equation}

Combining Equations \ref{eq:eq3p5p1p3} and \ref{eq:eq3p5p1p4} by both adding and subtracting them, we find  
\begin{align} \label{eq:eq3p5p1p5}
\begin{split}
\omega_{stretch}' + \omega_{squish}' \,&=\, 2(\omega_{clk} + \Delta^{(0)}), \\
\omega_{stretch}' - \omega_{squish}' \,&=\, 2(7\Delta_{B} + \Delta^{(1)}). \\
\end{split}
\end{align}

We can also measure the unshifted frequencies by turning the trap off, then immediately performing spectroscopy before the atoms fall too far. In this case, Equations \ref{eq:eq3p5p1p5} reduce to
\begin{align} \label{eq:eq3p5p1p6}
\begin{split}
\omega_{stretch} + \omega_{squish} \,&=\, 2\omega_{clk}, \\
\omega_{stretch} - \omega_{squish} \,&=\, 14\Delta_{B}. \\
\end{split}
\end{align}

By combining the four measurements of the shifted and unshifted frequencies (Eqns. \ref{eq:eq3p5p1p5} and \ref{eq:eq3p5p1p6}, respectively), we find that the SLS and VLS are given by 
\begin{align} \label{eq:eq3p5p1p7}
\begin{split}
\Delta^{(0)} \,&=\, \frac{1}{2}((\omega_{stretch}' + \omega_{squish}') - (\omega_{stretch} + \omega_{squish})) \\
\Delta^{(1)} \,&=\, \frac{1}{2}((\omega_{stretch}' - \omega_{squish}') - (\omega_{stretch} - \omega_{squish})) \\
\end{split}
\end{align}

\indent Based on this idea, we can prepare the atoms in either of the $\ket{4,\pm4}$ states, make the four spectroscopic measurements in rapid succession via TOF, estimate the light shifts according to \ref{eq:eq3p5p1p7}, make an adjustment to the dipole trap polarizers if necessary, and then iterate the process until the VLS has been satisfactorily nulled. We find that the scalar light shift is typically on the order of 400 Hz red detuned for the roughly 30 W combined dipole trap, which agrees with direct calculation from the light shift Hamiltonian given in \ref{eq:eq2p35}, and the vector light shift for each beam can be nulled to within a few tens of Hz. 

\indent This works well for the trap despite it being very far off resonance because the trap intensity is so high. Unfortunately it does not work for the probe, not only because the probe power is 3 orders of magnitude lower, but because the probe light is confined to a region roughly 1/3 the size of the trap, so the spectroscopic signal using the TOF is much smaller.

\indent Instead, we rely on the fact that the VLS will be inhomogeneous across the cloud, resulting in dephasing of the collective spin under continuous rotations. We drive the atoms with continuous RF during polarimetry, resulting in a decaying oscillatory signal. The VLS will cause this oscillation to damp faster, and so we simply rotate the polarizer until the coherence time is maximized. In practice, we can only make small adjustments because doing so also shifts the zero level of the polarimeter, but we can iterate between a polarization adjustment and balancing the polarimeter.

\subsection{Tensor Light Shift - Two-Color Probe Scheme}
\label{chapter:ch3p5p3}
\paragraph{}
The tensor light shift, like the others, depends on the local intensity of the probe, but because this dependence is quadratic, the effect of inhomogeneities across the cloud is amplified. In order to eliminate the TLS we use a two-color probe, one near the D2 line and the other near the D1 line, with equal and opposite detunings, a scheme that has been successfully implemented in prior work by our group \cite{Hemmer2020, HemmerThesis2020}. By choosing the relative powers of the two probe colors appropriately, we can null either the SLS or the TLS, but not both simultaneously without sacrificing measurement strength. We typically prefer to null the TLS as completely as possible in order to maximize the lifetime of the state.

\indent The D2 probe is typically red-detuned by approximately 3 GHz, and the D1 probe is blue-detuned by the same amount. The power for the D2 probe is roughly $20 \mu W$, and we typically find that the TLS is nulled when the D1 probe power is around 20\% of the D2 power. The wavelength detunings are set by measuring directly with a Burleigh wavemeter with a resolution of $\pm 0.1 \, pm$.

\begin{figure}[ht]
\centering
\includegraphics[width=0.7\columnwidth]{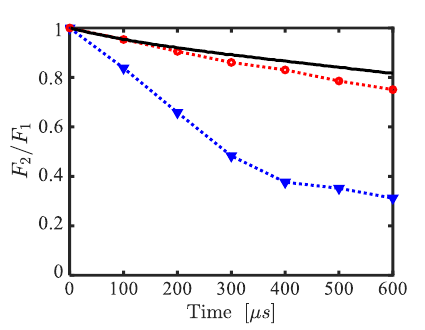}
\caption[Tensor light shift cancellation via two color probe]{Relative to the baseline decay of the spin when it is up along $\hat{z}$ (black), when the spin is rotated to the equator and back again we see that the TLS greatly increases the decay rate with only a single probe (blue). With a two color probe, the decay from the TLS can be effectively cancelled (red).} \label{fig:fig3p5p3}
\end{figure}

Since the action of TLS is essentially a rotation about the $\hat{z}$-axis, we only see the reduced lifetime when the SCS has been rotated away from $\hat{z}$. The power in the D2 probe is typically set such that the time for the spin to decay by half is around 2 ms. To set the D1 probe power, we measure the initial spin size, then apply an RF $\pi/2$ pulse to rotate the spin to the equator, wait for some time (typically 1.5 ms is sufficient), and then apply continuous RF rotations with the same relative phase to the initial pulse. We then adjust the D1 power to maximize the amplitude of the recovered spin relative to the initial size.  

\chapter{Fiducial State Preparation and Control}
\label{chapter:ch4}
\paragraph{} In this chapter we will lay out the procedures followed for the initialization and control of a pure spin coherent state of the maximum possible size. We start by using optical pumping to spin polarize the atoms, resulting in a nearly pure spin coherent state (SCS). We then follow a state purification procedure designed around the microwave transition on the stretched state. To manipulate the collective SCS, we use a combination of RF and DC fields to induce arbitrary classical rotations within the $f = 4$ manifold. It is vital that these control fields be as noise-free as possible, so we will discuss our methods to prevent and diagnose these sources of classical noise. Finally, we will briefly touch on the preparation of a state used for verifying our calibration of the atom number and the amount of quantum projection noise.

\section{Preparation of a Collective Spin Coherent State}
\label{chapter:ch4p1}
\paragraph{} After trapping with the MOT and cooling with optical molasses, the ensemble of atoms are randomly distributed throughout the $f=4$ manifold. Ideally, all atoms would occupy the same state. In particular, for these experiments we want as many atoms as possible in the $\ket{f=4, m_f=+4}$ state, since it is the most magnetically sensitive state. We achieve this through the use of optical pumping. 

\indent In order to optically pump the atoms, however, we need to establish a preferred direction in space to orient the atomic spins. This is by turning on a constant magnetic field, hereafter referred to as the bias, after optical molasses has finished. The magnetic field coils are discussed in detail in section \ref{chapter:ch4p3}.

\indent Once the bias field is turned on, optical pumping is performed by illuminating the atoms with $\sigma^+$ light tuned to the $f = 4 \rightarrow f' = 4$ hyperfine transition on the D2 line. From the $f' = 4$ excited state, atoms can decay down to either the $f = 3$ or $4$ ground state, with a net increase of $m'_f  +1$. In order to prevent loss of atoms to the $f = 3$ ground state, a second beam resonant with the $f = 3 \rightarrow f' = 3$ transition is added. This sets up a pumping cycle in which the majority of atoms are driven to the desired $\ket{f=4, m_f=+4}$ state. 

\indent In practice, this process takes about 5 ms, and the 3 light is left on for 0.5 ms longer than the 4 light to recover any atoms leftover in $f=3$. The main reason for inefficiency in the pumping process is that the pump beams propagate along a line through the trap that is not parallel to the bias, so there is a small $\pi$-polarization component to the pumping field. The angle between the pump and bias is around 10 degrees.

\section{State Purification}
\label{chapter:ch4p2}
\paragraph{} The process laid out in the previous section is not perfect, typically resulting in $~95\%$ of atoms in the desired state with the rest in $\ket{4, m_f < +4}$. In order to remove these atoms, we perform a state purification protocol as follows. 

\indent First a microwave $\pi$ pulse tuned to the $\ket{f=4, m_f = +4} \leftrightarrow \ket{f=3, m_f = +3}$ transition, also known as the stretched state transition, is applied to the atoms. This transfers the atoms to the $f=3$ manifold. A beam resonant with the $f=4 \rightarrow f'=5$ transition is then applied to the ensemble, blowing away any atoms in the $f = 4$ manifold via radiation pressure. Once all the $f=4$ atoms are gone, a second $\pi$ pulse is applied, transferring the population back to the desired state. This procedure does leave a small amount of atoms in the $f = 3$ manifold, but these atoms contribute to very little of the polarimetry measurement signal, typically on the order of $1\%$ or less. 
	
\begin{figure}[H]
\renewcommand{\baselinestretch}{1}
\centering
\includegraphics[width=0.85\columnwidth]{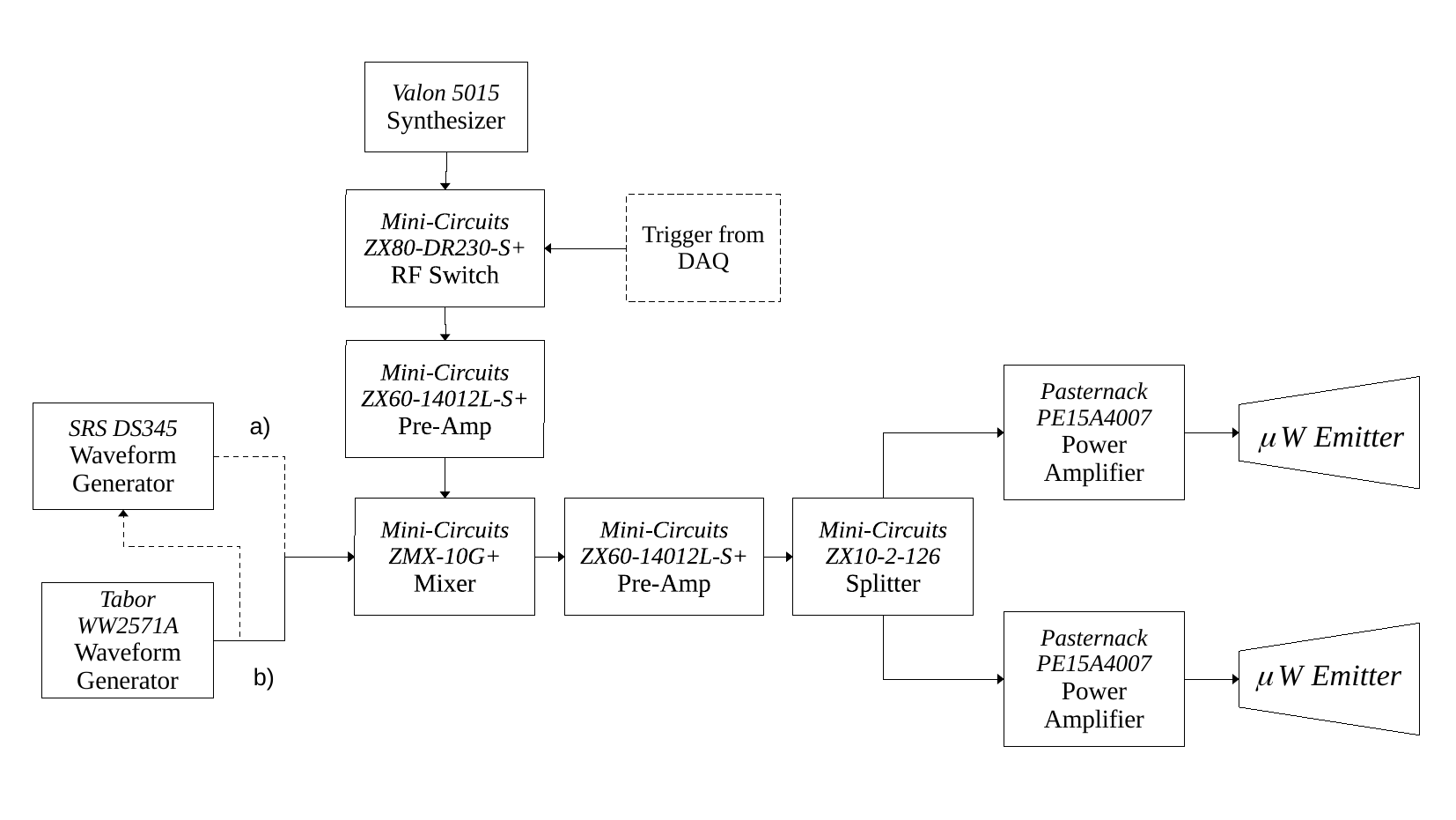}
\caption[Microwave signal chain for state purification]{$\mu$W drive generation connection diagram. A Valon 5015 frequency synthesizer supplies the 9.2 GHz carrier wave, which is modulated by an arbitrary waveform generator. In configuration (a), used for spectroscopy, The DS345 generates the modulation frequency, which is amplitude modulated by a Gaussian envelope supplied by the WW2571A. In configuration b), the WW2571A supplies the $\pi$ - $\pi$ pulse sequence directly for state purification. }
\label{fig:fig4p2_uwChain}
\end{figure}

\indent The microwave signal chain is shown in figure \ref{fig:fig4p2_uwChain}. A 9.2 GHz carrier signal from a Valon 5015 frequency synthesizer is mixed with a intermediate frequency on the order of 10s of MHz. We use two devices to supply this modulation signal. The first is an SRS DS345 arbitrary waveform generator, which we programmatically control over GPIB to perform microwave spectroscopy on the trapped atoms. Once the proper frequency is found, a Tabor Electronics WW2571A arbitrary waveform generator is used to apply the 2-pulse sequence discussed above. When performing spectroscopy we also use the WW2571A to amplitude modulate the DS345 with a Gaussian envelope. The width is 1.5 ms, so the Fourier limited bandwidth is 210 Hz.

\section{Magnetic Fields for Internal State Control}
\label{chapter:ch4p3}
\begin{figure}[H]
\renewcommand{\baselinestretch}{1}
\centering
\includegraphics[width=0.85\columnwidth]{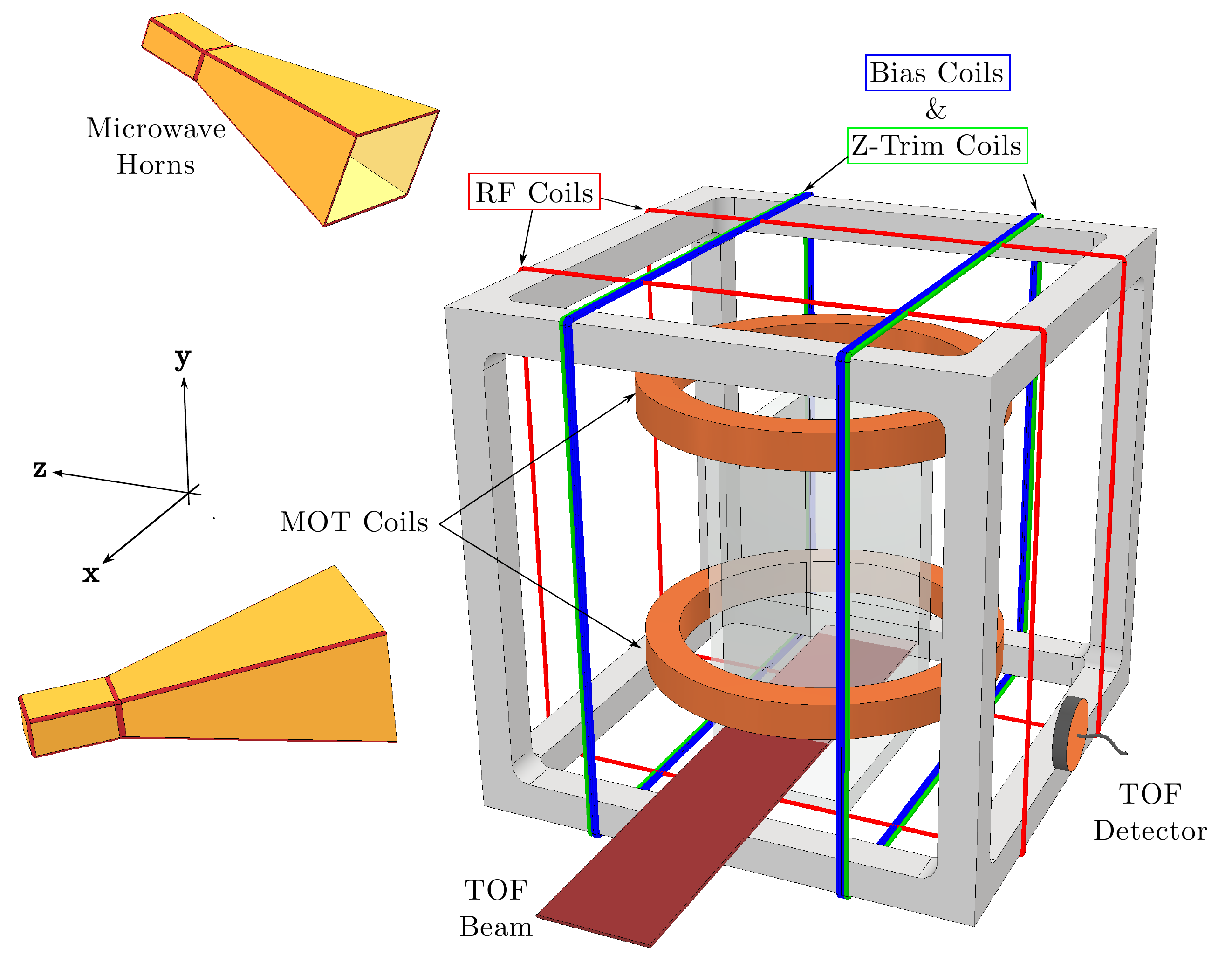}
\caption[Magnetic field coil and microwave emitter arrangement]{ Diagram of the physical arrangement of the magnetic field coils used for state preparation and control. The coils are wrapped around an acrylic frame centered on the location of the atoms, oriented such that the bias field and probe beam direction are parallel. Also shown are the two $\mu$W emitter horns used for state purification, placed at optimal locations above and to the side of the frame to produce a homogeneous drive field at the atoms.}
\label{fig:fig4p3_Coils}
\end{figure}

\paragraph{} 
Once the pure spin coherent state has been formed, the next step is to coherently manipulate it. This is done by using both static and RF magnetic fields to induce rotations on the collective spin. each of these magnetic fields is created by running current through a set of square Helmholtz coils wrapped around a clear acrylic frame, 7" on a side. A diagram of the coils is shown in figure \ref{fig:fig4p3_Coils}, as well as the placement of the microwave emitters discussed in the previous section.

As mentioned in section \ref{chapter:ch4p1}, the first and perhaps most important magnetic field is the bias, establishing the $\hat{z}$-direction in space. Stability of the bias field strength is vital, and special care has been taken to ensure it is as quiet as possible. The coils are driven by an Arroyo Instruments 4300 current source, specially modified to drive inductive loads, and the drive current is passed through a separate RLC filter. The Bias field strength is $\sim 715$ mG, stable to nearly $20\, \mu$G between experimental cycles. The bias lifts the degeneracy of the magnetic sublevels via the Zeeman effect by an energy shift of $\Delta E_m/h = 250$ kHz. This is equivalent to saying the atoms undergo Larmor precession in the bias at that same frequency. 

An RF field orthogonal to the bias is used then to induce geometric rotations in each hyperfine manifold. The frequency of the RF fields is fixed to the 250 kHz Larmor precession frequency. We typically set the amplitude of the RF so that a 40 $\mu$s pulse rotates the spin through an angle of $\pi/2$, which translates to a Rabi frequency of $\Omega \= 6.25$ kHz. The RF coil is driven in one of two ways. 

First, to set the bias and RF amplitudes and establish that they are sufficiently noise-free, we directly drive by a Tabor Electronics WW2572A dual channel arbitrary waveform generator. The WW2572A is programmed with various pulses needed to produce either simple rotations or more complicated composite pulses which give us information about the presence of noise in the magnetic fields. These composite pulses have been studied in previous work by our group \cite{Hemmer2020}, and we will briefly discuss how we make use of them in the next section.

Second, the output of the WW2572A is fed into a custom built analog voltage multiplier, along with a second signal derived from a programmable FPGA which modulates the RF amplitude. The FPGA serves as the control element in the closed loop control applications discussed in this thesis. The feedback loop is discussed in detail in chapter \ref{chapter:ch5}.

One final coil, also used for the feedforward emulation experiment, is a single loop Helmholtz pair placed directly over the Bias coils. This coil is used to trim the bias field, generating rotations around the z axis. This field is modulated in the same way as the previous coil, by mixing the output of the second WW2572A channel with a control signal from the FPGA. For maximum frequency stability (minimum phase noise), both the Tabor WW2571A for the RF and the WW2572A for the $\mu$W modulation have their 10 MHz reference clocks derived from the Valon 5015.

\section{Classical Control Errors}
\label{chapter:ch4p4}
\paragraph{}
If we wish to accurately manipulate the SCS, it is necessary that all of the magnetic fields we use be noise-free and correctly calibrated. Moreover, the system is highly sensitive to background magnetic fields, and so we must prevent any such influence, especially if we wish to explore the effect that the quantum nature of the system has on the driven dynamics. To put bounds on the allowable error, we can consider the uncertainty in the pointing direction for a spin coherent state, given by
\begin{equation}
\Delta\theta_{SCS} \p{=} \frac{\Delta F_{\vec{n}_{\perp}}}{\qty|\bk{\vec{F}}|} \p{=} \frac{\sqrt{N_{eff}^{(2)} f/2}}{N_{eff}^{(1)} f}.
\end{equation}

For the work in this thesis, we typically operate with $N_{eff}^{(1)} \sim 10^6$ atoms with $f=4$, and the fixed probe-cloud geometry yields $N_{eff}^{(2)} \approx 0.5 N_{eff}^{(1)}$. This gives an allowable error of $\Delta\theta \p{<} 2.5 \times 10^{-4} \, rad$. This is an extremely tight tolerance to meet, but typically we are able to achieve $\Delta\theta_{rot} \p{<} 0.2 \times 10^{-4} \, rad$ through a combination of magnetic shielding and measurement techniques that allow us to diagnose the presence of classical noise in the experiment. Such problems can stem from a number of different issues, including improper impedance matching in the electrical path for a  magnetic coil, poor isolation from external signals, inappropriate signal filtering, ground loops, and even mechanical vibrations through physically unsecured wires. 

The errors themselves boil down to incorrect field amplitudes. For the RF fields, this translates to an error in the rotation angle. For the bias, it leads to detuning errors, where the actual Larmor frequency differs from the programmed frequency of the RF fields. In the rotating frame, detuning errors generate a component of torque along the z-axis, lifting the torque vector out of the equatorial plane. This leads to pointing errors away from the intended plane of rotation. 

We can further categorize these errors by whether they are a) a systematic error that is constant over shots, b) a static error that is constant over the duration of a given rotation, or c) random noise during the rotation itself. For the bias, only the first two categories matter, since the bias current is filtered enough to prevent significant noise during a single cycle of the experiment. We refer to systematic errors in the bias as a 'fixed' detuning and errors that vary between experiment cycle as a 'static' detuning. For the RF, the first two categories are similarly called 'fixed' and 'static' amplitude errors, but the third can be reinterpreted as phase noise, which is typically not a concern for us.

\subsection{Magnetic Shielding}
\label{chapter:ch4p4p1}
\begin{figure}[H]
\renewcommand{\baselinestretch}{1}
\centering
\includegraphics[width=1.0\columnwidth]{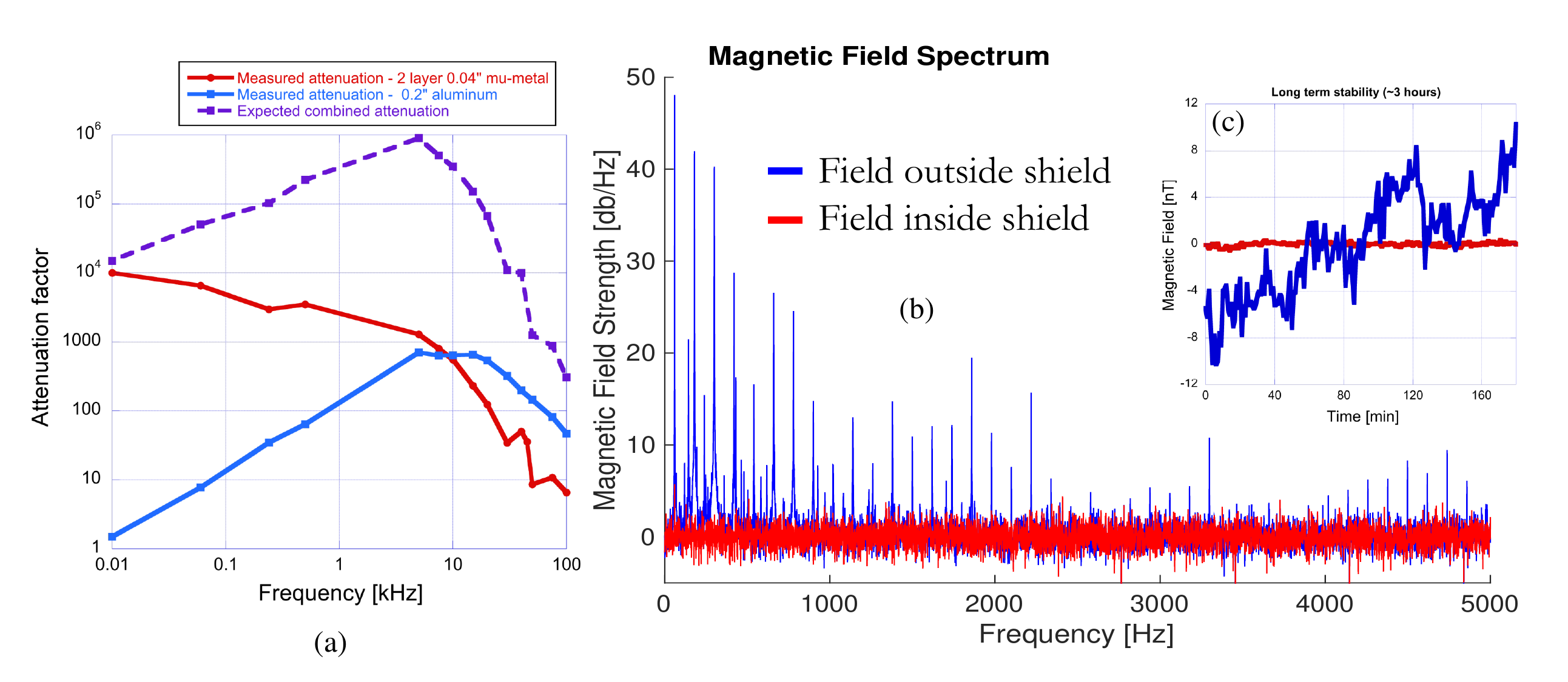}
\caption[Background magnetic field attenuation from shield]{ (a) Measured and expected attenuation from the mu-metal and aluminum layers of the magnetic shield. (b) Comparison of measured magnetic field spectra outside (blue) and inside (red) the shield. (c) Long term comparison of the background fields inside and outside the shield. The slow drift of the fields outside are attenuated by the shield to a level below our measurement capabilities. Figure taken from \cite{HemmerThesis2020}.}
\label{fig:fig4p4a}
\end{figure}

\paragraph{}
The core of the experiment is contained within a carefully designed 3-layer magnetic shield. The inner two layers are made from a mu-metal alloy composed of approximately 77\% Ni, 16\% Fe, 5\% Cu, and 2\% Cr, with a relative permeability of $10^5$. The mu-metal alloy attenuates magnetic fields from DC to a few tens of kHz by at least a factor of $10^3$. The outer aluminum layer assists by attenuating higher frequency components beyond this range (see Figure \ref{fig:fig4p4a}). 

The shield was designed to optimize the background field isolation while still allowing maintenance and optical access. It is also relatively large to prevent gradual magnetization of the inner shield layer, which encompasses a volume of $(0.76 \times 0.76 \times 0.76) \, m^3$. In combination with water cooling of the MOT coils, this also helps ensure that the temperature inside the shield is relatively constant.

\subsection{Noise Scaling with Averaging Time}
\label{chapter:ch4p4p2}
\paragraph{}
With background fields addressed, we turn our attention to the control fields themselves. We would like a way to measure the amount of classical noise in our system so that we can diagnose problems and make the appropriate changes. There are several methods we can use, but we will briefly discuss only the three most important. 

\begin{figure}[H]
\renewcommand{\baselinestretch}{1}
\centering
\includegraphics[width=0.7\columnwidth]{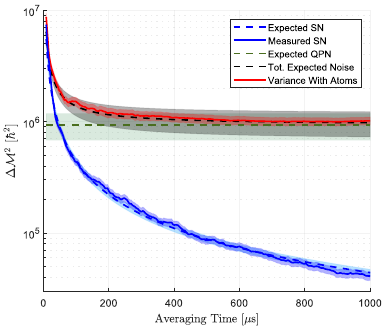}
\caption[Scaling of noise with averaging time]{Polarimetry measurement variance scaling with averaging window duration after rotating spin to the equator with a single $\pi/2$ pulse, showing a good agreement with theoretical calculations (dashed lines) due to a lack of classical noise. Solid lines are experimental data, calculated from 300 cycles of the experiment. We can see that the probe laser is shot noise limited (blue), and the variance with atoms present (red) asymptotes to the predicted noise floor for the QPN (green). Shaded transparent regions represent calculated standard uncertainty.}
\label{fig:fig4p4p2}
\end{figure}

The first method is the easiest to check because it does not require us to change any of the experimental conditions. We start by rotating the prepared SCS to the equator with a $\pi/2$ pulse, and then calculate the average of the resulting polarimetry signal over a window starting at $t\=0$ and lasting for a time $T$. We then repeat this process over many cycles of the experiment and look at the variance of the averaged signal as a function of $T$. In theory, the expected noise should be a sum of the probe shot noise and the QPN as given by Eqs. \ref{eq:eq2p4_qpn} and \ref{eq:eq2p4_sn}, respectively. The QPN does not scale with averaging time, but the shot noise, being normally distributed, goes as $1/T$. However, if there is any excess noise from any source, even those not caused by control errors (such as the presence of stray light or an impure state), we will see deviations from this expectation. This excess noise is referred to as classical projection noise (CPN).

\subsection{Composite Pulses}
\label{chapter:ch4p4p3}
\begin{figure}[H]
\renewcommand{\baselinestretch}{1}
\centering
\includegraphics[width=0.8\columnwidth]{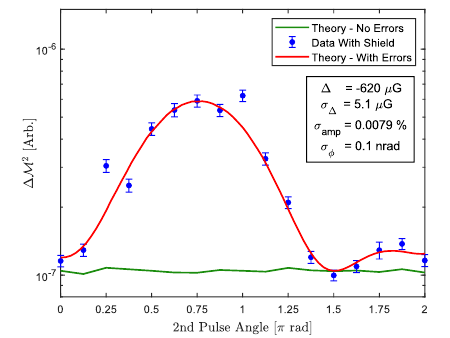}
\caption[Noise variation with second rotation angle of composite pulse]{Variation of total noise in polarimetry measurement with the second angle $\theta$ of the two-part composite pulse $R(\hat{x},\theta)R(\hat{y}, \pi/2)$. The red and green lines were generated from a numerical model based on noisy geometric rotations of a classical angular momentum vector. The noise parameters which generate this fit are listed in the upper right hand box. From top to bottom, they are the fixed detuning in the bias, the static detuning which is constant over the pulse duration but varies between experiment cycles, amplitude noise on the RF, and phase noise on the RF. We can see that at $\theta = 3\pi/4$, the classical noise is amplified, providing a means to diagnose control errors. Also of interest is $\theta = 3\pi/2$, a pulse sequence which completely eliminates CPN. Such a pulse is useful for initial state preparation, but not for real-time feedforward control.}
\label{fig:fig4p4p3}
\end{figure}

\paragraph{}
In the second method we use a specially designed composite pulse sequence consisting of multiple consecutive RF pulses with different phases and amplitudes which allows us to control the effect of classical noise in the system. The use of such composite pulses are well established in the NMR community \cite{Levitt1985, Tycko1985, Levitt1986}, and can be designed for any number of purposes. The composite pulse sequence we wish to consider is quite simple, consisting of just two rotations. The first is a $\pi/2$ rotation about $y$, which is then followed by a rotation about $x$ by some angle $\theta$. Such a sequence can be written as $R(\hat{x},\theta)R(\hat{y}, \pi/2)$, with $\theta$ in radians. Figure \ref{fig:fig4p4p3} shows how the total noise is affected by this pulse sequence as a function of $\theta$.

If $\theta = 3\pi/4$, then one can show that $\bk{\hat{F}_{z}} \propto \Delta / \Gamma$, meaning a measurement of $\hat{F}_{z}$ is a measurement of the static detuning of the bias magnetic field, which can then be nulled. For that same reason, it also allows us to diagnose problems related to noise in the bias. The minimum step size of the bias current driver is 0.1 mA, and the geometry of the coils dictates that a Larmor precession rate of 250 kHz is reached at a current of 1310 mA, implying that we can set the bias detuning directly with an accuracy of 20 Hz. A single step of this magnitude results in an under or over rotation of the spin by $\sim 2$\%, which is easily detectable, and any residual detuning below this resolution can be nulled using the single loop trim coils. 

If we look at Figure \ref{fig:fig4p4p3}, there is another interesting point to consider. If $\theta = 3\pi/2$, then it appears that any residual CPN is eliminated. This is indeed the case \cite{Hemmer2020}, but unfortunately, it is only useful for the initial state preparation in any feedforward experiment we wish to perform. This composite pulse is designed specifically to bring a state prepared up along $\hat{z}$ to the equator, and any other rotation we might want to perform would require a different pulse sequence to accomplish it. Moreover, such composite pulses require extra time to complete, which would reduce the available simulation time, especially for discrete time models.

\subsection{Noise Scaling with Number of Atoms}
\label{chapter:ch4p4p4}
\begin{figure}[H]
\renewcommand{\baselinestretch}{1}
\centering
\includegraphics[width=1.0\columnwidth]{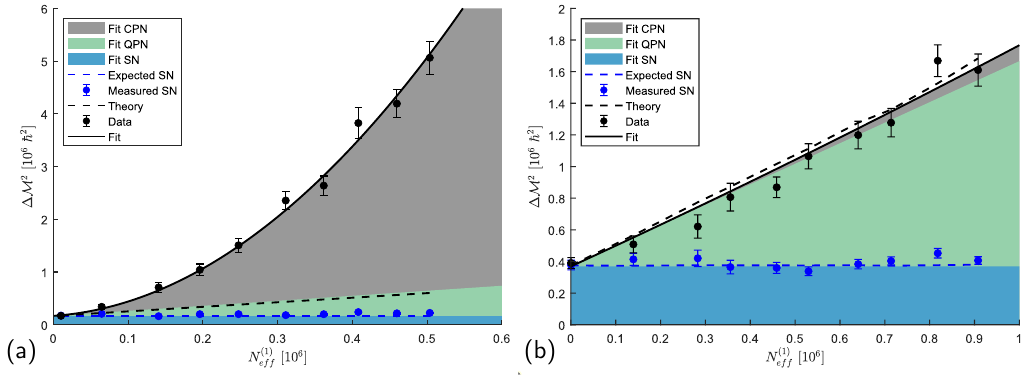}
\caption[Scaling of noise with number of atoms]{Polarimetry measurement variance scaling with $N_{eff}^{(1)}$ at fixed probe power and averaging time (140 $\mu$s). In each image, the solid back line is a quadratic fit to the black data points, and the dashed lines represent the expected shot noise (blue), quantum projection noise (green), and their sum (black). The blue, green, and grey shaded regions represent the relative amounts of shot, quantum, and classical noise respectively, based on the fit. (a) Data taken in the presence of a large amount of classical noise. The quadratic dependence can clearly be seen. (b) Data free of classical noise.}
\label{fig:fig4p4p4}
\end{figure}

\paragraph{}
The third and most comprehensive method for detecting classical noise is by examining how it scales with the number of atoms. For this method we follow the same procedure as in Section \ref{chapter:ch4p4p2}. Over many experimental cycles, the total noise in our measurements will be given by Eq. \ref{eq:eq2p4_measout} with an additional term for the CPN. In the limit of small errors, the CPN scales with the square of the atom number, so we can write
	\begin{align}
	\begin{split}
		\label{eq:eq4p4_CPN}
		\Delta\mathcal{M}^{2} &\= \Delta\mathcal{M}_{SN}^{2} \p{+} \Delta\mathcal{M}_{QPN}^{2} \p{+} \Delta\mathcal{M}_{CPN}^{2} \\
		&\= C_{SN} \p{+} C_{QPN} N_{eff}^{(1)} \p{+} C_{CPN} \qty(N_{eff}^{(1)})^{2}
	\end{split}
	\end{align}	 
where the $C_{i}$ are constants. Note that the linear scaling of the QPN with the atom number is derived from combining Eq. \ref{eq:eq2p4_qpn} with Eq. \ref{eq:eq2p4_F3dVarN2eff} along with the fact that $N_{eff}^{(2)} \p{/} N_{eff}^{(1)}$ is constant for fixed probe-trap geometry. By varying the number of atoms then, we can determine the relative contributions of each noise source in Eq. \ref{eq:eq4p4_CPN} and verify that we have eliminated any CPN. In Figure \ref{fig:fig4p4p4}, we show data collected from the experiment showing the noise budget determined from such a procedure. In Figure \ref{fig:fig4p4p4}b we show our most recent dataset, where it can be seen that the classical noise contribution is now negligible when compared to the QPN and SN.

\section{Preparation of the Maximally Mixed Spin State}
\label{chapter:ch4p5}
\paragraph{}
In order to perform the feedforward emulation experiments, we need a way to accurately calibrate both the size of the collective spin and the expected QPN. Of course, we can directly calculate the effective atom number $N^{(1)}_{eff}$ based on the measured trap and probe geometry, the probe power, and all other fixed parameters as laid out in section \ref{chapter:ch2p4}, but it is good to have a secondary method to confirm our estimates. One way to do this is by creating a maximally mixed spin state (MMSS) \cite{Koschorreck2010, Enrique2015}, which is a rotationally symmetric state in which the atomic populations are evenly distributed among the magnetic sublevels of a particular hyperfine state. The quantum projection noise of such a state is 10/3 that of a spin coherent state of equal atom number which is oriented orthogonally to the probe, and since it is rotationally symmetric, it is totally insensitive to classical noise from the magnetic control fields, such as bias fluctuations and RF amplitude errors. 

\indent To prepare a MMSS, we start with the unpolarized atoms cooled out of optical molasses and held in the dipole trap. The atoms are already somewhat well distributed throughout the ground state, but there is some asymmetry that needs to be corrected. After cooling, the bias is immediately turned on, and we apply a large continuous RF magnetic field, causing the atoms to rapidly precess. During the rotation, we apply 4 ms pulses from the optical pumping beams, alternating between the $f=4 \rightarrow f'=4$ and $f=3 \rightarrow f'=3$ light. Since the rotation rate through the hyperfine sublevels is much larger than the optical pumping scattering rate, the atoms are effectively randomized throughout both the $f = 3$ and $f = 4$ manifolds. After 20 ms, we apply a final burst of repump light to push all atoms into the $f=4$ manifold. 

\indent We can then switch to polarimetry and estimate the variance in an averaging window of duration $T$ across many shots. The QPN from the MMSS will then be the difference between the measurement noise and the shot noise. However, some atoms are lost in the randomization process, and so to correct for this we make a TOF measurement of the dipole trap with and without randomization. Since the area under the TOF signal is proportional to the number of atoms, the ratio between the TOF area measured before and after randomization gives us the correction factor we need. We typically find that the QPN measurements for a maximally mixed state agrees with our calculated expectations to within 2-5\%.

\chapter{Feedback Control System}
\label{chapter:ch5}
\paragraph{}
Here we will discuss the final ingredient in this experiment: the feedback controller itself. We will begin by briefly reviewing some of the theory of feedback control, focusing on how we can characterize properties and performance of the control loop.  Then we will give the technical details for the controller used in our experiments as well as the custom mixers used to modulate the control signals. We will also provide details for the system under control in this context. We will follow with a discussion of our implementation of the control laws for the Hamiltonians to be emulated. Finally we will wrap up with a discussion of the two major types of dynamics we can attempt to drive: discrete-time vs. continuous-time. 
 
\section{Feedback Control Background}
\label{chapter:ch5p1}
\begin{figure}[ht]
\centering
\includegraphics[width=0.9\columnwidth]{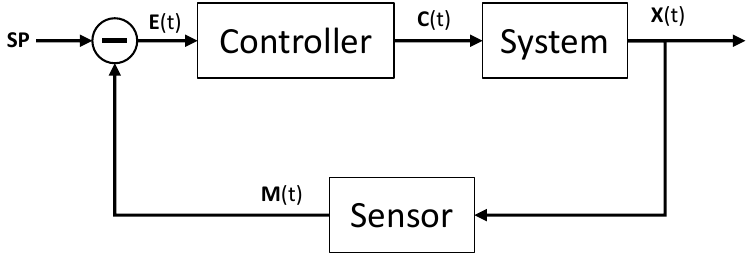}
\caption[Diagram of a generic feedback loop.]{General diagram of a feedback loop. The system state, $\mathbf{X}(t)$, is measured by the sensor, whose output $M(t)$ is compared against some desired condition called the setpoint. The difference, called the error signal $E(t)$, is transformed into a control signal $C(t)$ by the controller, which then affects the system properties. Typically the action of the control is designed to drive the error signal to zero as quickly as possible with limited overshoot.} \label{fig:fig5p1_FeedbackLoop}
\end{figure}

\paragraph{}
A traditional closed-loop control system consists of three elements: a system with some property to be controlled, a sensor used to measure that property, and a controller which translates the measurement into some action that affects the system. This basic arrangement is shown in figure \ref{fig:fig5p1_FeedbackLoop}, with the signal direction indicated by arrows. The most common use of closed loop control is to drive the system to a desired state, often referred to as a setpoint. The difference between the setpoint and the measurement yields the error signal, which the controller works to null out.

The most ubiquitous form for the feedback controller is known as a PID, or Proportional, Integral, Derivative controller. It is so named because the control signal is the sum of those three operators acting on the error signal. Such a controller is used to drive the system to a steady state for a fixed setpoint, with the P term directly driving the error signal to zero, the I term eliminating residual offsets, and the D term suppressing signal spikes and noise in the measurement. For a given system and sensor, the performance of a feedback loop with this type of controller is determined by the relative contributions of each term, typically quantified by their coefficients, which we refer to as their gains. Control systems design is often concerned with finding the optimal set of gain parameters, where design criteria might be to minimize the settling time of the error signal, to reduce sensitivity to perturbations, or to constrain the size of signal overshoot. 

This classic framework is useful for control in the sense that the system state is held wherever we wish it to be held with tight, quantifiable tolerance. However, it is not the right type of control for our purposes. Instead of driving the system to a fixed state, we want to drive specific dynamics which are characterized by a Hamiltonian function. More specifically, we want to apply a torque to our collective spin which depends in some way on the current state of the collective spin, or more specifically on the measurement of $J_{z}$. The new torque rotates the atoms, which then yields a new measurement, and so on. Such a controller is called a feedforward controller, because the measurement is fed forward to produce a new state to be measured in the next step in time. 



\section{Core Platform \& System Model}
\label{chapter:ch5p2}
\begin{figure}[p]
\centering
\includegraphics[width=0.9\columnwidth]{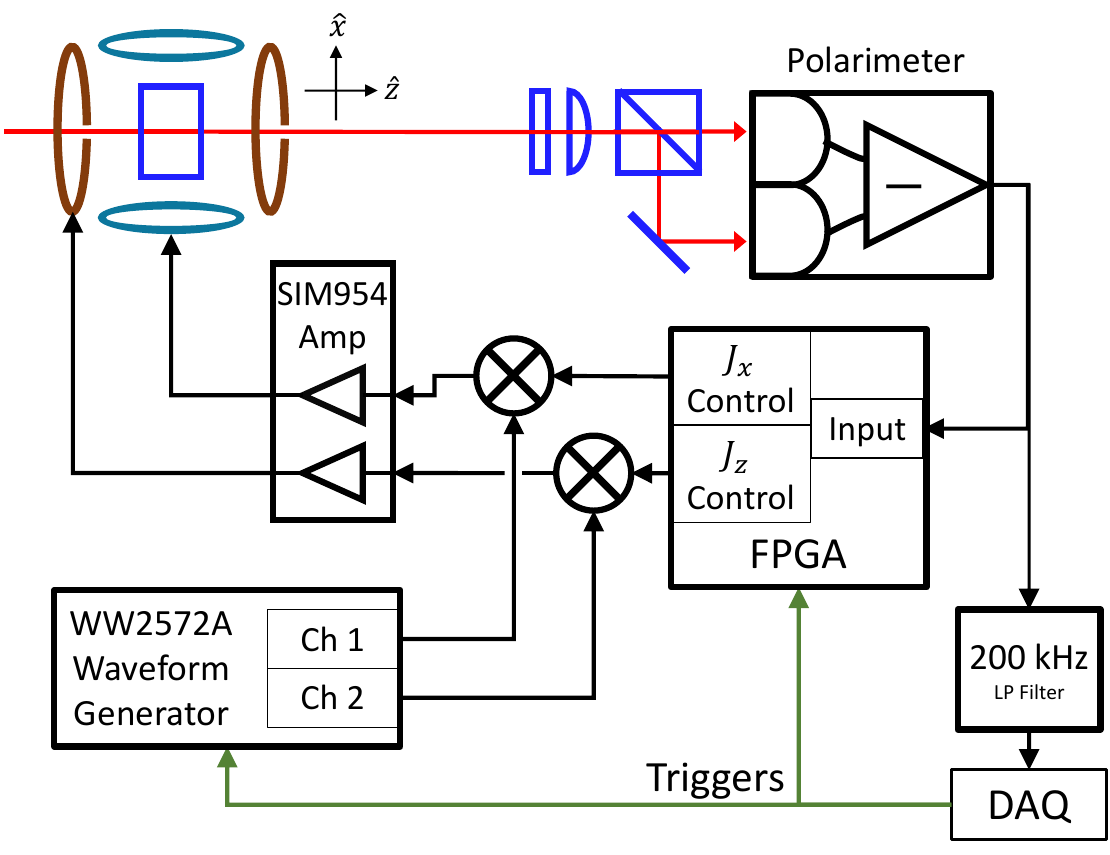}
\caption[Detailed diagram of the feedforward emulation control loop]{Detailed diagram of the control loop for the feedforward emulation experiments. The polarimetry signal is fed directly into the FPGA for processing, which then output two control signals for $\hat{x}$ and $\hat{z}$ rotations of the spin. The controls modulate the amplitude of synthesized waveforms sourced from the WW2572A signal generator, and the modulated signals are amplified to drive a set of coils surrounding the atoms. The entire procedure is orchestrated with triggers from the central DAQ system.}  \label{fig:fig5p2_DetailedLoop}
\end{figure}

\paragraph{}
We discussed some parts of the feedback loop in Section \ref{chapter:ch4p3}, but here we will give a more detailed overview of the components in its signal chain. A full diagram of the loop is shown in Figure \ref{fig:fig5p2_DetailedLoop}. 

The loop starts at the atoms, where we measure $J_{z}$ using the polarimetry measurement as detailed in Sections \ref{chapter:ch2p5} and \ref{chapter:ch3p4}. The resulting electrical signal is then routed directly to the controller, a National Instruments USB-7855R Multifunction RIO device. The 7855R has an array of analog and digital IO ports, as well as a configurable FPGA, all programmed in LabVIEW. The FPGA, a Kintex-7 70T, has a base clock speed of 40 MHz, which can be increased via PLLs up to 200 MHz. Increasing the speed should in principle improve the control calculation time, but we found that our programs ran best at 80 MHz when using the analog IO ports. Its analog IO ports have a maximum update rate of 1 MHz and a resolution of 16 bits (0.6 mV step at full $\pm$10 V range). 

One analog input is used for the polarimetry measurement, and two more ports are used for one time measurements of the two monitor ports of the balanced detector. Once calibrated, the sum of the two monitor signals is a measurement of the total probe laser power, taken after the atoms are dropped. The control programs use two analog outputs, one which serves as the control signal to modulate the z rotation rate, and the other to modulate the RF amplitude, which sets the rotation rate about axes in the xy-plane. We primarily use the latter control to make sure the RF amplitude is set to a desired level and to switch it on and off as needed. The waveforms that generate the $\hat{x}$ and $\hat{z}$ rotations have dead-time hard-coded during times when there should be no signals to prevent any unwanted magnetic influence, but sometimes when setting up an experiment or during calibration, it is convenient to have the ability to quickly turn off a signal without waiting several tens of seconds to load a new waveform. Moreover, for stroboscopic discrete-time systems like the QKT, we need to be able to pulse each control signal at a specified rate, which can be easily modified and coordinated by the controller.

Regardless of the Hamiltonian which we aim to emulate, the control programs are designed to perform several necessary functions. Once all the configuration parameters are set, the basic flow of a program proceeds as follows:
\begin{enumerate}
 \item When triggered by the central DAQ system, begin the control procedure.
 \item In sequence, perform the following actions:
 	\begin{enumerate}
 		\item Wait until the probe laser is turned on.
 		\item Measure the initial magnitude of the collective spin. 
 		\item Apply an RF pulse to prepare the desired initial state for emulation.
 		\item Wait for 8-12 $\mu$s and then initialize the main control loops.
	\end{enumerate} 	  
 \item There are 3 main loops working in parallel, synchronized at a specified rate. The output of each loop at time $t_{i}$ becomes the input of the next loop at $t_{i+1}$. 
 	\begin{itemize}
 		\item The first loop measures the signal and optionally passes it through a built-in FIR filter.
 		\item The second loop performs the control calculations. It also calculates the expected spin length at the beginning of the next time step based on the initial spin size and the elapsed time since that first measurement. This estimate is fed back into this loop at the next time step to be used for the control signal calculation. 
 		\item The third loop outputs the control signals to their respective analog outputs. It also records all important signals at each time step into a FIFO (First-In, First-Out) data connection for transfer back to the computer interface.
 	\end{itemize}
 \item Once the trigger to stop the control loop is detected, all three loops exit and the analog outputs are reset to their default values.
 \item Finally, 10 ms after the atoms have been dropped a second trigger is sent on a separate line to the FPGA. Once triggered, the FPGA averages the balance detector output for 30 ms. This value is used as an offset for the polarimetry measurement during the next cycle of the experiment to account for low-frequency common-mode noise in the balance detectors differential output. 
\end{enumerate}

\begin{figure}[H]
\centering
\includegraphics[width=0.9\columnwidth]{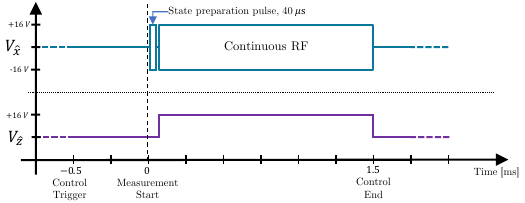}
\caption[Waveform timing for feedforward emulation procedure]{Waveform timing for feedforward emulation procedure. At t = -0.5 ms, the DAQ triggers the FPGA and the WW2572A waveform generator. The outputs from the latter are amplitude modulated by the FPGA control signals through a custom voltage multiplier. Rotations about $\hat{z}$ (purple) are done with a DC signal and rotations about $\hat{x}$ (blue) with an RF signal at 250 kHz. At t = 0, the controller measures the initial spin size, then applies an RF pulse to prepare the initial state for emulation. The pulse amplitude determines the polar angle, and the phase of the continuous RF segment relative to the initial pulse dictates the azimuthal angle. After 1.5 ms, the FPGA is triggered again to stop control. The atoms are dropped at t = 30 ms, and then at 40 ms the FPGA averages the balance detector signal to determine the measurement offset to use during the next experiment cycle.}  \label{fig:fig5p2_WaveformTiming}
\end{figure}

A few details about this general procedure warrant further discussion. First, the RF pulse used for initial state preparation lasts 40 $\mu$s, and its amplitude is adjusted to control the polar angle. This duration was chosen partially as a matter of convention, and partially because at the 250 kHz Larmor frequency this duration has an integer number of cycles, so the start and end phase is the same, and the necessary driving amplitude for a $\pi$/2 pulse is a reasonable value. To control the azimuthal angle, we change the relative phase between this initial pulse and the subsequent continuous RF rotation. This phase delta must be hard-coded into the RF waveform because we do not have hardware that would allow us to manipulate it on the fly.

All calculations performed by the FPGA are done natively using fixed-point math. In LabVIEW, the convention for representing a number is (s, n, m), where s specifies whether the number is signed or unsigned, n the total number of bits used to store the value (including the sign bit, if signed), and m is the number of bits in the integer portion of the value. Due to the finite resources available to us in the FPGA, much of the effort in design of our program went into careful consideration of the fixed-point representation of numbers at all stages of internal calculation, with the goal of minimizing the internal resources used while keeping the fixed point error well below the 16-bit resolution of the analog IO.

Based on the measured initial size, spin up decay rate, and the total elapsed time since the initial measurement of the spin magnitude, the decay of the mean spin is tracked by a calculation involving simple exponential decay. Since the LabVIEW's FPGA module lacks a built-in exponential function, we implemented its Pad\`{e} approximation of order [5,5], given by 

\begin{equation}
	\label{eq:eq5p2_padeExp}
	exp(x) \approx \frac{1 + \frac{1}{2}x + \frac{1}{9}x^2 + \frac{1}{72}x^3 + \frac{1}{1008}x^4 + \frac{1}{30240}x^5}
	{1 - \frac{1}{2}x + \frac{1}{9}x^2 - \frac{1}{72}x^3 + \frac{1}{1008}x^4 - \frac{1}{30240}x^5}  .
\end{equation}

The error in this approximation and a calculation using floating point arithmetic is at least 2 orders of magnitude smaller than the analog resolution of the USB-7855R. However, there is at least a 2\% error between the actual spin up decay signal and the exponential decay estimation due to minute oscillations that likely stem from center-of-mass motion caused by the probe scalar light shift (see Section \ref{chapter:ch2p6}). We could have used a record of the spin up signal as a look-up-table to incorporate those oscillations, but if the probe power or the atom number fluctuates from shot to shot, the magnitude of the polarimetry signal would change, and thus the error in the spin magnitude would also fluctuate. Using the exponential decay calculation allows the FPGA to incorporate these shot to shot fluctuations in the measurement, rendering this source of error consistent across a run of many shots. 

Another major issue is the design choice to parallelize the functions of the main control loop. We could have set the program up to perform all of these actions sequentially in a single loop, but this way the minimum sampling period for control would be limited to twice the analog signal update time (2$\times$1 $\mu$s), plus whatever time is needed for all the other functions described (control signal calculation and loading data into memory). Instead, the parallelization pattern, known as pipelining, allows the controller to operate at the theoretical limit of 1 MHz for a single analog port at the cost of increased signal latency. That said, in practice we found that the control program is only stable up to a sample rate of 500 kHZ. 

In retrospect, it may have been better to use the serial design pattern, since, as we shall see in Section \ref{chapter:ch6p1p3}, latency appears to be the primary driver of dynamical decay towards fixed points in continuous time evolution models like the LMG. LabVIEW does have a mechanism called ``Occurences" that act as internal triggers that, if used correctly, would allow the Control and Output loops to run asynchronously, only activating at the end of each iteration of the previous step. Unfortunately, for unknown reasons that may have to do with timing and routing constraints on compilation of the FPGA programs, we were unable to get this mechanism to work. It may also be possible to achieve better timing performance by using an array of digital ports with custom high-speed ADC/DAC circuitry, or by designing a purely analog controller to replicate the desired control function. There are ADC/DAC chips on the market capable of GHz sample rates with low latency, such as the DAC5670-SP from Texas Instruments. On the other hand, while a dedicated fully analog controller would be the most faithful and lowest latency option, we would need a new controller for every Hamiltonian system we wish to model. This is where the FPGA truly shines, in its ability to be reconfigured as needed. 

The control signals produced by the 7855R are sent to a pair of custom designed analog signal mixers based around the VCA824 voltage-controlled variable gain amplifier. The circuit schematic for the mixers can be found in Appendix \ref{chapter:rfMixer}. Each mixer takes in two impedance matched inputs, one from the FPGA for modulation, and one from the Tabor WW2572A discussed in Section \ref{chapter:ch4p3} for the desired type of magnetic control (DC or RF). The maximum drive amplitude for the latter is $\pm16$ V into 50 $\Omega$. The output of the VCA824 is amplified through a variable gain stage, which allows us to make effective use of the full voltage range of the FPGAs analog outputs, and is then buffered to match the input impedance of the Stanford Research Systems SIM954 inverting amplifier used to drive the control coils. 

The SIM954 has a maximum output current of 1 A, which translates to a maximum rotation rate of 63 kHz for both RF and z rotations (in the 250 kHz rotating frame). The RF rate is typically an order of magnitude smaller than this limit, and we try to keep the drive current for the z-rotations below 90\% of this value to avoid deviations from linearity. 

Since the property of the atoms we measure is directly related to the polar angle by $J_{z} = J \cos(\theta)$,  we can model the system response to magnetic influence as an integrator with gain. We precisely measure all of the gain factors that relate the FPGA's control signal voltage to the rotation frequency and incorporate them into the FPGA program as calibration factors so that we can directly input a desired rotation rate or descriptive parameter, such as the $s$ parameter of the LMG. This calibration gain factor which converts the control voltage to physical rotation rate is given by
\begin{equation}
	\label{eq:eq5p2_ctlgain}
	g_{ctl} \= \frac{f_{rot}}{V_{ctl}} \= \frac{2N\gamma G g}{R}, 
\end{equation}
where $N = 1$ is the number of loops in the coil, $\gamma = 3.5$ kHz/$\mu$T is the gyromagnetic ratio, $G = 4.5$ $\mu$T/A is a factor determined by the coil geometry, $g$ is the gain between the FPGA analog output voltage and the voltage across the coil, and $R = 5.75 \Omega$ is the coil resistance.

\section{Feedforward Control Law Synthesis}
\label{chapter:ch5p3}
\paragraph{}
As we saw in Section \ref{chapter:ch2p4}, under certain straightforward conditions we can emulate the dynamics of a given Hamiltonian by linearizing each term according to a mean field approximation. The emulation is then acheived by conditioning the strength of each influence by the result of our measurement. While the mean field approximation to both the LMG and QKT models are essentially the same, the details of how the control laws are implemented have major differences.

In both cases, the control signal strength is proportional to the measurement outcome. The first major difference comes from the fact that the LMG is parametrized according to a rotation rate, while the QKT is parametrized by a rotation angle per step. This comes directly from the fact that the LMG is a continuous time model, while the QKT is a stroboscopic, discrete-time model. We will discuss the differences between these two types of emulation more generally in the next section, but for now we will give the specifics of how the control law programs for each model were designed.

For the LMG, the control voltage at time step $i$ is 
\begin{equation} \label{eq:eq5p3_LMGctl}
	V_{ctl} \= \frac{1}{g_{ctl}} \qty(s \Lambda \frac{ \bk{J_{z}(i)}}{\bk{J(i)}} ).
\end{equation}

The control is updated at each time step, which iterates at a sample rate of 500 kHz. Since $s$ encodes the relative rates of the $\hat{x}$ and $\hat{z}$ rotations, we found it easiest to fix the linear rotation rate, $\alpha = (1-s)\Lambda$, input a desired $\hat{z}$-rotation strength $k = s \Lambda$, and then calculate $s$ as
\begin{equation} \label{eq:eq5p3_LMGsCalc}
	s \= \frac{k}{\alpha + k}.
\end{equation}
For $\alpha = 6.25$ kHz, we are able to reach $s = 0.8$, which is well past the dynamical phase transition point. 

\begin{figure}[H]
\centering
\includegraphics[width=0.9\columnwidth]{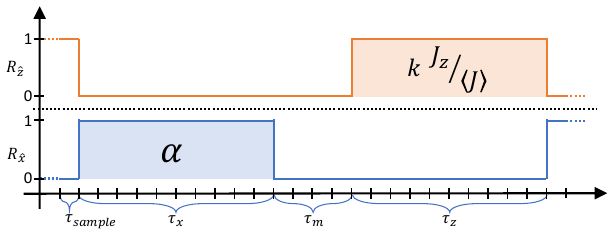}
\caption[Rotation pulse timing for QKT emulation]{Rotation pulse timing for QKT emulation program. Two internal digital signals dictate when the control signals for rotation about $\hat{x}$ (blue) and about $\hat{z}$ (orange) are enabled. Tick marks on the time axis are spaced in units of the sampling period for the control loops. A single emulation time step consists of an $\hat{x}$ rotation, a measurement gap, and a $\hat{z}$ rotation. The measurement gap, lasting a time $\tau_{m}$, allows the FPGA to update the stored value of $J_{z}$ used to calculate the nonlinear angle with the most recent measurement. The timing is fixed at the beginning of a full emulation run, and so the net rotation angle for each pulse is controlled by adjusting the pulse amplitudes.}  \label{fig:fig5p3_QKTtiming}
\end{figure}

The QKT program is more complicated. Since the two models have essentially the same form, we started with the program for the LMG. However, instead of specifying a rotation rate, we need to rotate through a specific angle at each step, alternating between the linear $\hat{x}$ rotation and the nonlinear $\hat{z}$ rotation `kick' whose strength is determined by the measurement outcome at the end of the preceding linear rotation. We chose to add two internal digital square wave signals which serve to toggle the two controls. The general idea is illustrated in Figure \ref{fig:fig5p3_QKTtiming}. During a single update step the linear rotation happens first, followed by a short gap that allows the program to update the value of $J_{z}$ used to calculate the kick strength, and then the nonlinear rotation occurs. Fundamentally, the program still loops at the 500 kHz sample rate, but these two internal signals determine which rotation occurs at what time, and the timing parameters for each segment are specified by the user. We found that 20, 8, and 20 $\mu$s for the linear rotation, measurement update, and nonlinear rotation segments respectively worked well, allowing us to emulate up to 25 steps reliably.

The other core difference between the QKT and the LMG is that by leveraging the fact that in the former case we are aiming for a specific rotation angle instead of a rotation rate, we can reach much higher values for the nonlinear strength than would otherwise be possible if we simply drove the rotations directly. If, for a given value of $k$, the nonlinear rotation angle happens to lie outside the range $[0, 2\pi)$, we can instead rotate through the angle corresponding to $k J_{z}/J$ modulo $2\pi$. In principle, the net result should be the same. 

As the FPGA does not have a native modulo function for fixed point numbers, we again need to implement it ourselves. As we are working with binary numbers that represent angles, there is a very nice trick we can use to perform the modulo calculation. Using the properties of modulo arithmetic, it can be shown that
\begin{equation}
	\label{eq:eq5p3_QKTmod}
	\text{mod} \qty(k \frac{J_{z}}{J}, 2\pi ) \= \pi \cdot \text{mod} \qty(\frac{k}{\pi} \frac{J_{z}}{J}, 2 ).
\end{equation}
The trick lies in the fact that, for any fixed point number x, the result of calculating mod(x,n) is the same as retaining only the fractional part of that number when expressed in base n. Then mod(x,2) can be performed by retaining the fractional bits when x is expressed in binary. We can easily accomplish this by taking the logical AND between the binary fixed-point representation of the quantity $\frac{k}{\pi} \frac{J_{z}}{J}$ with a masking value that has the same representation, consisting of all zeros for the sign bit and integer part, and all ones for the decimal part. This operation is far less expensive than a method that uses division, which is very desirable. The only thing we need to account for is the sign of the original value, since $J_{z}$ is a signed quantity. Block diagrams for the LabVIEW FPGA program which accomplishes this operation can be found in Appendix \ref{chapter:labview}.

\section{Continuous vs Discrete Time Evolution}
\label{chapter:ch5p4}
\paragraph{}
As we have mentioned previously, the key difference between the two models under consideration is whether the processes are continuous or happen in discrete steps. It is important to remember that, in the implementations described in the previous chapter, the physical control of the atoms is always stepwise continuous, since the controller is limited to operate at a finite sample rate. So, in this context, what do we mean by a continuous process?

For the LMG, when we say continuous, we mean two things: that the two sources of torque ($J_{x}$ and $J_{z}^{2}$) encoded by the Hamiltonian are acting simultaneously, and that the control sample rate is fast relative to the evolution timescale. The sample rate, as we have mentioned previously, was designed to be 500 kHz, and the linear rotation rate was fixed to be $6.25$ kHz. The latter was a more or less arbitrary choice, based on the precedent of having 40 $\mu$s $\pi/2$ pulses. We can see then that, in this regard, the sample rate is just under 2 orders of magnitude larger than the evolution timescale.

For the discrete KT model, on the other hand, we look at the evolution stroboscopically. The two rotations happen independently of each other for a fixed duration, rotating the collective spin through a target angle, rather than at a target rate. The controller still operates at the 500 kHz sample rate, and records the full measurement record at that rate, but the evolution happens in discrete chunks. 

What we do then is to extract the stroboscopic, discrete evolution from the measurement record in post processing. This is done by down-sampling the full measurement record, where the down-sampling period is the time to complete both rotations, plus the gap between them (see Figure \ref{fig:fig5p3_QKTtiming}), and the measurements are taken in that gap, since it is that value of $J_{z}$ that is used to calculate the $z$-rotation control amplitude.

\chapter{Mean Field Feedforward Emulation - Results}
\label{chapter:ch6}
\paragraph{}
In this chapter, we will discuss the results of the series of feedforward emulation experiments we have carried out. We start with the Lipkin-Meshkov-Glick (LMG) Hamiltonian as introduced in Section \ref{chapter:ch2p3p1}, which is a continuous time evolution model. The key feature of this model is a phase transition wherein one of its stable fixed points bifurcates into a pair of points symmetrically placed above and below the equator, with an unstable fixed point taking its place. In the first experiment, we produce one of the more striking results of this work by initializing the SCS to be up along x, on top of the unstable fixed point. We find that beyond the phase transition the collective spin randomly ends up in either of the two bifurcated fixed points, whose apparent location agrees well with theory. Next, starting with an initial state up along z, we measure the infinite time averages of the z-magnetization and the two-body correlation function, which act as order parameters for the dynamical phase transition. We will then wrap up with a discussion of some numerical results that seem to indicate that the observed decay of the spin is not primarily due to quantum projection or shot noise, but rather due to latency in the control loop. 

The second model is the Quantum Kicked Top (QKT). This model exhibits chaotic behavior above a certain threshold, and also has some interesting regular dynamical phases. Here we start by discussing our efforts to directly estimate the maximal Lyapunov exponents which describe the sea of chaos. We will also explore an experimental realization of a driven time crystal, a relatively new phase of matter of interest due to the robustness of this phase to perturbations.

\section{Emulation of the LMG}
\label{chapter:ch6p1}
\paragraph{}
The LMG is a good first model to test the QMF paradigm with, for a number of reasons. Its form is relatively simple, it has wide body of literature surrounding it and, as such, it is well-understood. Its main feature is a symmetry-breaking phase transition whose characteristics are easily measured. Here we will show the results of two experiments centered around the phase transition feature.

\subsection{Spontaneous Symmetry Breaking at the Unstable Fixed Point}
\label{chapter:ch6p1p1}
\begin{figure}[hp] 
\renewcommand{\baselinestretch}{1}
\centering
\includegraphics[width=1.0\columnwidth]{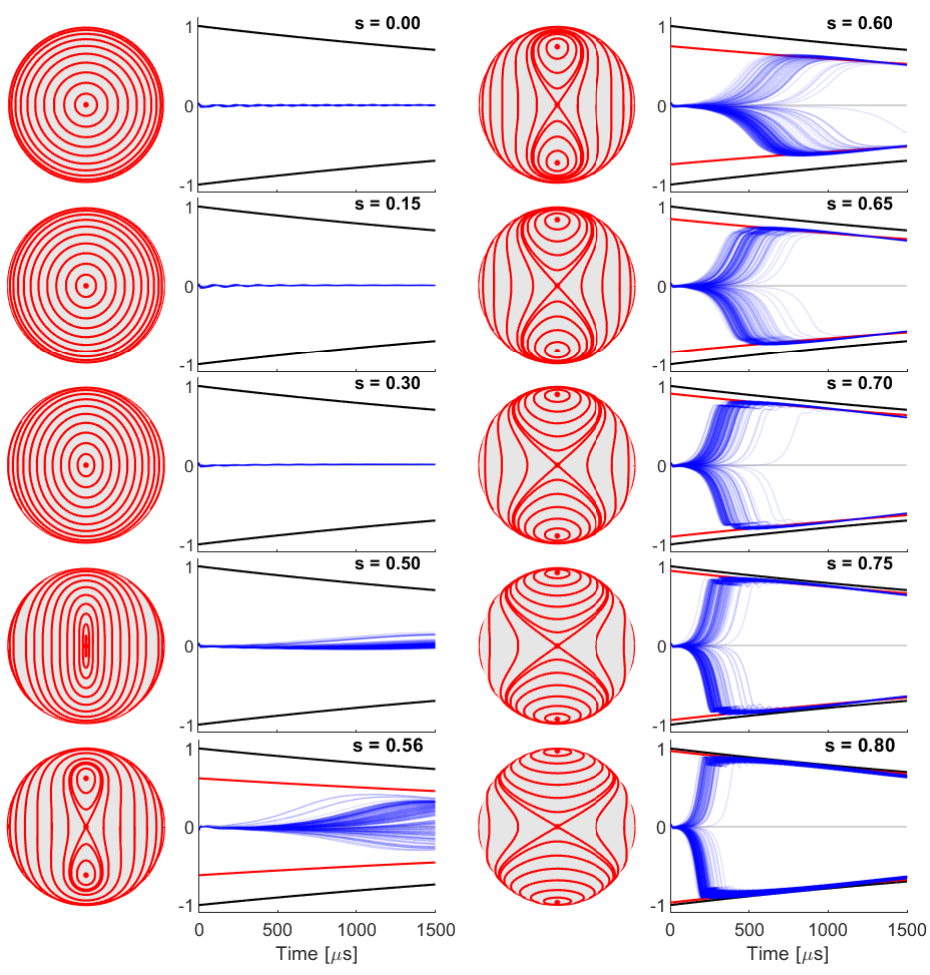}
\caption[LMG emulation symmetry breaking data, spin up along x]{LMG emulation experimental data with initial state spin up along $\hat{x}$ for several values of $s$ between 0 and 0.8. Phase space portraits showing the $x \, > \, 0$ hemisphere for each value of $s$ are shown to the left of each time series plot, with $+\hat{z}$ pointed up. The time series plots themselves consist of 300 shots of data (blue) which are made slightly transparent so that the density of trajectories can be visualized. The trajectories are bounded by the expected length of the mean spin $\bk{J}$ (black), and for $s \, > \, 0.5$, the estimated location of the fixed points are shown in red.} \label{fig:fig6p1p1_LMG_UpX_decay1}
\end{figure}

\paragraph{}
For the first experiment, we start by preparing an initial spin coherent state (SCS) oriented up along $\hat{x}$. As discussed in Section \ref{chapter:ch2p3p1}, this is the location of the unstable fixed point for $s \, >\, 0.5$. As such, we expect that the initial uncertainty in the pointing direction will break the symmetry, and this initial condition will randomly end up in an orbit centered about either the upper or lower stable fixed point. The data for this experiment is shown in Figure \ref{fig:fig6p1p1_LMG_UpX_decay1}.

\begin{figure}[H] 
\renewcommand{\baselinestretch}{1}
\centering
\includegraphics[width=1.0\columnwidth]{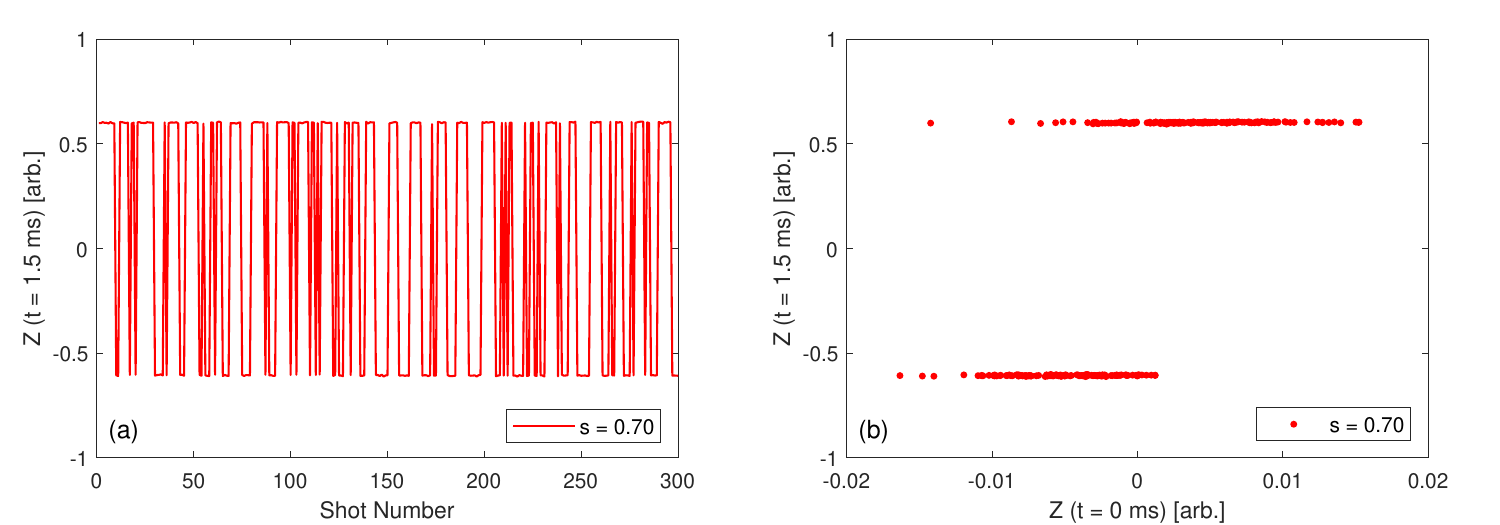}
\caption[Shot-to-shot randomness in LMG symmetry breaking experiment]{(a) Final value of $J_{z}$ on a shot-to-shot basis for $s \,=\, 0.7$ in LMG emulation experiment with inital state up along $\hat{x}$. (b) Correlation plot between the initial (x-axis) and final (y-axis) values of $Z$. The choice of the upper or lower well is clearly random, and the correlation plot suggests the initial measurement outcome is a strong determining factor for the direction in which symmetry is broken.}
\label{fig:fig6p1p1_LMG-ssb} 
\end{figure}

At a glance, the data looks very promising. Starting at $s = 0.6$, the initial state ends up moving into one of the two fixed points, with good agreement between the data and the fixed point's expected location. The symmetry breaking is also random on a shot-to-shot basis, which can be seen in Figure \ref{fig:fig6p1p1_LMG-ssb}a. It also seems apparent that the quantum projection noise, which primarily contributes to the result of the first measurement, is a clear determining factor in which of the two fixed points the spin ends up in. Figure \ref{fig:fig6p1p1_LMG-ssb}b shows a correlation scatter plot between the initial and final values of the spin. While the distribution of initial values is quite small, and the end result is essentially binary, there is still a clear correlation between the two.

\begin{figure}[H] 
\renewcommand{\baselinestretch}{1}
\centering
\includegraphics[width=0.8\columnwidth]{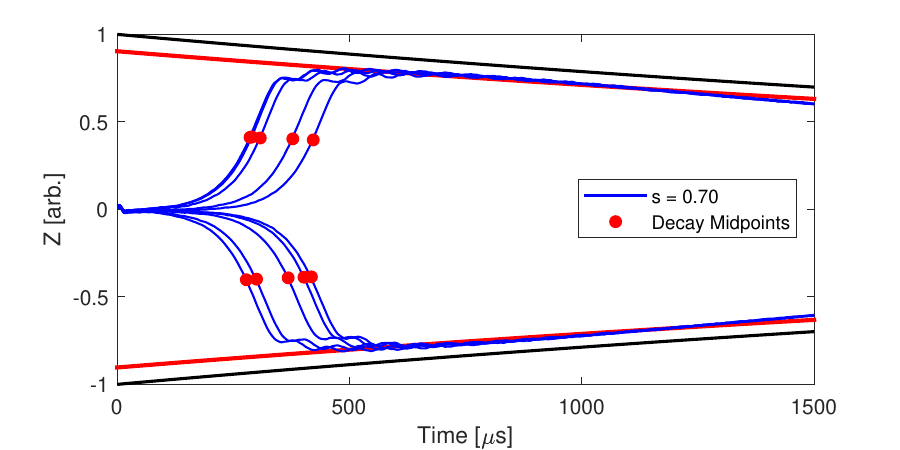}
\caption[Extraction of dynamical decay time information]{Some example trajectories (blue) for $s \,=\, 0.7$ with the midpoints $(t_{DD}\, , Z(t_{DD}))$ marked (red points). Also shown are the upper and lower bounds corresponding to the expected mean spin $J(t)$ (black) and the estimated location of the stable fixed points (red lines).} 
\label{fig:fig6p1p1_LMG_ddHalfEx}
\end{figure}

\begin{figure}[h] 
\renewcommand{\baselinestretch}{1}
\centering
\includegraphics[width=0.8\columnwidth]{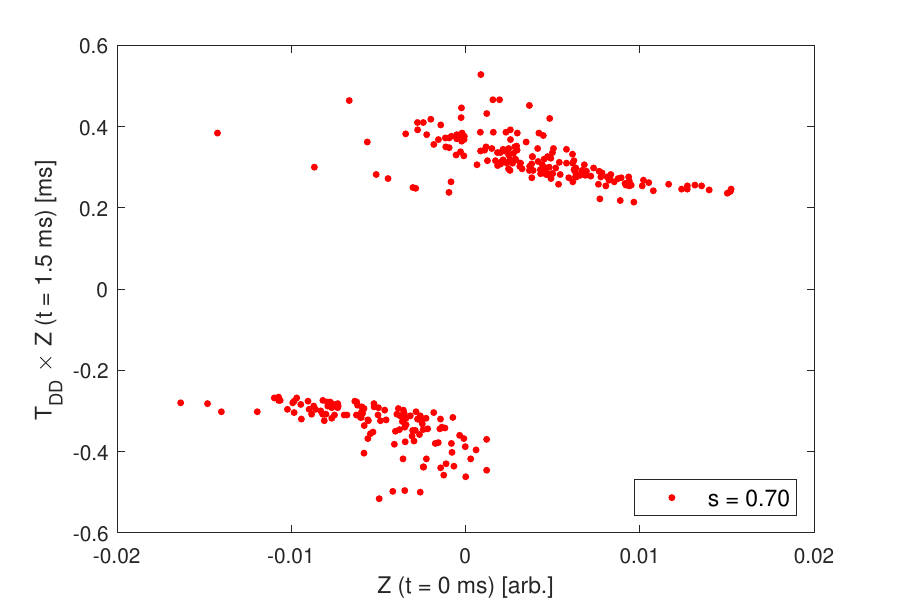}
\caption[Correlation between $Z(0)$ and dynamical decay time $t_{DD}$]{Correlation plot between the initial values of $Z$ (x-axis) and signed dynamical decay time $t_{DD}$ (y-axis). $t_{DD}$ is the time at which a trajectory reaches the midpoint between its initial and final values. If a trajectory evolves into the lower well, we take $t_{DD}$ to be negative.} 
\label{fig:fig6p1p1_LMG_ddHalfCorr}
\end{figure}

We can actually make that correlation a little clearer if we quantify the time it takes for a particular initial condition to decay to whichever fixed point it ends up in. In Figure \ref{fig:fig6p1p1_LMG_ddHalfEx} we show our method for extracting this information for a handful of example trajectories. We first calculate $Z(t) = J_{z}(t) / J(t)$, and then we find the time $t_{DD}$ at which $Z$ is equal to the halfway point between its initial and final values. Finally, if a given trajectory evolves into the lower well, we take $t_{DD}$ for that run to be negative. The results of this analysis can be seen in Figure \ref{fig:fig6p1p1_LMG_ddHalfCorr}.

All of this sounds rather promising, but there is one issue. The form of the LMG Hamiltonian implies that energy should be conserved. However, rather than being confined to an orbit of constant energy as shown in the phase space plots, it appears that the dynamical evolution is dissipative. For $s\, >\, 0.5$, the state we initialize the spin into, up along $\hat{x}$, is higher in energy than a state localized at one of the fixed points.

\begin{figure}[hp] 
\renewcommand{\baselinestretch}{1}
\centering
\includegraphics[width=1.0\columnwidth]{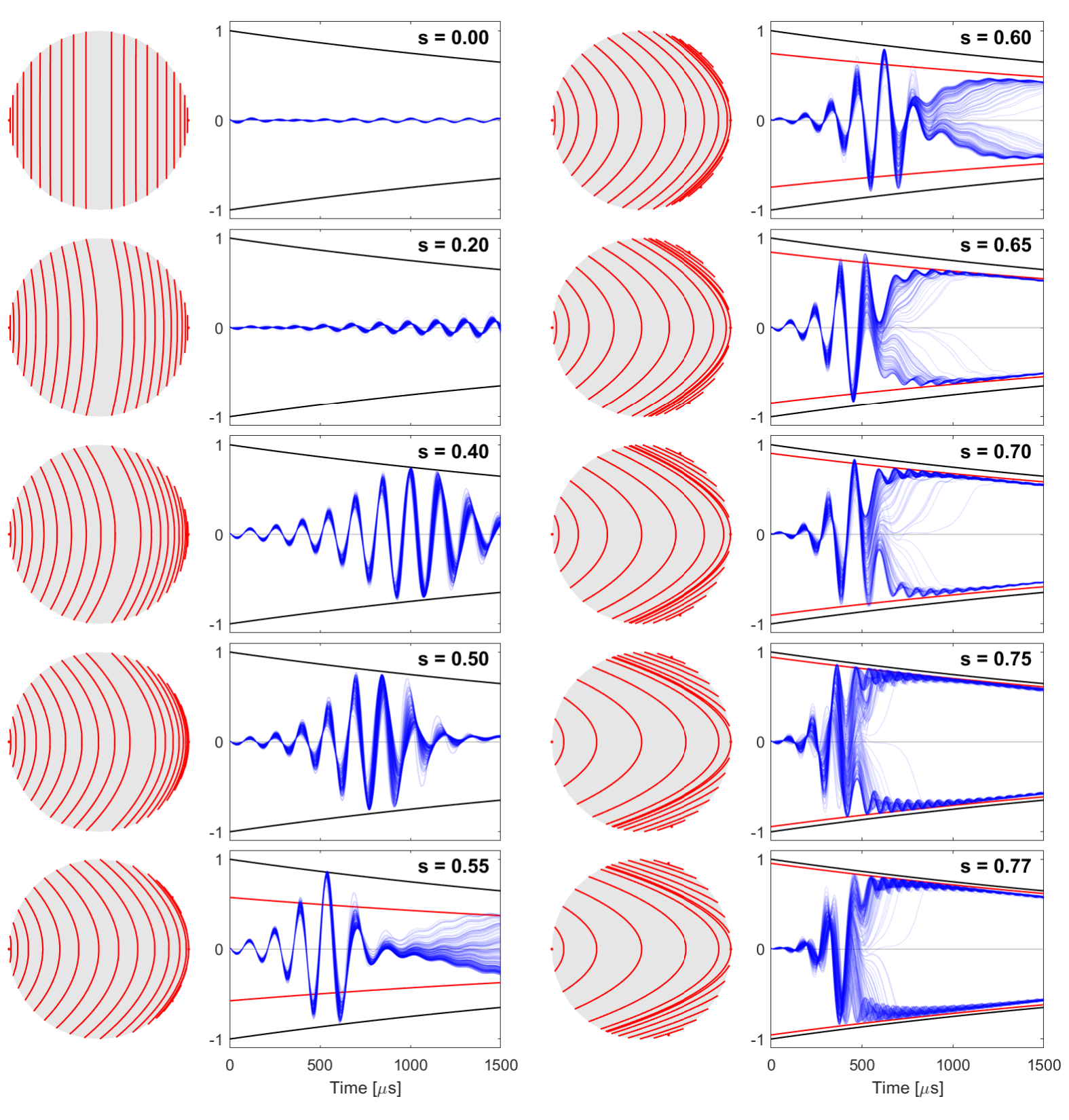}
\caption[LMG emulation symmetry breaking data, spin down along x]{LMG emulation experimental data with initial state spin down along $\hat{x}$ for several values of $s$ between 0 and 0.8. Phase space portraits showing the $y \, < \, 0$ hemisphere for each value of $s$ are shown to the left of each time series plot, with $+\hat{z}$ pointed up. The time series plots themselves consist of 300 shots of data (blue) which are made slightly transparent so that the density of trajectories can be visualized. The trajectories are bounded by the expected length of the mean spin $\bk{J}$ (black), and for $s \, > \, 0.5$, the estimated location of the fixed points are shown in red. Even though $-\hat{x}$ is a stable fixed point for all $s$ and $H_{LMG}$ is energy conserving, we see dynamical decay towards the fixed points with minimum energy.} \label{fig:fig6p1p1_LMG_dnX_allData}
\end{figure}

Actually, the problem is much more readily apparent if we instead start with an initial state pointing in the $-\hat{x}$ direction. Even though this point is the global maximum for the energy, it should still be a stable fixed point for all values of $s$. Despite this, the data for this experiment (Figure \ref{fig:fig6p1p1_LMG_dnX_allData}) still exhibits the same kind of dynamical decay toward the fixed points with minimum energy.

When it became clear to us that the dynamical decay was not expected behavior, we put together a semi-classical numerical model of the feedforward experiments which combines the expected noise model for our system with geometric rotations to calculate the expected trajectories. We will discuss this model in depth in Section \ref{chapter:ch6p1p3}, but the bottom line is that we found the primary driver of dynamical decay to be the signal delay, or latency, in the control loop. 

As one final point of discussion, looking at Figure \ref{fig:fig6p1p1_LMG_UpX_decay1}, we can also see that the mean time until the symmetry is broken decreases as $s$ is increased. This fact, however, is to be expected. Instead of fixing the timescale $\Lambda$ in Eqn. \ref{eq:eq2p3p1_LMGham} and directly changing $s$, it was much easier to fix the linear strength $(1-s)\Lambda$ and vary the nonlinear strength $s\Lambda$, and then calculate $s$ after the fact. As a consequence, $\Lambda \, =\, k + \alpha$ is increasing as $s$ increases, so the time scale $1/\Lambda$ should be decreasing.

\subsection{Dynamical Phase Transition Signatures}
\label{chapter:ch6p1p2}
\begin{figure}[H]
\renewcommand{\baselinestretch}{1}
\centering
\includegraphics[width=1.0\columnwidth]{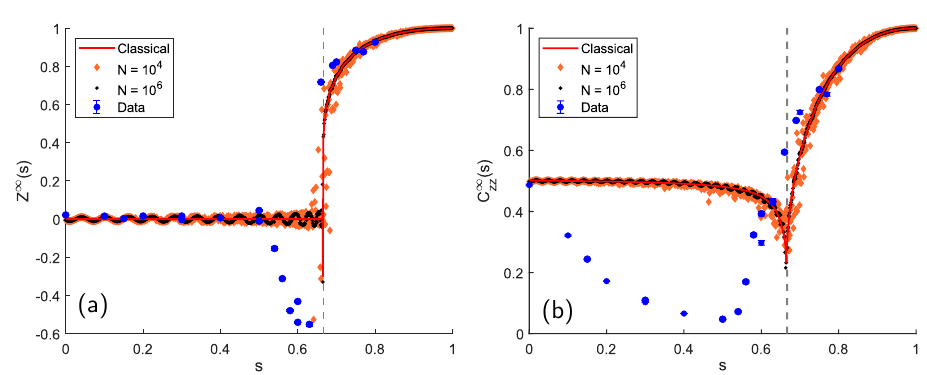}
\caption[Dynamical phase transition signatures in infinite-time order parameters.]{Dynamical phase transition comparison between experimental data (blue), numerical master equation simulations (orange, black), and classical simulation (red). The phase transition is expected to occur at s = 2/3 (grey dashed). (a) Time-averaged magnetization along Z. (b) Time-average of 2-body correlation function $C_{zz}^{\infty}$. The phase transition occurs at the expected location for $Z^{\infty}$, but there are notable deviations from the expected behavior in both parameters. This is likely due to the presence of non-trivial latency in the feedback loop. Simulated data reproduced with permission from \cite{Munoz2020_pSpinSim}.}  \label{fig:fig6p1p2_LMG-DPT}
\end{figure}

\paragraph{}
In the second experiment, we made measurements of a set of observables which constitute order parameters for the dynamical phase transition in the LMG \cite{Munoz2020_pSpinSim}. These observables are the long term averages of the z-magnetization $Z^{\infty}$ and the two-body correlation function $C_{zz}^{\infty}$, given respectively by

\begin{align} \label{eq:eq6p1p2_LMGorderparams}
	Z^{\infty} &\= \lim_{T \rightarrow \infty} \frac{1}{T} \int_{0}^{T} \frac{\bk{\hat{J}_{z}}_{t}}{J} dt, \\ 
	C_{zz}^{\infty} &\= \lim_{T \rightarrow \infty} \frac{1}{T} \int_{0}^{T} \frac{\bk{\hat{J}_{z}^{2}}_{t}}{J} dt,
\end{align}
where $\bk{\hat{O}}_{t} \,=\, \expval{\hat{O}}{\psi (t)}$ for any observable $\hat{O}$. We start with an initial state up along $\hat{z}$, and then we let the state evolve as in the previous experiment, and repeat the process to collect many shots. We then vary $s$, and then calculate the long time average order parameters, averaged again across all shots. The results of this experiment are shown in Figure \ref{fig:fig6p1p2_LMG-DPT}. 

The good news is that the location of the critical point for the phase transition is where we expect it to be, as seen in Figure \ref{fig:fig6p1p2_LMG-DPT}a. The critical point occurs when the initial state, in this case $\ket{+\hat{z}}$, lies on the separatrix. We can find it by finding the value of s at which the energy (Eqn. \ref{eq:eq2p3p1_LMGenergy}) at the poles equals the energy at the unstable fixed point ($+\hat{x}$). Doing so, we find the critical point should occur at $s \= 2/3$.

The bad news is that there are notable deviations from the expected behavior, especially in the 2-body correlation function. Both parameters fall below the expected value before the critical point, though $Z^{\infty}$ only does so beyond the bifurcation at $s \= 0.5$. Why? If we look at the trajectories, we can see that for $s \,> \, 0$ the trajectories seem to be decaying towards the fixed points. The reason for this was rather surprising, but in hindsight makes sense. For continuous time models, latency in the control loop is extremely important. 

\vspace{10pt}

\begin{figure}[h]
\renewcommand{\baselinestretch}{1}
\centering
\includegraphics[width=1.0\columnwidth]{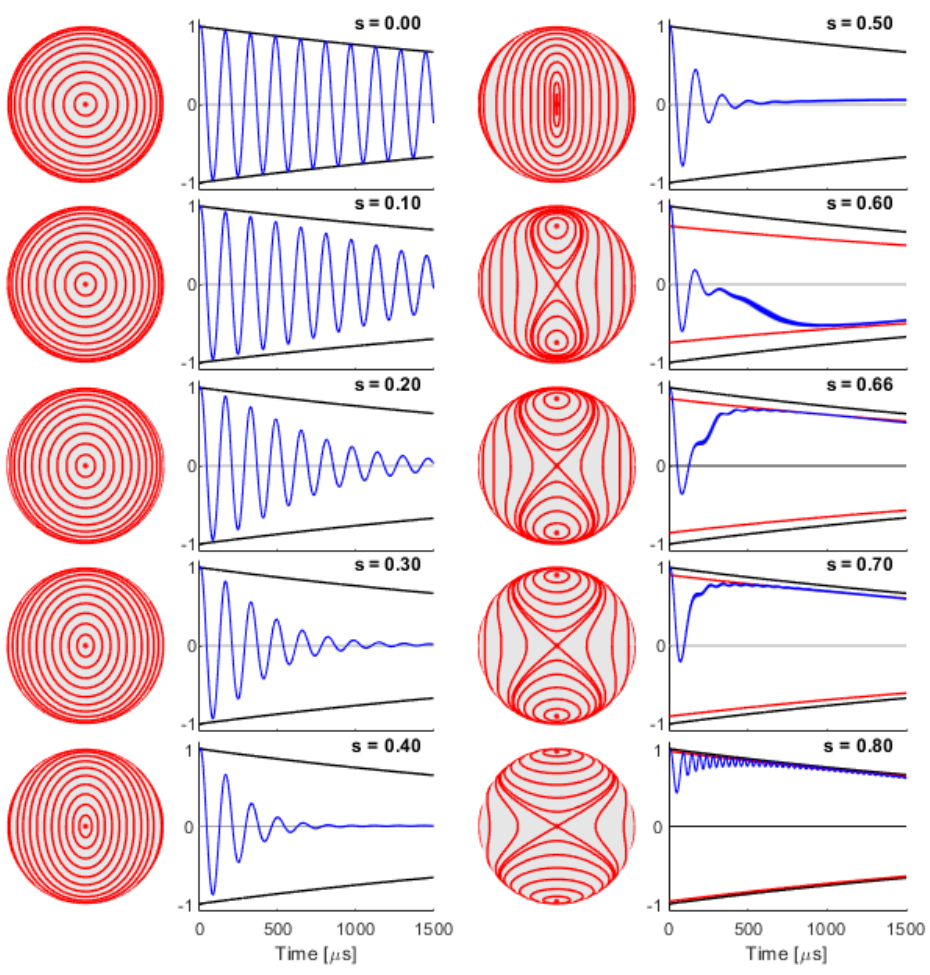}
\caption[Experimental data for LMG dynamical phase transition experiment.]{Experimental data for LMG dynamical phase transition experiment.}  \label{fig:fig6p1p2_LMG_upZ_DD}
\end{figure}

\FloatBarrier
\newpage

\subsection{Effect of Latency on Emulated Dynamics}
\label{chapter:ch6p1p3}
\paragraph{}
As we have seen in the previous sections, there is some unanticipated behavior in the dynamical evolution of our state. We know that it is not due to an excess of classical noise because of all the tests we can run that were discussed in Section \ref{chapter:ch4p4}. As such, it must be something to do with the feedforward control itself. 

After some time, we realized that signal latency is the most probable cause for the observed dynamical decay. Broadly speaking, the idea is that if the loop latency is large enough, the  control signal $C(t)$ is being applied at the wrong time. To put it another way, if $C(t)$ is the correct control to apply at time $t$, it is instead acting at time $t + \tau_{delay}$, at which point the state has evolved away. The torque seen by the spin would then not point in the correct direction with the correct strength to reproduce the intended mean-field dynamics. 

If this is so, why then does the state evolve towards the correct fixed points? Since those points represent steady-state solutions to the EoM for a given model, by definition the state never changes at those points, and so the delay time does not matter; the control always uses the correct measurement there.

The first thing to sanity-check this argument is to look at the relevant timescales. We can work out an approximate metric to compare the emulation and latency timescales as follows. The basic idea is that the measurement of the state, and, therefore, the control signal dependent on it, should change slowly within an interval of signal delay $\tau_{delay}$ so that the control error is minimized. For the LMG, the strength of the nonlinear field under control at each timestep is
\begin{equation} \label{eq:eq6p1p3_LMGjzW}
	\frac{k}{2}\frac{\bk{J_{z}}^{2}}{J^{2}} \= \frac{k}{2}\cos^{2}(\theta) \equiv \omega_{k}(\theta),
\end{equation}
where $k \= s\Lambda$. Rotations about $\hat{z}$ do not change the measurement outcome, so over a time $\tau_{delay}$, $\theta$ will change by at most $\alpha \tau_{delay}$, where $\alpha \= (1-s)\Lambda$. 

So our emulation procedure should work well if $\Delta\omega_{k}$ is small over $\tau_{delay}$ when compared to its full range. Writing it out, we have
\begin{equation} \label{eq:eq6p1p3_wkdelta}
\begin{split}
	\Delta\omega_{k}(\theta) &= \omega_{k}(\theta + \alpha \tau_{delay}) - \omega_{k}(\theta) \\
	&= \qty[\omega_{k}(\theta + 0) - \frac{k}{2}\sin(\theta)(\alpha \tau_{delay})] - \omega_{k}(\theta) \\
	&= - \frac{k}{2}\sin(\theta)(\alpha \tau_{delay}).
\end{split}
\end{equation}

Note that in the middle step we expanded out $\omega_{k}$ to first order using its Taylor series. Now, $\omega_{k}$ ranges between $\pm k/2$, and so we require
\begin{equation} \label{eq:eq6p1p3_ineq}
	\qty|\frac{k}{2}\sin(\theta)(\alpha \tau_{delay})| \p{\ll} \frac{k}{2}.
\end{equation}

Since $\qty|\sin(\theta)|$ is at most 1, this reduces to the requirement that $\alpha \tau_{delay} \ll 1$. For the LMG experiments shown in the previous sections, we exclusively operated with $\alpha \= 2\pi\times 6.25$ kHz. The latency was also measured to be $\tau_{delay} \p{\approx} 6\,\mu$s. We therefore have $\alpha \tau_{delay} \p{\approx} 0.23$. This isn't exactly negligible, but it is at least on the right side of the inequality. 

\begin{figure}[hp]
\renewcommand{\baselinestretch}{1}
\centering
\includegraphics[width=1.0\columnwidth]{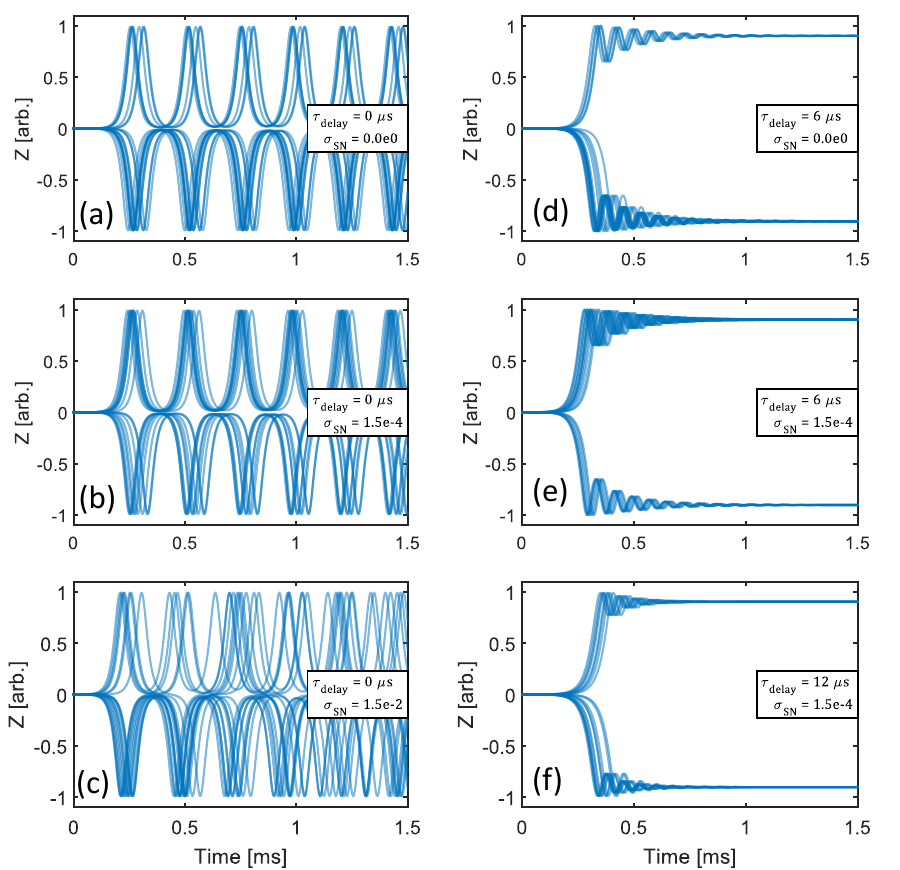}
\caption[Numerical simulations comparing the effects of latency and noise]{Numerical simulations comparing the effects of latency and noise in feedforward emulation of the LMG. Each plot shows 20 runs with $s \,=\, 0.7$. The initial state is a SCS up along $\hat{x}$. (a-c) No latency and varying amounts of shot noise. There is no dynamical decay, but the shot noise clearly affects the run-to-run coherence. (d) No noise, but latency typical of our system. (e) Typical shot noise and typical latency. (f) Typical shot noise and twice the typical latency. There is a noticeable decrease in the time to decay to the fixed points with increased latency. }  \label{fig:fig6p1p3_latencySims}
\end{figure}

The preceding argument was unfortunately developed late in the experiment, long after our data was taken, and so we did not have time to test it experimentally, which would involve decreasing $\alpha$. However, we expect that there would be two side effects of such a change. First, since the LMG $s$ parameter is a measure of the relative strength of the two rotations, and the maximum absolute strength of the nonlinear rotation is constrained by our equipment, we would be able to reach a higher value of $s$ than 0.8, our previous limit. The second effect, however, is that the effective emulation time should decrease, since the overall rotation strengths for a given $s$ have been lowered. While the former is a clear benefit, it remains to be seen whether the latter is a problem. If it is, it would interfere with experiments like the observation of spontaneous symmetry breaking in Section \ref{chapter:ch6p1p1} which center around the effects of quantum noise.

We did, however, develop a set of numerical simulations which do lend much credibility to our arguments. In Figure \ref{fig:fig6p1p3_latencySims}, we show several plots of our simulations with various amounts of latency and shot noise for $s \= 0.7$. The initial state in all our simulations is a spin coherent state oriented up along $\hat{x}$, which accounts for the expected amount of quantum projection noise by adding a Gaussian random number to the initial orientation angles. 

In \ref{fig:fig6p1p3_latencySims}(a), there is no shot noise and no latency, and also no dynamical decay. From this we can say that the QPN is clearly responsible for breaking the symmetry, but this isn't really a surprise. In \ref{fig:fig6p1p3_latencySims}(b), we incorporate probe shot noise typical in our system, and in (c) we increase the SN by a factor of 100. We can see that the coherence of the individual trajectories is decreased as the shot noise is increased. In \ref{fig:fig6p1p3_latencySims}(d) we remove the shot noise and add in 6 $\mu$s of latency, which is typical. With that, the decay has returned, and we can see that if we increase the latency, as in \ref{fig:fig6p1p3_latencySims}(f), the settling time decreases noticeably. 

As a final point, it should be noted that the finite update rate also plays a similar, albeit lesser role in the observed dynamical decay. Using essentially the same argument that led us to Equation \ref{eq:eq6p1p3_ineq}, if the sampling period of the controller is too large, then the piece-wise constant nature of the control signal produces an error at each step that grows in a sawtooth-like manner. These errors would accumulate over time, leading to the same sort of decay, but to a much lesser degree than an equivalent duration of latency.

\section{Emulation of the QKT}
\label{chapter:ch6p2}
\paragraph{}
The second model we explore, the Quantum Kicked Top, is notable because its classical counterpart exhibits global chaos for sufficiently large values of $k$, the nonlinear driving strength \cite{Haake1987}. It is this aspect we would like to explore in the third experiment, by attempting to estimate the maximal Lyapunov exponents which characterize the degree of chaos. Then, in the final experiment, we will initialize a so-called Floquet time crystal (FTC) state, notable because of its robustness to perturbations. We will attempt to quantify this robustness by looking at the average power spectrum and varying the driving strength.

Control loop latency may be a critical challenge to overcome when emulating continuous time models such as the LMG, but this is not the case for discrete time models like the QKT. Since each step happens in isolation and the control depends only on the value of the previous step, we can always add time in between control steps to ensure that the measurement used for control is accurate. This assumes that the state does not change in between steps, which should be true if the control fields are disabled during the measurement. Unfortunately, that does not mean the KT emulation experiments we ran were trouble-free.

\subsection{Chaotic Dynamics - Estimating Lyapunov Exponents}
\label{chapter:ch6p2p1}
\paragraph{}
The first of the Kicked Top experiments we would like to attempt, and really the main motivating factor for this work, is to try to make an observation of chaotic dynamics in a semi-classical system. For classical systems, we often characterize chaotic regions using the Lyapunov exponents, which are a measure of how quickly infinitesimally separated neighboring trajectories diverge \cite{Skokos2010}. For two points in a chaotic region of phase space, separated by an initial vector $\mathbf{\delta_{0}}$, the maximal Lyapunov characteristic exponent (mLCE) is defined as
\begin{equation}
	\lambda_{max} \= \lim_{t\rightarrow\infty} \lim_{\qty|\mathbf{\delta_{0}}|\rightarrow 0} \ln(\frac{\qty|\mathbf{\delta(t)}|}{\qty|\mathbf{\delta_{0}}|}).
	\label{eq:eq6p2p1_lyapMaxDef}
\end{equation}

We say \textit{maximal} because if you fix one of the two points and the size of the initial separation, the Lyapunov exponent at that point actually depends on the direction of the separation vector. We can calculate the Lyapunov spectrum, which has the same dimensionality as the underlying phase space, by looking at the Jacobian as follows. For a dynamical system with evolution equations $\dot{x}_{i} = f_{i}(\mathbf{x})$, the Jacobian is defined as 
\begin{equation}
	J_{ij}(t) = \frac{ df_{i}(\mathbf{x})}{dx_{j}} \eval_{ \mathbf{x}(t) }.
\end{equation}
The Jacobian defines the evolution of the tangent vectors at a point, given by a matrix $Y$, according to $\dot{Y} \= JY$. The Lyapunov spectrum is given by the eigenvalues of the matrix
\begin{equation}
	\Lambda \= \lim_{t \rightarrow \infty} \frac{1}{2t} \ln( Y(t) Y^{T}(t) ).
	\label{eq:eq6p2p1_lyapSpecDef}
\end{equation}

\begin{figure}[H]
\renewcommand{\baselinestretch}{1}
\centering
\includegraphics[width=0.9\columnwidth]{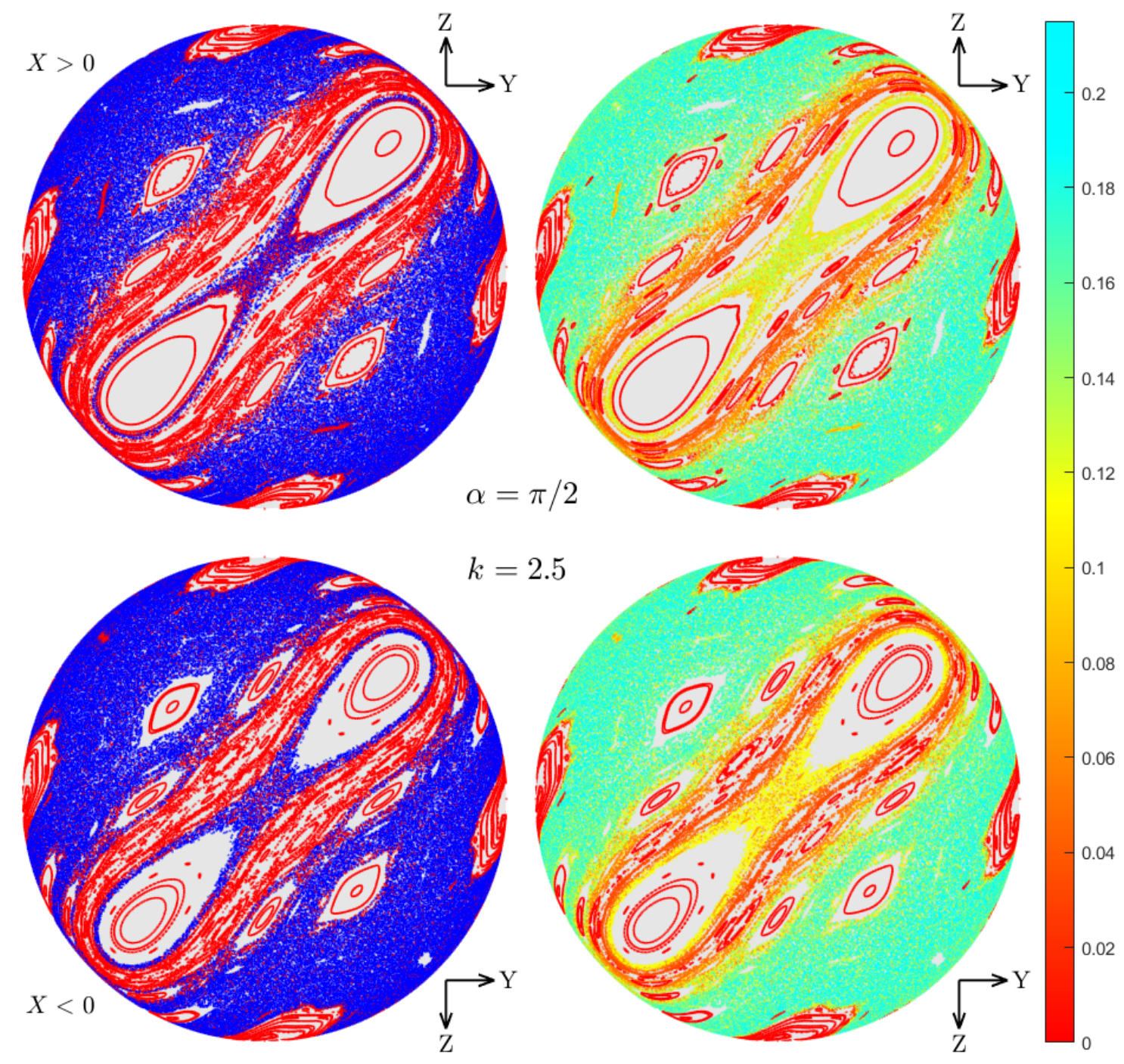}
\caption[Kicked Top phase space colored by local maximal Lyapunov exponents]{Kicked Top phase space portraits for $\alpha \, = \, \pi/2$ and $k \, = \, 2.5$. (Right) Regular trajectories are colored red and chaotic motion is blue, with the threshold being $\lambda_{max} = 0.01$. (Left) Trajectories are colored by their local maximal Lyapunov exponent.}  \label{fig:fig6p2p1_KTphaseLyapComp}
\end{figure}

The largest of these eigenvalues at a given point is, of course, $\lambda_{max}$. Typically, a connected region of phase space characterized by positive Lyapunov exponents undergo ergodic mixing, so the maximum exponent is also the global maximum. So, given the equations of motion for a system, we can calculate the spectrum and find the mLCE. Figure \ref{fig:fig6p2p1_KTphaseLyapComp} shows a set of phase space portraits for the Kicked Top, which on the left is colored red for regular orbits and blue for chaotic orbits ($\lambda > 0$), and on the right is colored by the local maximal lyapunov exponents.

This is all well and good, but how do we estimate the Lyapunov exponents based on data from the experiment? Currently, there exists a sizable body of research concerning the extraction of information about a chaotic system from a time series measurement record. As early as 1980, methods for the reconstruction of phase space geometry from observed data were produced \cite{Packard1980, Fraser1989}, and later in 1990 researchers started to come up with methods to calculate Lyapunov exponents from time series \cite{Bryant1990, Abarbanel1992}. Nowadays, there is even a full software package known as TISEAN (for TIme SEries ANalysis) \cite{TISEAN1999, Kantz_Schreiber_2003} geared toward this exact problem. In particular, it has methods for the estimation of both local and global maximum Lyapunov exponents.

However, there are two serious problems standing in the way of the use of such tools. For one, the minimum number of steps in the time series must be much larger than what we are currently capable of in this experiment, north of several hundred. TISEAN does not seem to be capable of providing accurate answers with only a few tens of points, which is currently all we have available to us here. The other, much worse problem is that we do not have a complete measurement record, since we can only measure $J_{z}$. As such, we cannot estimate the angular separation between two trajectories after they have sufficiently diverged. To put it another way, we cannot tell the difference between two vectors on opposite sides of the unit sphere if they have the same z-component. 

\begin{figure}[H]
\renewcommand{\baselinestretch}{1}
\centering
\includegraphics[width=1.0\columnwidth]{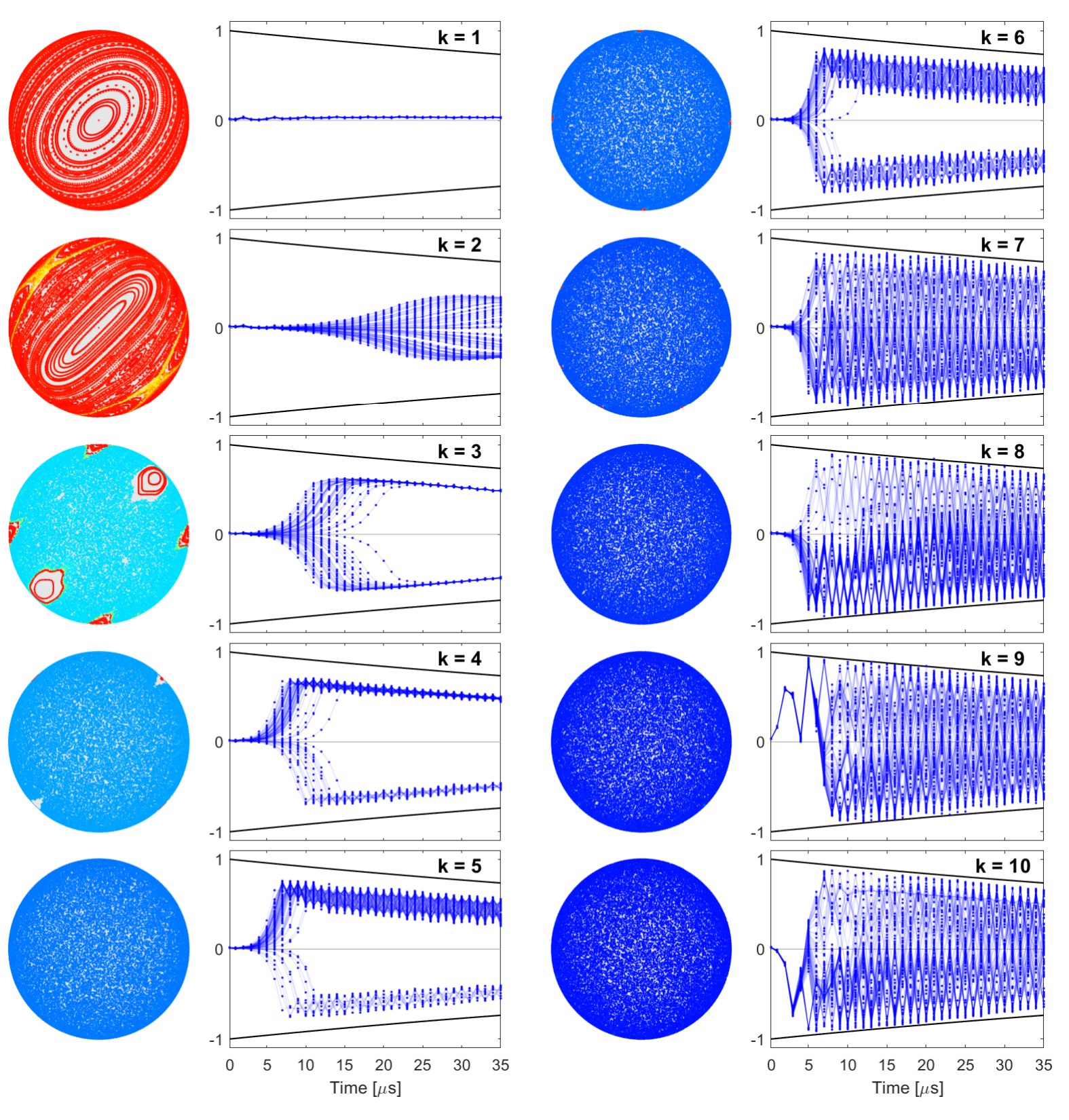}
\caption[Experimental data for QKT showing chaotic trajectories.]{Experimental data for KT emulation with $\alpha \, =\, \pi/2$. Phase space portraits show the $x \, > \, 0$ hemisphere and are colored according the the local mLCE using the same color scale as in Figure \ref{fig:fig6p2p1_KTphaseLyapComp}.}  \label{fig:fig6p2p1_QKT_UpX_chaos_allData}
\end{figure}

\newpage
That said, there is still a way for us to make a rough estimate of the mLCE. If we start with the spin oriented up along $\hat{x}$ and make many repetitions of the experiment, for at least a few points at the beginning of the time series, the trajectories will be close enough that we can simply calculate the standard deviation of the elevation angle $\theta_{n} \, =\, \acos(J_{z}(t_{n}) /J(t_{n}))$ across shots. We divide out the running value of $J$ in order to remove the spin decay component. Consider an ensemble of phase space points $x_{i}$ that includes its mean $\bk{x}$. Writing down the standard deviation and manipulating it a bit, we find 
\begin{align}
\begin{split} \label{eq:eq6p2p1_lyapStdDev}
	\sigma_{x}(t)  \, &= \, \sqrt{\frac{1}{N} \sum \qty(x_{i}(t) - \bk{x(t)}) } \\
	              &= \, \sqrt{\frac{1}{N} \sum \qty|\delta(t)|^{2} } \\
	              &= \, \sqrt{\frac{1}{N} \sum \qty|\delta(0)|e^{2\lambda t} } = \, \sigma_{x}(0) \, e^{\lambda t}.
\end{split} 
\end{align}

\begin{figure}[hp]
\renewcommand{\baselinestretch}{1}
\centering
\includegraphics[width=1.0\columnwidth]{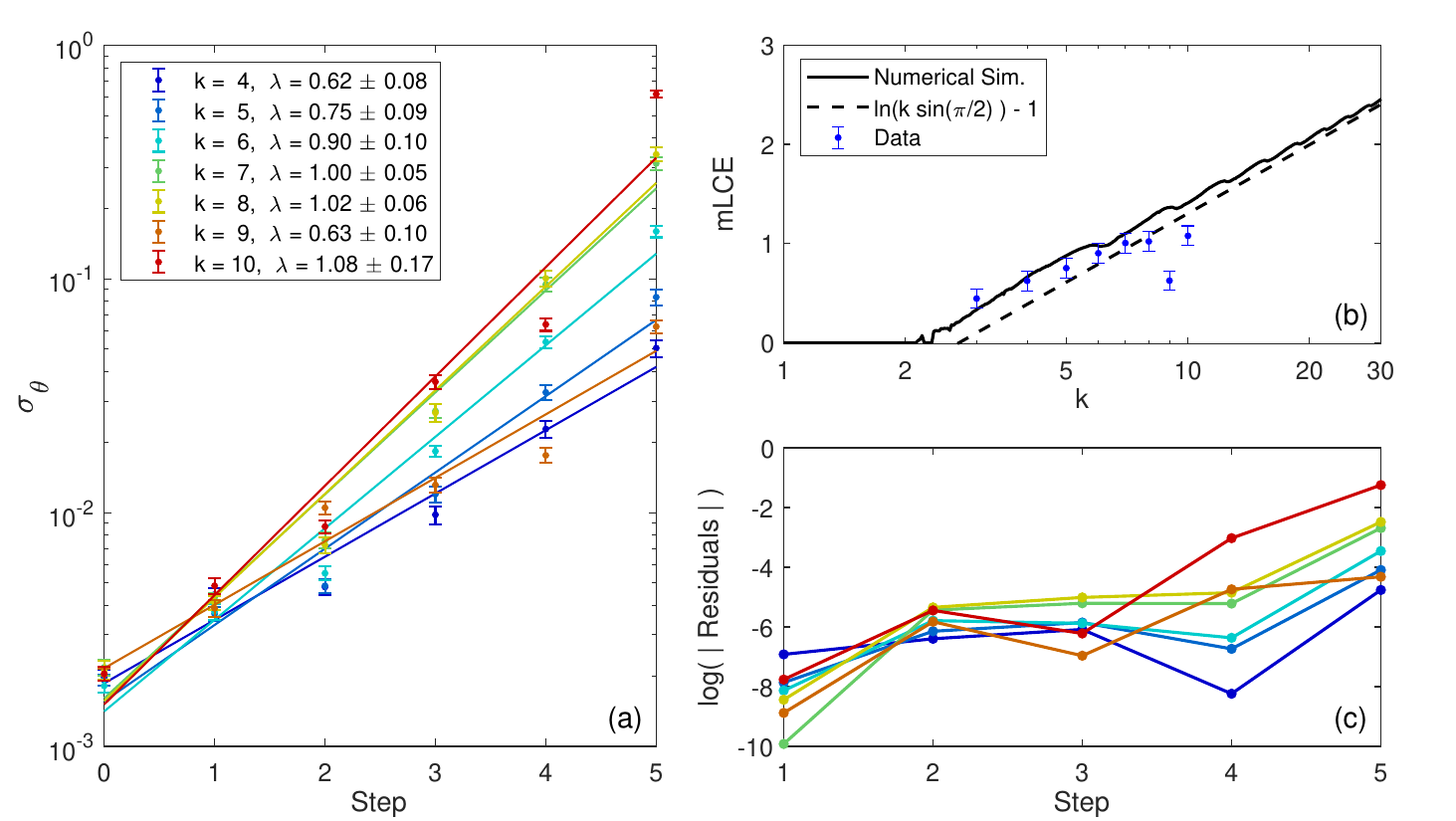}
\caption[Estimation of KT Lyapunov exponents for $\alpha \, = \, \pi/2$ from experimental data. ]{Statistical estimation of the KT mLCE for $\alpha \, = \, \pi/2$. (a) Standard deviation of the elevation angle $\theta{n} \, =\, \acos(J_{z}(t_{n}) /J(t_{n}))$ across all trajectories for the first 5 emulation steps (points), and exponential fits for each $k$ (lines). (b) Numerical calculation of the global mLCE (solid), asymptotic form for the mLCE for large $k$ (dashed), and experimental estimates for the mLCE (points). (c) log of the absolute values of the fit residuals from plot (a).}  \label{fig:fig6p2p1_KT_LyapStdData_AmpFree}
\end{figure}

The problem of not being able to measure the azimuthal angle will still hurt us here, but we can at least say that if we do this for only $\theta$, the results will be a lower bound on the true mLCE.  The raw stroboscopic data for this experiment can be found in Figure \ref{fig:fig6p2p1_QKT_UpX_chaos_allData}. The phase space portraits for each value of $k$ are colored by the local mLCE according the scale in Figure \ref{fig:fig6p2p1_KTphaseLyapComp}. In Figure \ref{fig:fig6p2p1_KT_LyapStdData_AmpFree} we show our statistical analysis following the preceding argument. 

After calculating the standard deviation of the elevation angle across all trajectories for the first 5 steps, we fit the data to an simple exponential function $A e ^{\lambda\, n}$, with no y-offset. We should note, however, that we found that the fitting results were more stable if we did a log-linear fit, using a fitting function $y = \ln(\sigma_{\theta}) \,=\, \ln(A) + \lambda \, n$.

Much to our surprise, we can see in Figure \ref{fig:fig6p2p1_KT_LyapStdData_AmpFree}(b) that there is actually decent agreement between our extremely rough estimates of the mLCE and the numerical calculation. Not only that, our estimates fall squarely below the numerical results, which is what we should expect from neglecting one of the two degrees of freedom in our system. The actual angular separation can only be growing faster than we think it is as we take more steps. 

\newpage
\FloatBarrier

\subsection{Time Crystal Phase}
\label{chapter:ch6p2p2}
\paragraph{}
The second KT experiment that we would like to attempt is to initialize the spin into a relatively new type of out-of-equilibrium phase know as a time crystal phase. Time crystals, like their spatial counterparts, are characterized by a broken symmetry. For everyday crystals, it is continuous translational symmetry that is broken due to the periodic structure of its atoms. Likewise, for a time crystal, it is time-translational symmetry that is broken. 

Time crystals can be categorized according to whether they arise from discrete or continuous time systems. As one might guess, our system falls into the former category. In fact, nearly all experimental realizations of time crystal are derived from periodically driven systems such as ours\cite{Ball2018}. We call such systems Floquet time crystals (FTC) because the periodic drive naturally gives rise to a stroboscopic description characterized by a Floquet operator. When we say that the discrete time symmetry is broken in a FTC, we mean that the system begins to exhibit subharmonic oscillations at a multiple of the driving period. 

To put things more rigorously, we consider a periodically driven Hamiltonian system with period $T$, so that $\hat{H}(t+T) \, =\, \hat{H}(t)$. An FTC can be defined as a class of intitial states $\qty{\ket{\psi_{0}}}$ and an observable $\hat{O}$ such that the time-depend expectation value in the limit of large system size $N$, given by \cite{Russomanno2017, Huang2018}
\begin{equation}
	f_{O}(t) \= \lim_{N \rightarrow \infty} \expval{\hat{O}}{\psi(t)},
\end{equation}
satisfies the following conditions: \\
\begin{enumerate}
	\item \textbf{Time-translation symmetry breaking}: $f_{O}(t+T) \,\neq\, f_{O}(t)$ while $\hat{H}(t+T) \, =\, \hat{H}(t)$.
	\item \textbf{Rigidity}: $f_{O}(t)$ has a fixed oscillation period, without needing to fine-tune parameters in $\hat{H}$.
	\item \textbf{Persistence}: the oscillations of $f_{O}(t)$ persist for an infinitely long time.
\end{enumerate}

Our FTC experiment is mainly based on the work in \cite{Munoz2022_TC}, which focuses on FTC phases in kicked p-spin models, an extension of the Quantum Kicked Top to $p$-body interactions. The QKT is the $p \,=\, 2$ case of the kicked p-spin Hamiltonian. For the QKT, the spin is in the FTC phase when $\alpha$ is nominally equal to $\pi$, with the state initially pointing along the $\hat{z}$-axis. Under these conditions, the spin exhibits subharmonic oscillations with a period twice that of the drive, flipping from spin up to spin down along $\hat{z}$.

\begin{figure}[h]
\renewcommand{\baselinestretch}{1}
\centering
\includegraphics[width=0.7\columnwidth]{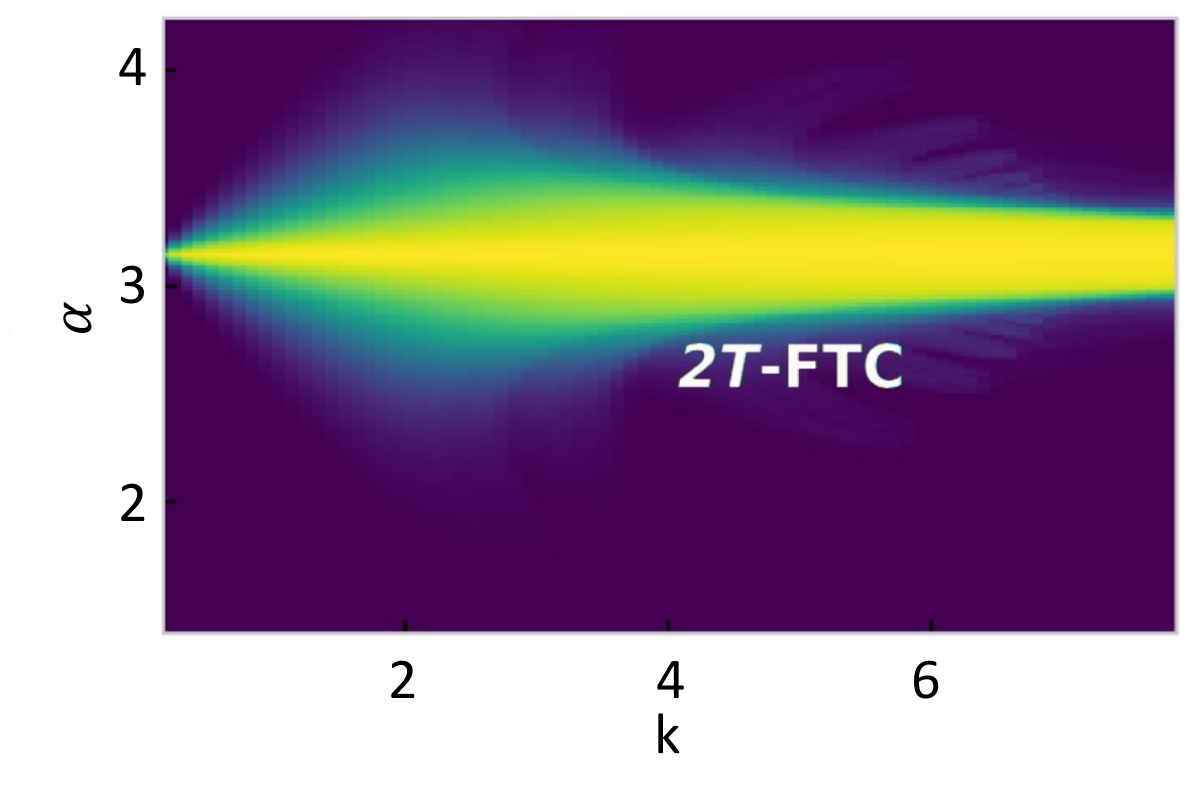}
\caption[KT time crystal phase rigidity]{Power spectral density of $J_{z}$ as a function of the linear rotation strength $\alpha$ and nonlinear strength $k$ for a state initially up along $\hat{z}$ in the Kicked Top model. The lighter regions correspond to a Floquet time crystal phase. The expansion of the FTC phase region as $k$ grows is a manifestation of the rigidity property. Reproduced with permission from \cite{Munoz2022_TC}.}  \label{fig:fig6p2p2_FTCphase}
\end{figure}

The second property, rigidity, implies that we can adjust either $\alpha$ or $k$ to some degree and still remain in the FTC phase. In Figure \ref{fig:fig6p2p2_FTCphase}, we reproduce a phase diagram for the Kicked Top FTC from \cite{Munoz2022_TC} based on the power spectrum of $J_{z}$. The system is in the FTC phase in the lighter region. The fact that the width of the FTC phase broadens as $k$ increases up to around $\pi$ is a reflection of the rigidity property.

\begin{figure}[ht]
\renewcommand{\baselinestretch}{1}
\centering
\includegraphics[width=0.7\columnwidth]{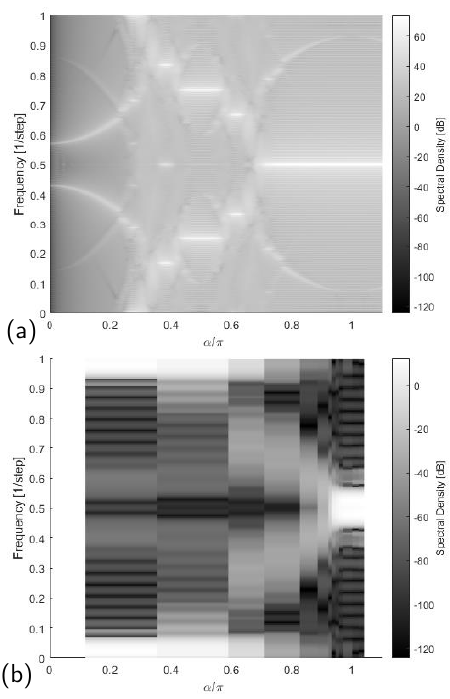}
\caption[KT time crystal phase rigidity, comparing theory and data.]{KT time crystal phase power spectral density as a function of the linear rotation strength $\alpha$, with $k \,=\, 2.7$. Shown are simulated data in (a) and experimental data in (b). The long solid bar around $\alpha \, = \, \pi$ is a consequence of the rigidity propery of a time crystal phase. While the simulation shows the TC phase is robust out to nearly $0.8 \pi$, in the experiment we only see rigidity to around $0.93\pi$. }  \label{fig:fig6p2p2_KT_timeCrystal}
\end{figure}

So, for our emulation experiment, we start with spin up along $\hat{z}$. Since $k$ does not appear to matter as long as it is non-zero, we fix $k \, =\, 2.7$, and vary $\alpha$. We then compute the power spectrum of $J_{z}$ and average it over many runs of the experiment. Our data is shown in Figure \ref{fig:fig6p2p2_KT_timeCrystal}(a). We quantify the rigidity by measuring the FWHM of the region over which the $2T$ subharmonic frequency is the dominant component of the spectrum.

Comparing (a) and (b), we see that the expected region of rigidity is much larger than the observed region, by about a factor of three. Despite this, it is clear that there is still a degree of rigidity in the data, with the subharmonic oscillations persisting out to around $(1 \pm 0.07)\pi$. It is also the case that if we turn off the nonlinear drive, the FTC phase disappears entirely when we deviate from $\alpha \, = \, \pi$.

\chapter{Summary and Outlook}
\label{chapter:ch7}
\paragraph{}
In this work, despite some experimental shortcomings, we have seen an encouraging amount of success in reproducing the dynamical behavior of the models we chose to implement. In the case of the LMG, we found that the collective spin settled in to the expected locations for the stable fixed points, and when we initialized our state at the unstable point for $s > 0.5$, it underwent spontaneous symmetry breaking towards the bifurcated fixed point pair. We also saw that the order parameter $Z^{\infty}$, the long time average magnetization along z, exhibited the dynamical phase transition at the expected value of $s\=2/3$. For the QKT, despite significant barriers in data analysis, we calculated Lyapunov exponents that were on the correct order of magnitude as our theoretical expectations. We also saw a hint of the rigidity expected for the Floquet time crystal state with $\alpha \=\pi/2$ and spin oriented up along $\hat{z}$.

That said, there is significant room for improvement, especially in the case of stroboscopic models like the QKT. We found that the amount of signal latency in the control loop caused unexpected problems, and indeed was the primary driver of dynamical decay for the continuous time LMG model. The decay of the collective spin towards the models fixed point also significantly impacted the behavior of the order parameters around the phase transition. We made an argument that latency in our system should be small when compared to the linear rotation rate, and showed that we were operating in a regime where $\alpha \cdot t_{delay} \p{\approx} 0.2$, which is less than unity but certainly is not small enough to be negligible.

For stroboscopic models, on the other hand, one can make an argument that latency should not play a role at all; all we need to do is wait the appropriate amount of time before the nonlinear update step so that we are using the correct measurement. However, latency combined with what appear to be signal transients in the measurement that occur when a magnetic drive is switched on and off by the controller contributed to measurement errors that more than halved the window of rigidity for the time crystal experiment. Also, because we were limited in the maximum drive strength for each type of rotation, a single evolution time step took a significant portion of the total emulation time available to us. The total number of evolution time steps available to us was thus limited to less than 25. Moreover, we were limited in that we could only measure a single projection of the collective spin, $J_{z}$. These two facts conspired to prevent us from using conventional analysis techniques for measuring Lyapunov exponents accurately, which typically require a full description of the state and several hundred evolution steps. It is quite surprising that estimating the exponents by considering the standard deviation of the evolution time series across many runs of the experiment yielded a result that was even remotely close. 

There are a few other problems with the experiment which ought to be addressed as well. For one, in our experiment there was a very small angular misalignment of the bias and the probe on the order of a tenth of a degree. If the probe and bias are not parallel, the polarimetry signal will be modulated by the Larmor precession of the atoms when they are oriented away from the z axis. Moreover, this misalignment introduces an error in our measurement of $J_{z}$, which leads to an error in estimation of the rotation strength we are applying to the system. We have seen both of these signatures when measuring the polarimetry signal with the FPGA. The DAQ system is limited to a sample rate of 250 kHz, which is equal to the Larmor frequency we operated at, and as such the Larmor signal was not within its Nyquist range. Because of this, we did not realize the misalignment was a problem until very late in the project, and we did not have the time required to physically correct it before the project ended. 

The other issue is of course, the relative time scales of the feedforward experiment. Decoherence of the collective spin puts a hard limit on the total emulation time available to us, and the loop latency, the controller sample rate, and the linear Rabi frequency all contribute to limiting the minimum update time. Finding a good balance between all these parameters is a central challenge to overcome in the implementation of this technique. 

Despite the shortcomings of our demonstration, clearly this technique shows promise as a way to explore nonlinear dynamics in Hamiltonian systems. There are many interesting questions that can be explored with this method, and here we were only able to scratch the surface. If issues such as those discussed above can be addressed, the QMF protocol has much potential to supplement early quantum computation efforts by providing a means of verifying the results of large quantum simulations.

\phantomsection

\appendix

\chapter{LabVIEW Block Diagrams For Feedforward Control Programs}
\label{chapter:labview}
\paragraph{}
Here we show the back-panel block diagrams for the LabVIEW FPGA programs which run the feedforward control experiments. Most of the operative details are discussed in Sections \ref{chapter:ch5p2} and \ref{chapter:ch5p3}, so only brief comments are are given for each figure, with some minor details not discussed previously.

\newpage

\begin{figure}[ht]
\renewcommand{\baselinestretch}{1}
\centering
\includegraphics[width=1.0\columnwidth]{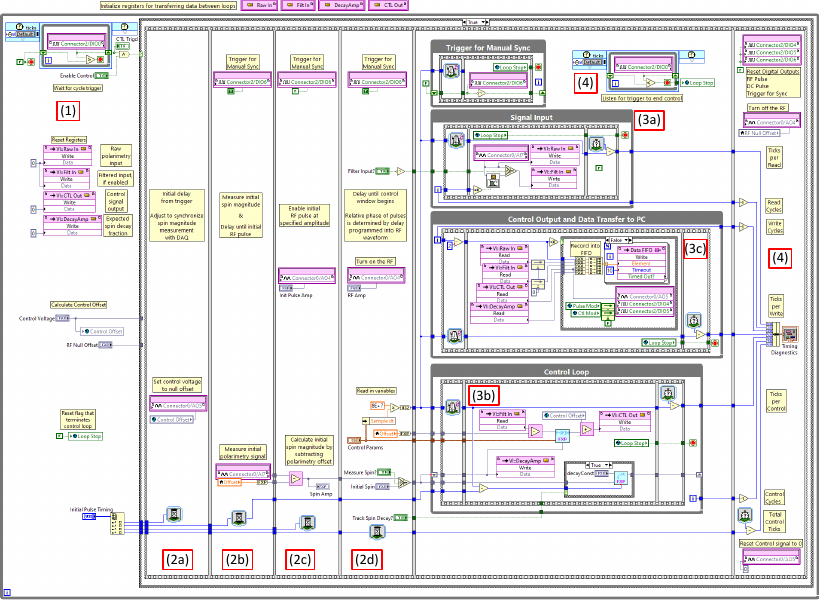}
\caption[LabVIEW block diagram for LMG emulation FPGA program]{LabVIEW block diagram for feedforward emulation FPGA program. Following the procedure detailed in Section \ref{chapter:ch5p2}, the program waits for a trigger (1), then executes a series of timed actions (2a-d) to measure the initial spin length (2b) and apply a state preparation pulse (2c). Control is then executed in three simultaneous timed loops: Input (3a), Control (3b) and Output (3c). A digital port is also toggled at each stage and then at the loop rate to aid in external synchronization. Finally, on a second trigger (4) the loops exit, the program resets the IO ports, and some diagnostic information is recorded.} \label{fig:fig9pa_lv_main}
\end{figure}

\begin{figure}[ht]
\renewcommand{\baselinestretch}{1}
\centering
\includegraphics[width=0.9\columnwidth]{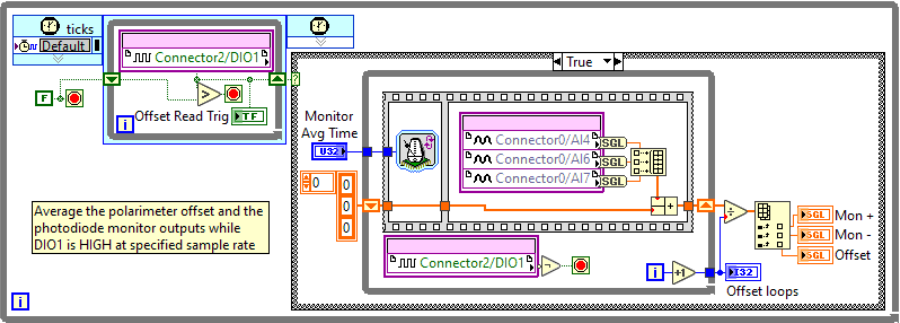}
\caption[LabVIEW block diagram for polarimetry offset measurement]{LabVIEW block diagram for polarimetry offset measurement, averaged over a duration controlled by an external trigger window sent by the DAQ. This loop also measures the balance detector monitor ports so that the total probe power can be monitored shot to shot. }  \label{fig:fig9pa_lv_mainOffset}
\end{figure}

\begin{figure}[hb]
\renewcommand{\baselinestretch}{1}
\centering
\includegraphics[width=1.0\columnwidth]{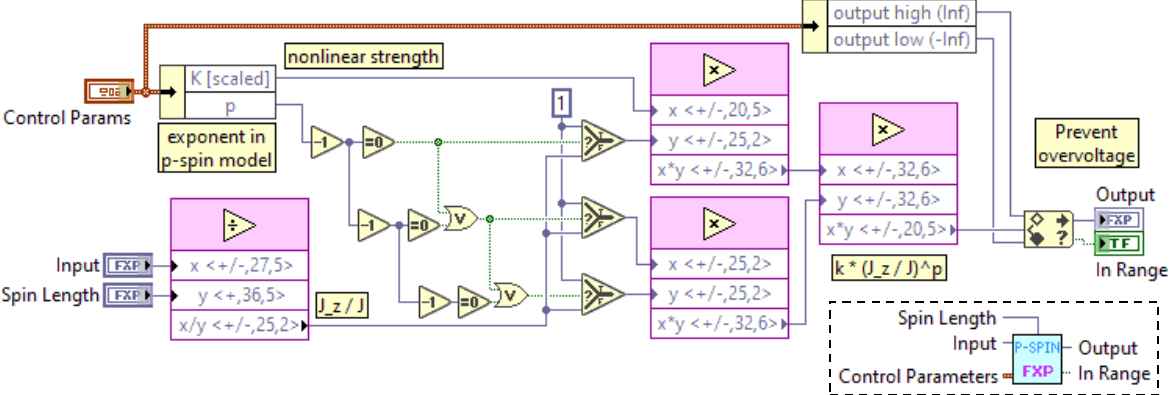}
\caption[LabVIEW block diagram for LMG control law]{LabVIEW block diagram for the LMG control law. The program was designed around a generalization of the LMG Hamiltonian known as the p-Spin model, in which the nonlinear term is an arbitrary power of $J_{z}$. Mainly due to resource constraints in the FPGA, only models with $p \, = \,$ 1, 2, 3, and 4 can be simulated with this program. This program's subVI logo is shown in the dashed box.}  \label{fig:fig9pa_lv_pSpin}
\end{figure}

\begin{figure}[htp]
\renewcommand{\baselinestretch}{1}
\centering
\includegraphics[width=1.0\columnwidth]{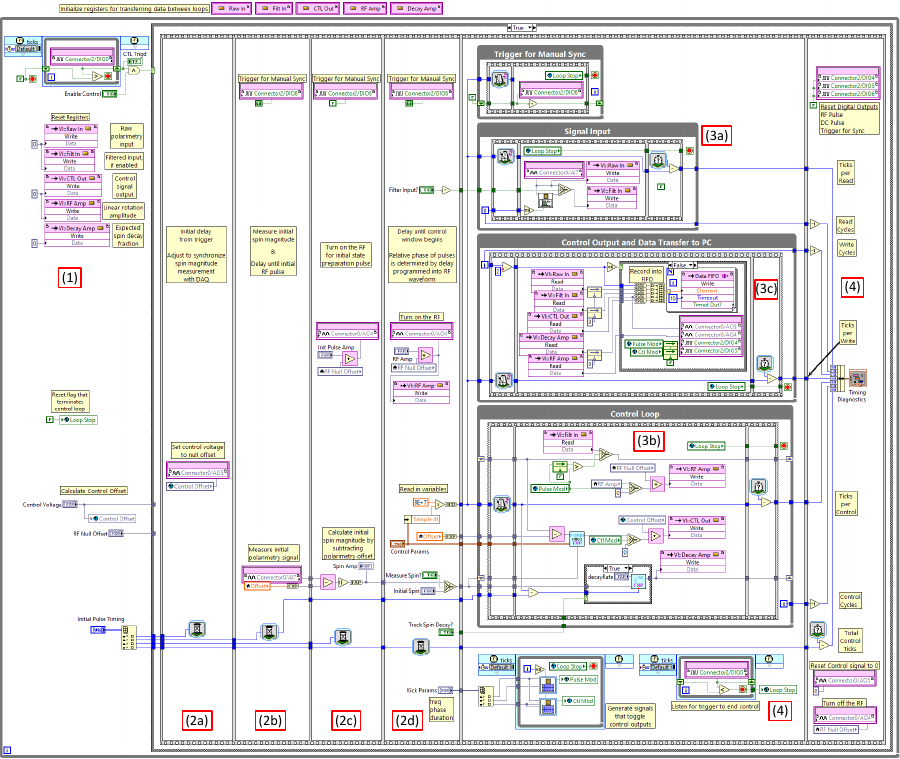}
\caption[LabVIEW block diagram for Kicked Top emulation FPGA program]{LabVIEW block diagram for the Kicked Top emulation FPGA program. The elements of this program are largely the same as in Figure \ref{fig:fig9pa_lv_main}, with three differences. First, a pair of square wave signals are generated in the timed loop near bottom center which toggle the control signals (See Figure \ref{fig:fig5p3_QKTtiming}). Second, the linear rotation $J_{x}$ is also modulated during control by one of the square wave signals. Finally, the nonlinear $J_{z}$ rotation is only updated during a short gap when both toggle signals are LOW. In this way, we execute a fixed rotation angle per step, rather than a target frequency.}  \label{fig:fig9pa_lv_mainKT}
\end{figure}

\begin{figure}[ht]
\renewcommand{\baselinestretch}{1}
\centering
\includegraphics[width=1.0\columnwidth]{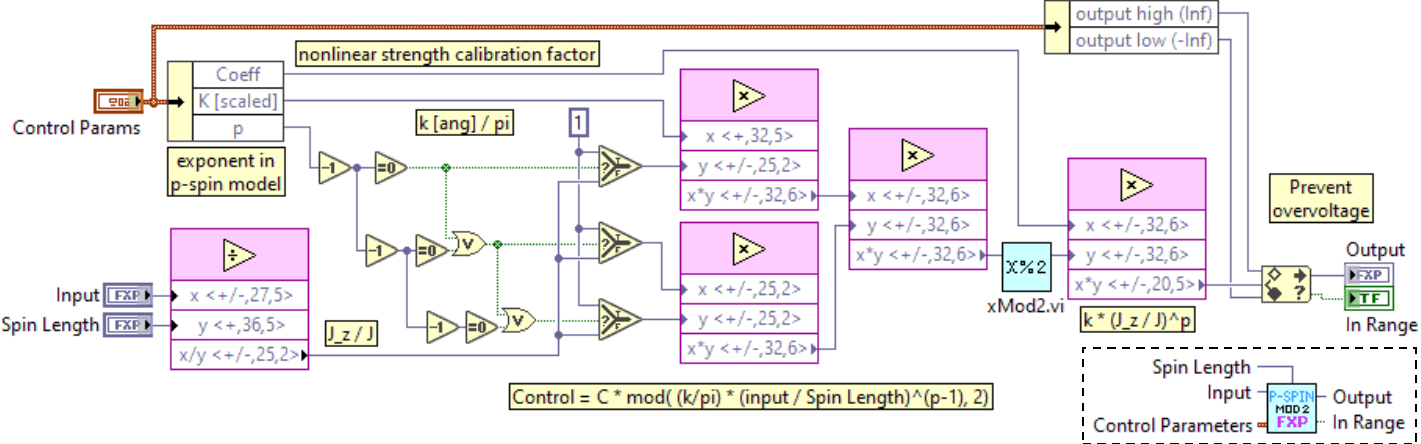}
\caption[LabVIEW block diagram for Kicked Top control law]{LabVIEW block diagram for QKT control law. Again, the program is largely the same as the LMG version (Figure \ref{fig:fig9pa_lv_pSpin}), but it incorporates the binary mod(x,2) operation discussed near the end of Section \ref{chapter:ch5p3} (See Figure \ref{fig:fig9pa_lv_xMod2}) that allows us to rotate through an 'arbitrary' angle with a finite output range. Note that in order to conserve resources, many of the calibration factors are pre-computed in the host computer before being sent to the FPGA. This program's subVI logo is shown in the dashed box.}  \label{fig:fig9pa_lv_pSpinMod}
\end{figure}

\begin{figure}[hb]
\renewcommand{\baselinestretch}{1}
\centering
\includegraphics[width=1.0\columnwidth]{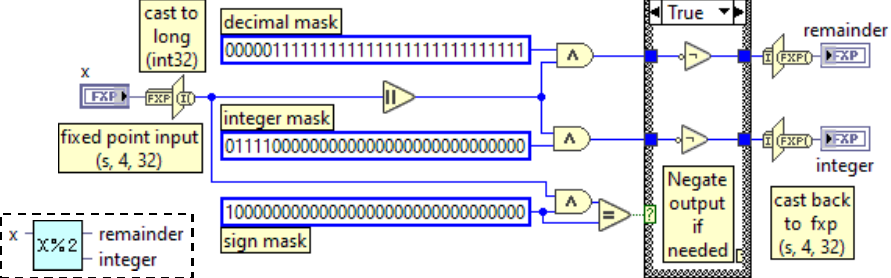}
\caption[LabVIEW block diagram for binary mod 2 operation]{LabVIEW block diagram for binary mod(x,2) operation. The input is a signed 32-bit fixed point number with 4 bits for the integer part. The masks which extract the decimal, integer, and sign parts are designed around this bitness. The final conditional matches the sign of the output to match the input. This program's subVI logo is shown in the dashed box.}  \label{fig:fig9pa_lv_xMod2}
\end{figure}

\begin{figure}[ht]
\renewcommand{\baselinestretch}{1}
\centering
\includegraphics[width=1.0\columnwidth]{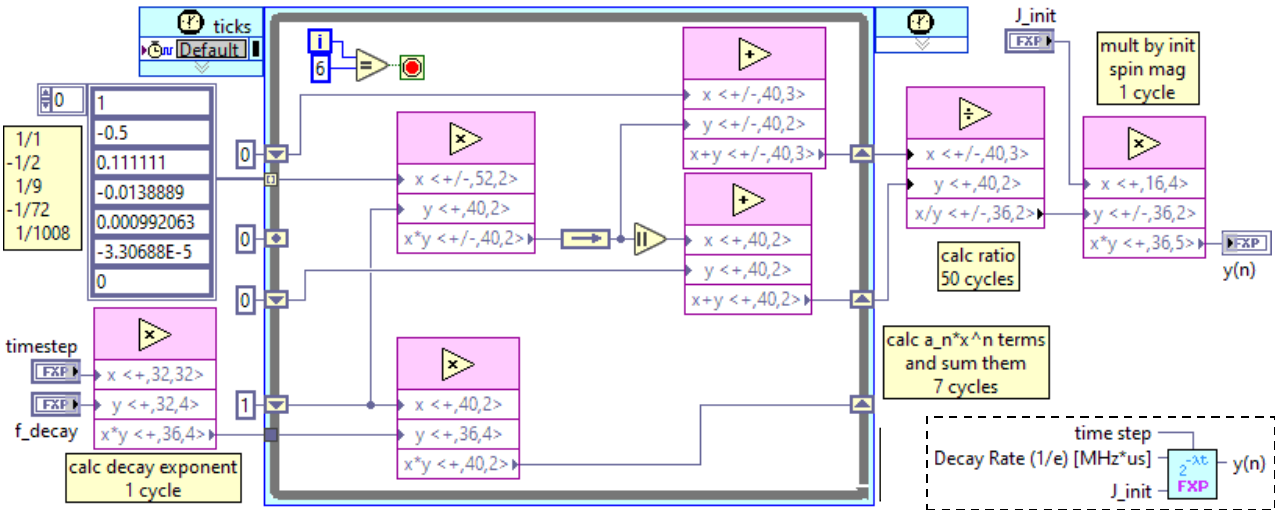}
\caption[LabVIEW block diagram for exponential decay Pad\`{e} approximant]{LabVIEW block diagram for the exponential decay Pad\`{e} approximant of order [5,5]. At each iteration $n \in [0,6]$ of the loop, the value $a_{n} (\lambda t)^{n}$ is calculated and then added to a running total for the numerator and denominator. The ratio is then calculated and scaled by the initial spin length. This program's subVI logo is shown in the dashed box.} \label{fig:fig9pa_lv_expPade}
\end{figure}

\chapter{RF Mixer Schematics}
\label{chapter:rfMixer}
The analog outputs of the USB-7855R RIO device controller are low impedance, but they can only source 2.5 mA of current. The SIM954 amplifiers we use to drive the magnetic coils, on the other hand, have an input impedance of 50 $\Omega$. If we drove the amplifiers directly, the maximum control voltage would be 0.125 V, nowhere near the rated maximum of 10 V. As such, we needed a way to impedance match the two. 

Instead of something simple, like a buffer, we chose to design a four-quadrant analog voltage multiplier so that we could amplitude modulate a carrier signal sourced from the WW2572A arbitrary waveform generator. There were several benefits in doing so, not least of which was the fact that we were already set up to drive the coils from the WW2572A anyways. The WW2572A has a much higher sample rate than the 7855R, up to 100 MHz, meaning we wouldn't have to worry about digitization noise. Also, we could incorporate a variable gain stage into the mixer to match the control voltage range to the maximum expected frequency range for a given set of parameters for any particular model being studied.

Over the next few pages we show the circuit schematics and board layout for the RF mixer. We based the design around the VCA824 variable gain amplifier from Texas Instruments, mainly due to availablility. The VCA824 requires input voltages of $\pm 1$ V, so both source (WW2572A) and control (7855R) inputs are scaled to this range with a voltage divider. The WW2572A maximum output amplitude is $\pm 16$ V, so after scaling this signal is buffered and then clamped with a pair of diodes to prevent accidental damage to the VCA824. The clamping voltage is set with a pair of potentiometers used as voltage dividers. The two signals are mixed in the VCA824, and then amplified with an inverting amplifier stage with gain controlled via an external potentiometer. Finally, the output is buffered to ensure sufficient driving current. There are also several voltage offset adjustment points, needed to make sure the output zero level is accurate across all the entire input range.

\begin{sidewaysfigure}[htp] 
\renewcommand{\baselinestretch}{1}
\centering
\includegraphics[width=1.0\columnwidth]{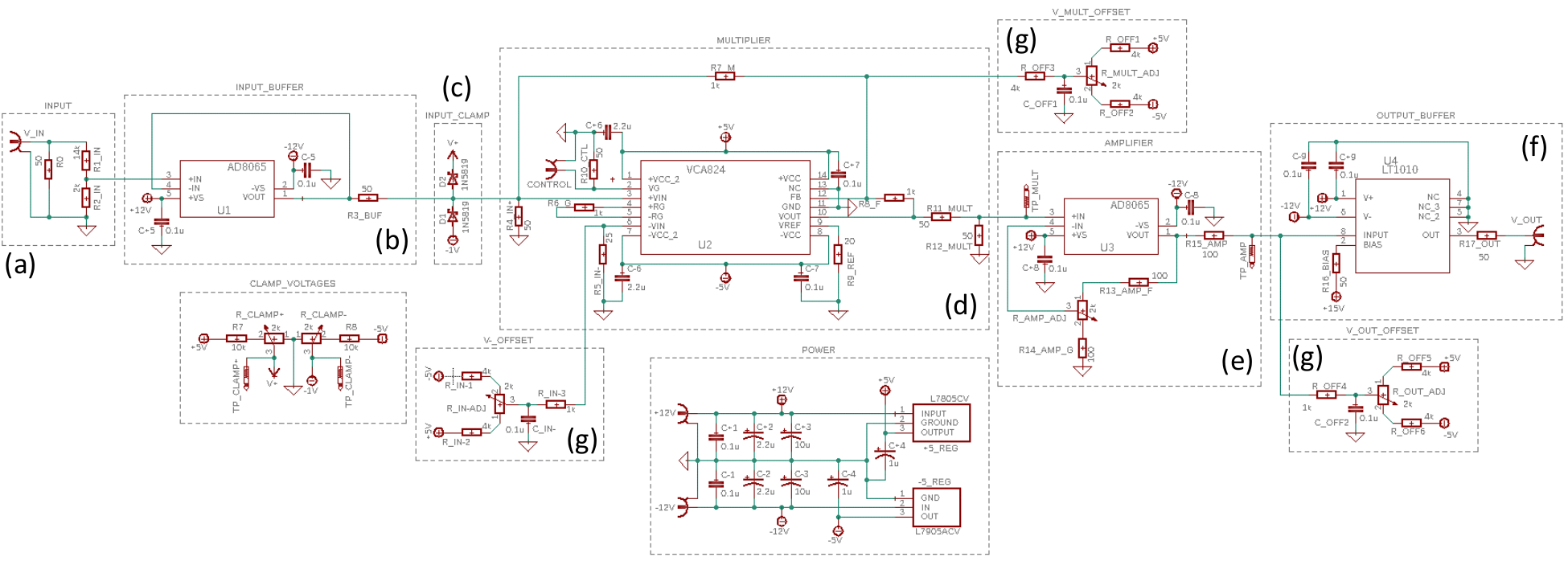}
\caption[Electrical schematic for RF mixer]{Electrical schematic for RF mixer. The carrier signal from the arbitrary waveform generator is scaled to $\pm 1$ V (a), and is then buffered to provide isolation (b). The scaled and buffered carrier signal is clamped to prevent over-voltage damage (c), and is then multiplied with the control signal by the VCA824 variable gain amplifier (d). The modulated output passes through a inverting amplifier with gain determined by an external potentiometer (e), and is then buffered to make sure there is enough output current(f). There are three offset adjustment points (g) for the multiplier inverting input, its output, and the amplifier output, to ensure an accurate zero level across the entire input range.} \label{fig:fig9pb_RFmixer}
\end{sidewaysfigure}

\begin{sidewaysfigure}[htp] 
\renewcommand{\baselinestretch}{1}
\centering
\includegraphics[width=1.0\columnwidth]{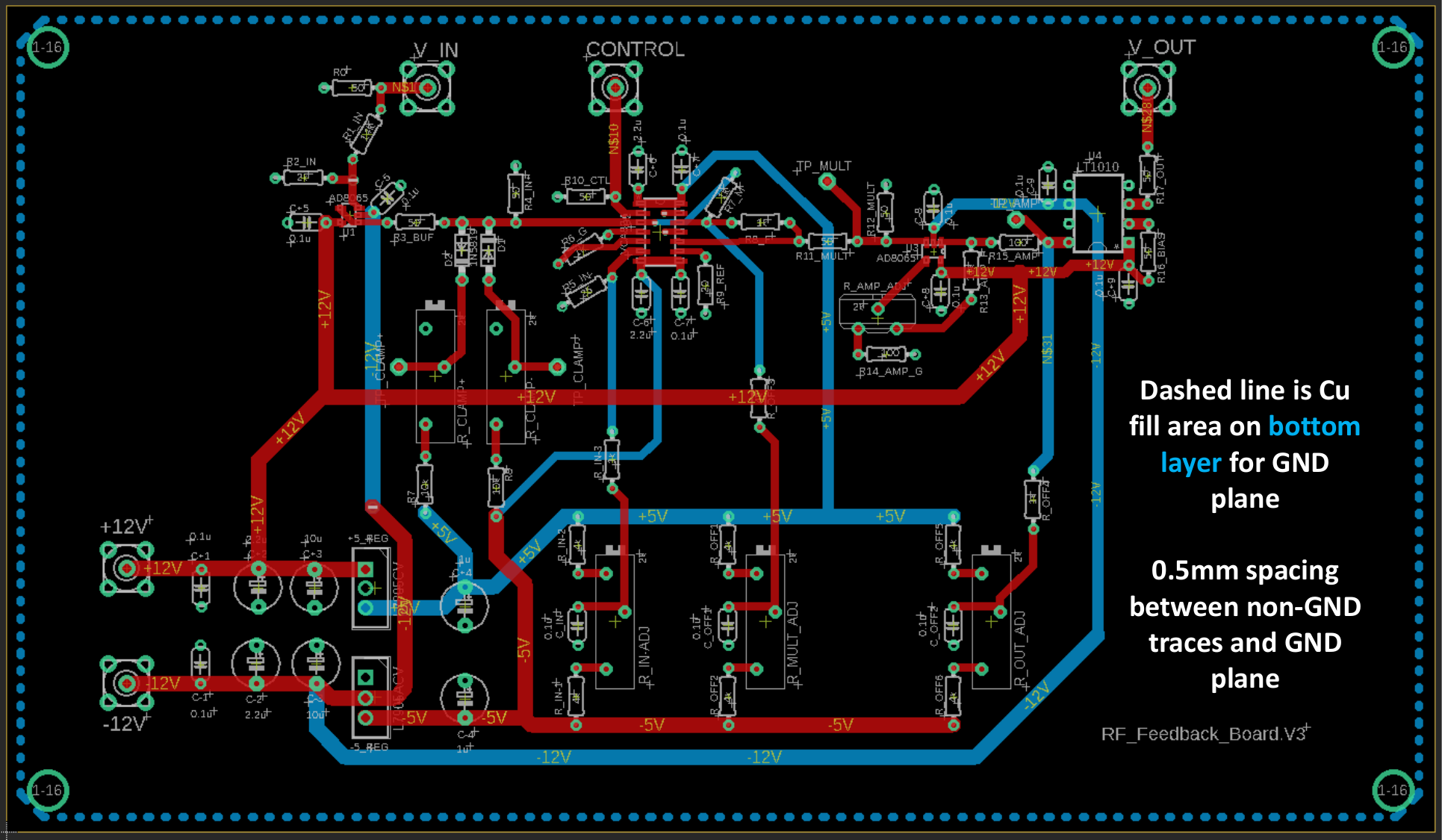}
\caption[Layout for RF mixer circuit board]{Layout for RF mixer circuit board} \label{fig:fig9pb_RFmixerBoard}
\end{sidewaysfigure}

\renewcommand{\baselinestretch}{1}		
\small\normalsize						

\newpage
\bibliographystyle{unsrt}		
\addcontentsline{toc}{chapter}{REFERENCES}
\bibliography{bibliography}

\begin{thebibliography}{10}

\bibitem{Moore1965}
G.E. Moore.
\newblock Cramming more components onto integrated circuits.
\newblock {\em Electronics}, 38:8, 1965.

\bibitem{Waldrop2016}
M.~Mitchell Waldrop.
\newblock The chips are down for moore’s law.
\newblock {\em Nature News}, 530:144–147, Feb 2016.

\bibitem{Kumar2015FundamentalLT}
Suhas Kumar.
\newblock Fundamental limits to moore's law.
\newblock {\em arXiv: Mesoscale and Nanoscale Physics}, 2015.

\bibitem{Feynman1982}
Richard~P. Feynman.
\newblock Simulating physics with computers.
\newblock {\em International Journal of Theoretical Physics}, 21(6):467--488,
  Jun 1982.

\bibitem{Lloyd1996}
Seth Lloyd.
\newblock Universal quantum simulators.
\newblock {\em Science}, 273(5278):1073--1078, 1996.

\bibitem{BrooksNature2023}
Michael Brooks.
\newblock Quantum computers: what are they good for?, May 2023.
\newblock [Online; accessed 23-June-2025].

\bibitem{Lloyd2000}
Seth Lloyd and Jean-Jacques~E. Slotine.
\newblock Quantum feedback with weak measurements.
\newblock {\em Phys. Rev. A}, 62:012307, Jun 2000.

\bibitem{Munoz2020_QMFsim}
Manuel~H. Mu\~noz Arias, Pablo~M. Poggi, Poul~S. Jessen, and Ivan~H. Deutsch.
\newblock Simulating nonlinear dynamics of collective spins via quantum
  measurement and feedback.
\newblock {\em Phys. Rev. Lett.}, 124:110503, Mar 2020.

\bibitem{Haake1987}
R.~Scharf F.~Haake, M.~Ku{\'{s}}.
\newblock Classical and quantum chaos for a kicked top.
\newblock {\em Zeitschrift f{\"u}r Physik B Condensed Matter}, 65(3):381--395,
  Sep 1987.

\bibitem{Lipkin1965}
H.J. Lipkin, N.~Meshkov, and A.J. Glick.
\newblock Validity of many-body approximation methods for a solvable model:
  (i). exact solutions and perturbation theory.
\newblock {\em Nuclear Physics}, 62(2):188 -- 198, 1965.

\bibitem{Sieberer2019}
L.~M. Sieberer, T.~Olsacher, A.~Elben, M.~Heyl, P.~Hauke, F.~Haake, and
  P.~Zoller.
\newblock Digital quantum simulation, trotter errors, and quantum chaos of the
  kicked top.
\newblock {\em npj Quantum Information}, 5(1):78, 2019.

\bibitem{Pavel2018}
Pavel Kos, Marko Ljubotina, and Toma\ifmmode \check{z}\else~\v{z}\fi{} Prosen.
\newblock Many-body quantum chaos: Analytic connection to random matrix theory.
\newblock {\em Phys. Rev. X}, 8:021062, Jun 2018.

\bibitem{Munoz2022_TC}
Manuel~H. Mu\~noz Arias, Karthik Chinni, and Pablo~M. Poggi.
\newblock Floquet time crystals in driven spin systems with all-to-all $p$-body
  interactions.
\newblock {\em Phys. Rev. Res.}, 4:023018, Apr 2022.

\bibitem{Steck2010}
Daniel~A. Steck.
\newblock {``Cesium D Line Data", available online at
  http://steck.us/alkalidata (revision 2.1.4, 23 December 2010)}.

\bibitem{BIPM2018}
BIPM.
\newblock {``SI base unit: second (s)", available online at
  https://www.bipm.org/en/si-base-units/second }.

\bibitem{Sachdev2011}
S.~Sachdev.
\newblock {\em Quantum Phase Transitions, 2nd ed.}
\newblock Cambridge University Press, 2011.

\bibitem{Malomed2016}
Boris~A. Malomed.
\newblock Spontaneous symmetry breaking in nonlinear systems: An overview and a
  simple model.
\newblock In Mustapha Tlidi and Marcel.~G. Clerc, editors, {\em Nonlinear
  Dynamics: Materials, Theory and Experiments}, pages 97--112, Cham, 2016.
  Springer International Publishing.

\bibitem{Else2016}
Dominic~V. Else, Bela Bauer, and Chetan Nayak.
\newblock Floquet time crystals.
\newblock {\em Phys. Rev. Lett.}, 117:090402, Aug 2016.

\bibitem{Chaudhury2009}
S.~Chaudhury, A.~Smith, B.~E. Anderson, S.~Ghose, and P.~S. Jessen.
\newblock Quantum signatures of chaos in a kicked top.
\newblock {\em Nature}, 461(7265):768--771, Oct 2009.

\bibitem{Gubin2012}
Aviva Gubin and Lea F.~Santos.
\newblock Quantum chaos: An introduction via chains of interacting spins 1/2.
\newblock {\em American Journal of Physics}, 80(3):246--251, 03 2012.

\bibitem{Munoz2021_pSpin}
Manuel~H. Mu\~noz Arias, Pablo~M. Poggi, and Ivan~H. Deutsch.
\newblock Nonlinear dynamics and quantum chaos of a family of kicked $p$-spin
  models.
\newblock {\em Phys. Rev. E}, 103:052212, May 2021.

\bibitem{Ray2016}
S.~Ray, A.~Ghosh, and S.~Sinha.
\newblock Quantum signature of chaos and thermalization in the kicked dicke
  model.
\newblock {\em Phys. Rev. E}, 94:032103, Sep 2016.

\bibitem{bajkova2025}
Anisa Bajkova, Anton Smirnov, and Vadim Bobylev.
\newblock Spectral analysis of the orbital dynamics of globular clusters in the
  central region of the milky way, 2025.

\bibitem{kathpalia2023}
Aditi Kathpalia and Nithin Nagaraj.
\newblock Compression spectrum: Where shannon meets fourier, 2023.

\bibitem{Tabor1989}
Michael Tabor.
\newblock {\em Chaos and integrability in Nonlinear Dynamics: An introduction},
  page 145.
\newblock Wiley, 1989.

\bibitem{Jacobs2010}
K.~K.~A. Jacobs.
\newblock {\em Stochastic Processes for Physicists: Understanding Noisy
  Systems}.
\newblock Cambridge University Press, Cambridge, England, 2010.

\bibitem{Vasilyev2012}
D.~V. Vasilyev, K.~Hammerer, N.~Korolev, and A.~S. Sorensen.
\newblock Quantum noise for faraday light--matter interfaces.
\newblock {\em J. Phys. B}, 45(12):124007, 2012.

\bibitem{Kupriyanov2005}
D.~V. Kupriyanov, O.~S. Mishina, I.~M. Sokolov, B.~Julsgaard, and E.~S. Polzik.
\newblock Multimode entanglement of light and atomic ensembles via off-resonant
  coherent forward scattering.
\newblock {\em Physical Review A}, 71(3):032348, 2005.

\bibitem{Smith2003}
Greg~A Smith, Souma Chaudhury, and Poul~S Jessen.
\newblock Faraday spectroscopy in an optical lattice: a continuous probe of
  atom dynamics.
\newblock {\em Journal of Optics B: Quantum and Semiclassical Optics},
  5(4):323, 2003.

\bibitem{Deutsch2010}
I.~H. Deutsch and P.~S. Jessen.
\newblock Quantum control and measurement of atomic spins in polarization
  spectroscopy.
\newblock {\em Opt. Comm.}, 283:681, 2010.

\bibitem{Hammerer2010}
Klemens Hammerer, Anders~S. S\o{}rensen, and Eugene~S. Polzik.
\newblock Quantum interface between light and atomic ensembles.
\newblock {\em Rev. Mod. Phys.}, 82:1041--1093, Apr 2010.

\bibitem{Enrique2015}
Enrique Monta{\~n}o.
\newblock {\em Quantum Control and Squeezing of Collective Spins}.
\newblock PhD thesis, University of Arizona, 2015.

\bibitem{Baragiola2014}
Ben~Q. Baragiola, Leigh~M. Norris, Enrique Monta{\~n}o, Pascal~G. Mickelson,
  Poul~S. Jessen, and Ivan~H. Deutsch.
\newblock Three-dimensional light-matter interface for collective spin
  squeezing in atomic ensembles.
\newblock {\em Physical Review A}, 89(3):033850, 2014.

\bibitem{Raab1987}
E.~L. Raab, M.~Prentiss, Alex Cable, Steven Chu, and D.~E. Pritchard.
\newblock Trapping of neutral sodium atoms with radiation pressure.
\newblock {\em Physical Review Letters}, 59(23):2631--2634, 1987.

\bibitem{Wineland1979}
D.~J. Wineland and Wayne~M. Itano.
\newblock Laser cooling of atoms.
\newblock {\em Physical Review A}, 20(4):1521--1540, 1979.

\bibitem{Migdall1985}
Alan~L. Migdall, John~V. Prodan, William~D. Phillips, Thomas~H. Bergeman, and
  Harold~J. Metcalf.
\newblock First observation of magnetically trapped neutral atoms.
\newblock {\em Physical Review Letters}, 54(24):2596--2599, 1985.

\bibitem{Lett1989}
P.~D. Lett, W.~D. Phillips, S.~L. Rolston, C.~E. Tanner, R.~N. Watts, and C.~I.
  Westbrook.
\newblock Optical molasses.
\newblock {\em J. Opt. Soc. Am. B}, 6(11):2084--2107, Nov 1989.

\bibitem{Kuppens2000}
S.~J.~M. Kuppens, K.~L. Corwin, K.~W. Miller, T.~E. Chupp, and C.~E. Wieman.
\newblock Loading an optical dipole trap.
\newblock {\em Phys. Rev. A}, 62:013406, Jun 2000.

\bibitem{OHara2001}
K.~M. O'Hara, S.~R. Granade, M.~E. Gehm, and J.~E. Thomas.
\newblock Loading dynamics of ${\mathrm{co}}_{2}$ laser traps.
\newblock {\em Phys. Rev. A}, 63:043403, Mar 2001.

\bibitem{SosaThesis2017}
Hector Sosa-Martinez.
\newblock {\em Quantum Control and Quantum Tomography on Neutral Atom Qudits}.
\newblock PhD thesis, The University of Arizona, June 2017.

\bibitem{LeeThesis2012}
Jae~Hoon Lee.
\newblock {\em Sub-Wavelength Resonance Imaging and Addressing of Cesium Atoms
  Trapped in an Optical Lattice}.
\newblock PhD thesis, The University of Arizona, April 2012.

\bibitem{AndersonThesis2013}
Brian~Eric Anderson.
\newblock {\em Unitary Transformations in a Large Hilbert Space}.
\newblock PhD thesis, The University of Arizona, May 2013.

\bibitem{SmithThesis2012}
Aaron~Coleman Smith.
\newblock {\em Quantum Control in the Full Hyperfine Ground Manifold of
  Cesium}.
\newblock PhD thesis, The University of Arizona, March 2012.

\bibitem{SoumaThesis2008}
Souma Chaudhury.
\newblock {\em Quantum Control and Quantum Chaos in Atomic Spin Systems}.
\newblock PhD thesis, The University of Arizona, November 2008.

\bibitem{OThesis2008}
Worawarong Rakreungdet.
\newblock {\em Quantum Information Science with Neutral Atoms}.
\newblock PhD thesis, University of Arizona, April 2008.

\bibitem{Miller1993}
J.~D. Miller, R.~A. Cline, and D.~J. Heinzen.
\newblock Far-off-resonance optical trapping of atoms.
\newblock {\em Phys. Rev. A}, 47:R4567--R4570, Jun 1993.

\bibitem{Takekoshi1996}
T.~Takekoshi and R.~J. Knize.
\newblock Co2 laser trap for cesium atoms.
\newblock {\em Opt. Lett.}, 21(1):77--79, Jan 1996.

\bibitem{Salomon1990}
C.~Salomon, J.~Dalibard, W.~D. Phillips, A.~Clairon, and S.~Guellati.
\newblock Laser cooling of cesium atoms below 3 $\mu$k.
\newblock {\em Europhys. Lett.}, 12(8):683--688, 1990.

\bibitem{Yavin2002ACO}
Itay Yavin, Matthew Weel, A.~Andreyuk, and A.~Kumarakrishnan.
\newblock A calculation of the time-of-flight distribution of trapped atoms.
\newblock {\em American Journal of Physics}, 70:149--152, 2002.

\bibitem{Hagman2009}
H.~Hagman, P.~Sj{\"o}lund, S.~J.~H. Petra, M.~Nyl{\'e}n, A.~Kastberg,
  H.~Ellmann, and J.~Jersblad.
\newblock Assessment of a time-of-flight detection technique for measuring
  small velocities of cold atoms.
\newblock {\em Journal of Applied Physics}, 105(8):--, 2009.

\bibitem{Marechal1998}
E.~Marechal, S.~Guibal, J.~L. Bossennec, M.~P. Gorza, R.~Barbe, J.~C. Keller,
  and O.~Gorceix.
\newblock Longitudinal stern-gerlach effect for slow cesium atoms.
\newblock {\em Eur. Phys. J. D.}, 2(3):195--198, June 1998.

\bibitem{Chormaic1994}
S.~Nic Chormaic, Ch. Miniatura, O.~Gorceix, B.~Viaris de~Lesegno, J.~Robert,
  S.~Feron, V.~Lorent, J.~Reinhardt, J.~Baudon, and K.~Rubin.
\newblock Atomic stern-gerlach interferences with time-dependent magnetic
  fields.
\newblock {\em Physical Review Letters}, 72(1):1--4, 1994.

\bibitem{HemmerThesis2020}
Daniel Hemmer.
\newblock {\em Spin Squeezing and Closed-Loop Magnetometry With A Collective
  Atomic Spin}.
\newblock PhD thesis, The University of Arizona, November 2020.

\bibitem{Hemmer2020}
D.~Hemmer, E.~Monta\~no, B.~Q. Baragiola, L.~M. Norris, E.~Shojaee, I.~H.
  Deutsch, and P.~S. Jessen.
\newblock Squeezing the angular momentum of an ensemble of complex multilevel
  atoms.
\newblock {\em Phys. Rev. A}, 104:023710, Aug 2021.

\bibitem{Levitt1985}
M.~H. Levitt and R.~R. Ernst.
\newblock Multiple‐quantum excitation and spin topology filtration in
  high‐resolution nmr.
\newblock {\em The Journal of Chemical Physics}, 83(7):3297--3310, 1985.

\bibitem{Tycko1985}
R.~Tycko, A.~Pines, and J.~Guckenheimer.
\newblock Fixed point theory of iterative excitation schemes in nmr.
\newblock {\em The Journal of Chemical Physics}, 83(6):2775--2802, 1985.

\bibitem{Levitt1986}
Malcolm~H. Levitt.
\newblock Composite pulses.
\newblock {\em Progress in Nuclear Magnetic Resonance Spectroscopy}, 18(2):61
  -- 122, 1986.

\bibitem{Koschorreck2010}
M.~Koschorreck, M.~Napolitano, B.~Dubost, and M.~W. Mitchell.
\newblock Sub-projection-noise sensitivity in broadband atomic magnetometry.
\newblock {\em Physical Review Letters}, 104(9), 2010.

\bibitem{Munoz2020_pSpinSim}
Manuel~H. Mu\~noz Arias, Ivan~H. Deutsch, Poul~S. Jessen, and Pablo~M. Poggi.
\newblock Simulation of the complex dynamics of mean-field $p$-spin models
  using measurement-based quantum feedback control.
\newblock {\em Phys. Rev. A}, 102:022610, Aug 2020.

\bibitem{Skokos2010}
Ch. Skokos, Jean~J. Souchay, and Rudolf Dvorak.
\newblock {\em The Lyapunov Characteristic Exponents and Their Computation}.
\newblock Springer Berlin Heidelberg, Berlin, Heidelberg, 2010.

\bibitem{Packard1980}
N.H. Packard, J.P. Crutchfield, J.D. Farmer, and R.S. Shaw.
\newblock Geometry from a time series.
\newblock {\em Physical Review Letters}, 45(9):712 – 716, 1980.

\bibitem{Fraser1989}
Andrew~M. Fraser.
\newblock Reconstructing attractors from scalar time series: A comparison of
  singular system and redundancy criteria.
\newblock {\em Physica D: Nonlinear Phenomena}, 34(3):391--404, 1989.

\bibitem{Bryant1990}
Paul Bryant, Reggie Brown, and Henry D.~I. Abarbanel.
\newblock Lyapunov exponents from observed time series.
\newblock {\em Phys. Rev. Lett.}, 65:1523--1526, Sep 1990.

\bibitem{Abarbanel1992}
H.~D.~I. Abarbanel, R.~Brown, and M.~B. Kennel.
\newblock Local lyapunov exponents computed from observed data.
\newblock {\em Journal of Nonlinear Science}, 2(3):343--365, Sep 1992.

\bibitem{TISEAN1999}
T.~Schreiber R.~Hegger, H.~Kantz.
\newblock Practical implementation of nonlinear time series methods: The tisean
  package.
\newblock {\em Chaos: An Interdisciplinary Journal of Nonlinear Science},
  9(2):413--435, Jun 1999.

\bibitem{Kantz_Schreiber_2003}
Holger Kantz and Thomas Schreiber.
\newblock {\em Nonlinear Time Series Analysis}.
\newblock Cambridge University Press, 2 edition, 2003.

\bibitem{Ball2018}
Philip Ball.
\newblock In search of time crystals.
\newblock {\em Physics World}, 31(7):29, jul 2018.

\bibitem{Russomanno2017}
Angelo Russomanno, Fernando Iemini, Marcello Dalmonte, and Rosario Fazio.
\newblock Floquet time crystal in the lipkin-meshkov-glick model.
\newblock {\em Phys. Rev. B}, 95:214307, Jun 2017.

\bibitem{Huang2018}
Biao Huang, Ying-Hai Wu, and W.~Vincent Liu.
\newblock Clean floquet time crystals: Models and realizations in cold atoms.
\newblock {\em Phys. Rev. Lett.}, 120:110603, Mar 2018.

\end{thebibliography}

\end{document}